\documentclass
[twocolumn,superscriptaddress,secnumarabic,amssymb,amsmath,nobibnotes,aps,prd,showpacs,nofootinbib]{revtex4}%
\usepackage{graphicx}
\usepackage{epsf}
\usepackage{multirow}
\usepackage{array}
\usepackage{bm}
\usepackage{booktabs}
\usepackage{amsmath}
\usepackage{amsfonts}
\usepackage{amssymb}
\usepackage{epstopdf}
\usepackage{natbib}
\usepackage{rotating}
\usepackage{pdflscape}
\usepackage{adjustbox}
\usepackage{color}%
\providecommand{\U}[1]{\protect\rule{.1in}{.1in}}
\newcommand{\be}{\begin{equation}}
\newcommand{\ee}{\end{equation}}

\newcommand{\mincir}{\raise
-3.truept\hbox{\rlap{\hbox{$\sim$}}\raise4.truept\hbox{$<$}\ }}
\newcommand{\magcir}{\raise
-3.truept\hbox{\rlap{\hbox{$\sim$}}\raise4.truept\hbox{$>$}\ }}

\begin{document}
\title{Future Constraints on Dynamical Dark-Energy using Gravitational-Wave Standard Sirens}

\author{Minghui Du}
\email{angelbeats@mail.dlut.edu.cn}
\affiliation{Institute of Theoretical Physics, School of Physics, Dalian University of Technology, Dalian, 116024, P. R. China}

\author{Weiqiang Yang}
\email{d11102004@163.com}
\affiliation{Department of Physics, Liaoning Normal University, Dalian, 116029, P. R.
China}

\author{Lixin Xu}
\email{lxxu@dlut.edu.cn}
\affiliation{Institute of Theoretical Physics, School  of Physics, Dalian  University  of  Technology,  Dalian,  116024, P.  R.  China}

\author{Supriya Pan}
\email{supriya.maths@presiuniv.ac.in}
\affiliation{Department of Mathematics, Presidency University, 86/1 College Street, Kolkata 700073, India}

\author{David F. Mota}
\email{mota@astro.uio.no}
\affiliation{Institute of Theoretical Astrophysics, University of Oslo, 0315 Oslo, Norway}


\begin{abstract}

The detection of gravitational waves (GW) by the LIGO and Virgo collaborations offers a whole new range of possible tests and opens up a new window which may shed light on the nature of dark energy and dark matter. In the present work we investigate how future gravitational waves data could help to constrain different dynamical dark energy models. In particular,  we perform cosmological forecastings of a class of well known and  most used dynamical dark energy models using the third-generation gravitational wave detector, the Einstein Telescope. We have considered 1000 simulated GW events in order to constrain the parameter space of the dynamical dark energy models.
Our analyses show that the inclusion of the GW data from the Einstein Telescope, significantly improves the parameter space of the dynamical dark energy models compared to their constraints extracted from the standard cosmological probes, namely, the cosmic microwave observations, baryon acoustic oscillations distance measurements, Supernove type Ia, and the Hubble parameter measurements.
\end{abstract}
\maketitle

%%%%%%%%%%%%%%%%%%%%%%%%%%%%%%%%%%%%%%%%%%%%%%%%%%%%%%%%%%%%%%%%%%
%%%%%%%%%%%%%%%%%%%%%%%%%%%%%%%%%%%%%%%%%%%%%%%%%%%%%%%%%%%%%%%%%%
\section{Introduction}

According to the latest claims by LIGO and Virgo collaborations, the gravitational
waves (GW) from a pair of two very massive black holes around 36 and 29 solar masses have been detected, known as the GW150914 event \cite{ligo01}. Subsequently, the investigations in a series of further works  \cite{Abbott:2016nmj,Gw03,Gw04,Gw05,Gw06} also claimed similar detection. Just after the detection of GW from binary black holes, again GW from a binary neutron star merger (known as GW170817 event \cite{Gw07}) together with an electromagnetic  counterpart known as GRB 170817A event \cite{GRB17} was also detected. Without any doubt, the detection of GW, if we avoid its counter attacks, is an appreciable event for modern cosmology that naturally thrilled the scientific community offering some new insights in the physics of dark energy and modified gravity theories at the fundamental level. Following this a lot of investigations have already been performed by many researchers in order to understand how GW could affect the cosmological theories of interest, see for instance \cite{Baker:2017hug,Creminelli:2017sry,Ezquiaga:2017ekz,Sakstein:2017xjx,DiValentino:2017clw,Oost:2018tcv,Casalino:2018tcd, Zhao:2018gwk, Liu:2018sia,DiValentino:2018jbh,Chakraborty:2017qve,Visinelli:2017bny, Wei:2018cov,Zhang:2018dxi,Ezquiaga:2018btd,Kase:2018aps,Creminelli:2018xsv, Lin:2018ken,Nunes:2018evm,Copeland:2018yuh,Casalino:2018wnc}. One of the most important properties is that GW propagate practically with the light speed, as reported the events GW170817 \cite{Gw07} and  GRB  170817A \cite{GRB17}. Thus, by using the extracted properties from GW, for instance its propagation speed, one can impose strong constraints on the cosmological models as well as exclude some  cosmological theories. Especially, GW data provide a novel approach of luminosity distance measurements, known as standard sirens.

Motivated by the earlier investigations, in the present work, we focus on the dynamical dark energy cosmologies through their parametrizations with an aim to examine how luminosity distances extracted from future GW data could affect the bounds on the aforementioned dark energy models.
The parametrizations of the dark energy sector is a well motivated area in cosmology where the primary content is the dark energy equation of state defined by, $w_x = p_x /\rho_x$, in which $p_x$, $\rho_x$ are respectively the pressure and energy density of the dark energy fluid. We note that in the context of modified gravity theories, such parametrizations can be viewed in terms of an effective dark energy equation of state. Alternatively, using different dark energy equation of state (in the context of Einstein's gravity theory) or effective dark energy equation of state (in the context of modified gravitational theories), one could be able to trace the expansion history of the universe, and test them using the observational data. In this work we consider that the underlying gravitational theory is described by the Einstein's gravity and the large scale structure of our Universe is homogeneous and isotropic, and hence, the geometry of the universe is described by the Friedmann-Lema\^{i}tre-Robertson-Walker metric. Now, concerning the dynamical dark energy parametrizations, we recall numerous parametrizations that have been investigated widely with the available observational data \cite{Chevallier:2000qy,Linder:2002et,Cooray:1999da,Astier:2000as,Weller:2001gf,Efstathiou:1999tm,Jassal:2005qc,Linder:2005ne,Linder:2006xb,Barboza:2008rh,DiValentino:2016hlg,Yang:2017amu,DiValentino:2017zyq,DiValentino:2017gzb,Yang:2017alx,Pan:2017zoh,Vagnozzi:2018jhn, Yang:2018qmz,Yang:2018prh}.
Some well known and most used dark energy parametrizations in this series are, Chevallier-Polarski-Linder (CPL) parametrization \cite{Chevallier:2000qy, Linder:2002et}, Logarithmic parametrization \cite{Efstathiou:1999tm}, Jassal-Bagla-Padmanabhan (JBP) parametrization \cite{Jassal:2005qc}, Barboza-Alcaniz parametrization \cite{Barboza:2008rh}. Here, considering these four well known dark energy parametrzations, namely, CPL, Logarithmic, JBP and BA, we perform a robust analysis by constraining their parameter space using the simulated GW data from the Einstein Telescope along with the standard astronomical probes such as cosmic microwave background radiation (CMB) \cite{Adam:2015rua, Aghanim:2015xee}, baryon acoustic oscillations (BAO) \cite{Beutler:2011hx,Ross:2014qpa,Gil-Marin:2015nqa}, Supernove Type Ia (SNIa) \cite{Betoule:2014frx} and Hubble parameter measurements from the cosmic chronometers (CC) \cite{Moresco:2016mzx}, in order to see how the data from GW improve the parameter space of these known parametrizations compared to their usual cosmological constraints availed from the known cosmological probes, CMB, BAO, SNIa and CC. We refer to some earlier works on dark energy with similar motivation, that means where the simulated GW data from the Einstein Telescope were taken into account \cite{Zhao:2010sz,Cai:2016sby,Zhang:2017sym,Cai:2017aea, Wang:2018lun,Zhang:2018byx}. We mention that it will be also interesting to use simulated GW data from other observatories like Laser Interferometer Space Antenna (LISA) \cite{Audley:2017drz}, Deci-hertz Interferometer Gravitational wave Observatory (DECIGO) \cite{Kawamura:2011zz,Sato:2017dkf}, TianQin \cite{Luo:2015ght}. However, in the present work we mainly concentrate, how  GW data from one particular source, namely from the Einstein Telescope, could affect {\it a class of well known and most used dynamical DE parametrizations}. One can equally apply other GW sources to a specific model in order to compare their constraining power.

However, apart from the GW data, a number of upcoming cosmological surveys, such as, Simons Observatory Collaboration (SOC) \cite{Ade:2018sbj}, 
Cosmic Microwave Background Stage-4 (CMB-S4) \cite{Abitbol:2017nao}, EUCLID Collaboration \cite{Scaramella:2015rra,Laureijs:2011gra}, 
Dark Energy Spectroscopic Instrument  (DESI) \cite{Aghamousa:2016zmz}, Large Synoptic Survey Telescope (LSST) \cite{Newman:2019doi,Hlozek:2019vjs,Mandelbaum:2019zej}, are all dedicated to explore more about the nature of the dark sector of our universe and to provide more precise constraints on the dark energy equation of state. So, it is expected that the upcoming cosmological surveys mentioned above will play a crucial role to understand the physics of the dark universe.
Along the similar lines, it is also important to understand the constraining power of different surveys by investigating the improvements of the cosmological parameters. This will enable us to understand how Einstein Telescope and other GW observatories perform with respect to other cosmological surveys, such as, SOC, CMB-S4, EUCLID, etc. Thus, for a better conclusion about the constraining power between the cosmological surveys, it is important to apply all of them on a specific cosmological model. Such an investigation is truly important in the context of cosmological physics. A systematic and dedicated analysis of the dynamical dark energy models taking all the future cosmological surveys mentioned above, is the subject of a forthcoming work.

The work has been structured in the following way. In section \ref{sec2} we briefly introduce the  background and perturbative evolutions for any dark energy parametrization as well as we introduce the parametrizations of our interest. After that in section \ref{sec-gw} we describe the method to simulate the GW data from the Einstein Telescope and show how to use the simulated GW data in order to constrain an underlying theory. In section \ref{sec-data} we introduce the standard astronomical probes as well as the methodology for constraining the model parameters. Then in section \ref{sec-results} we discuss the results of our analyses. Finally, we close the work in section \ref{sec-discuss} with a brief summary of all the results obtained.

%--------------------------------------------------------------------------------

\section{Dynamical Dark energy}
\label{sec2}

In this section we shall describe the general evolution laws of a dynamical dark energy component at the level of background and perturbations.

It is well known that at large scale, our Universe is perfectly homogeneous and isotropic. Such geometrical description of our Universe is characterized by the Friedmann-Lema\^{i}tre-Robertson-Walker (FLRW) line element given by

\begin{eqnarray}
d{\rm s}^2 = -dt^2 + a^2 (t) \left[\frac{dr^2}{1-kr^2} + r^2 \left(d \theta^2 + \sin^2 \theta d \phi^2 \right) \right],
\end{eqnarray}
where $a (t)$ (hereafter we shall denote it simply by $a$) is the expansion scale factor of the Universe and $k$ is the curvature scalar. For $k =0$, $+1$, $-1$, three different geometries, namely, the spatially flat, closed and the open Universe are described.  Further, we assume  that the gravitational sector of the Universe is described by the Einstein's general theory of relativity where the total matter sector of the Universe is minimally coupled to the Einstein gravity. This total matter sector comes from  radiation, baryons, pressureless dark matter and dark energy. Thus, with the above information, one can explicitly write down the Einstein's field equations as

\begin{eqnarray}
H^2 + \frac{k}{a^2} &=& \frac{8\pi G}{3} \rho_{tot}~, \label{efe1}\\
2\dot{H} + 3 H^2 + \frac{k}{a^2} &=& - 8 \pi G\, p_{tot}~, \label{efe2}
\end{eqnarray}
where $H \equiv \dot{a}/a$ is the Hubble factor of the FLRW Universe;
$\rho_{\rm tot} = \rho_r +\rho_b +\rho_c +\rho_x$, is the total energy density
of the Universe and $p_{\rm tot} = p_r + p_b + p_c + p_x$, is the total pressure coming from the individual fluid. Let us note that here 
$\rho_i$ ($i=r, b, c, x$) and $p_i$ are respectively the energy density and the pressure of the $i$-th component
where the subscripts $r$, $b$, $c$, $x$ respectively correspond to radiation, baryons, cold dark matter and the dark energy sector. Now, using the Bianchi's identity, the conservation law for the total fluid follows,
\begin{eqnarray}\label{tot-cons}
\dot{\rho}_{\rm tot} + 3 H (\rho_{\rm tot} +p_{\rm tot}) = 0~.
\end{eqnarray}
One can easily find that the conservation equation (\ref{tot-cons}) can be obtained if we simply use the field equations (\ref{efe1}) and (\ref{efe2}).
Since we do not have any interaction between the fluids, thus, the conservation equation of each fluid follows the evolution

\begin{eqnarray}\label{cons}
\dot{\rho}_i + 3 H (p_i +\rho_i) = 0 \Leftrightarrow \dot{\rho}_i + 3 H (1 +w_i ) \rho_i = 0~,
\end{eqnarray}
where $w_i = p_i/\rho_i$ is the equation of state of the $i$-th fluid and it takes $1/3$, $0$, $0$ for radiation, baryons and cold dark matter.
The equation of state of the dark energy fluid is unknown and in this work we consider that $w_x$ has a dynamical character and henceforth we shall consider some particular expressions for it.
We make a final comment regarding the geometrical shape of the Universe. As from the  observational sources, the Universe is almost flat \cite{Ade:2015xua}, and henceforth, throughout the present work we shall assume $k = 0$ in the Einstein's field equations (\ref{efe1}) and (\ref{efe2}). Now, let us get back to the the conservation equation (\ref{cons}), from which one can solve the evolution equations for the governing matter components.
In particular, the evolution of the dark energy fluid can be written in terms of its energy density as

\begin{eqnarray}\label{de-evol}
\rho_{x}=\rho_{x,0}\,\left(  \frac{a}{a_{0}}\right)  ^{-3}\,\exp\left(
-3\int_{a_{0}}^{a}\frac{w_{x}\left(  a'\right)  }{a'}\,da'
\right),
\end{eqnarray}
where $\rho_{x,0}$ is the present value of the dark energy density $\rho_x$, and here
$a_0$ is the present value of the scale factor where $1+z = a_0/a$.
Without any loss of generality we set the present value of the scale factor to be unity, that means, $a_0 =1$. Thus, with the above set of equations, for any prescribed dark energy equation of state, in principle, it is possible to determine the background evolution of the Universe.

However, at the same time, it is important to understand the behaviour of the model at the level of perturbations since that enables us to understand the formation of structure of the Universe.

Thus, in order to investigate the cosmological perturbations, we consider the perturbed FLRW metric that takes the following expression

\begin{eqnarray}
\label{perturbed-metric}
ds^2 = a^2(\tau) \left [-d\tau^2 + (\delta_{ij}+h_{ij}) dx^idx^j  \right],
\end{eqnarray}
where $\tau$ is the conformal time and the quantities
$\delta_{ij}$,  $h_{ij}$ respectively denote
the unperturbed and the perturbated metric tensors.
Now, for the above perturbed metric (\ref{perturbed-metric}),
one can conveniently write the Einstein's equations either in the
conformal Newtonian gauge or in the synchronous gauge in the Fourier
space $\kappa$. We choose the synchronous gauge and thus
using the energy-momentum
balance equation $T^{\mu \nu}_{; \nu}= 0$,
for the $i$-th fluid the continuity and the Euler  equations
for a mode can be written as  \cite{Mukhanov, Ma:1995ey,
Malik:2008im}

\begin{eqnarray}
\delta'_{i}  = - (1+ w_{i})\, \left(\theta_{i}+ \frac{h'}{2}\right) -
3\mathcal{H}\left(\frac{\delta p_i}{\delta \rho_i} - w_{i} \right)\delta_i \nonumber\\- 9
\mathcal{H}^2\left(\frac{\delta p_i}{\delta \rho_i} - c^2_{a,i} \right) (1+w_i)
\frac{\theta_i}
{{\kappa}^2}, \label{per1} \\
\theta'_{i}  = - \mathcal{H} \left(1- 3 \frac{\delta p_i}{\delta
\rho_i}\right)\theta_{i}
+ \frac{\delta p_i/\delta \rho_i}{1+w_{i}}\, {\kappa}^2\, \delta_{i}
-{\kappa}^2\sigma_i,\label{per2}
\end{eqnarray}
where any prime associated with each variable denotes the differentiation
with respect to the conformal time $\tau$; $\delta_i = \delta \rho_i/\rho_i$
is the density perturbation for the $i$-th fluid;
$\mathcal{H}= a^{\prime}/a$, is the conformal
Hubble factor; $h = h^{j}_{j}$ is the trace of $h_{ij}$, and
$\theta_{i}\equiv i \kappa^{j} v_{j}$ is the divergence of the $i$-th fluid
velocity. The quantity $c_{a,i}^2 = \dot{p}_i/\dot{\rho}_i$ denotes
the adiabatic sound speed of the $i$-th fluid whereas $c^2_{s} = \delta p_i / \delta \rho_i$, is the physical sound speed related with other as $c^2_{a,i} =  w_i - \frac{w_i^{\prime}}{3\mathcal{H}(1+w_i)}$.
Finally, we note that $\sigma_i$ is the anisotropic stress of the $i$-th fluid, however, we shall neglect its contribution for its minimal contribution as reported by some recent observational data \cite{Yang:2018ubt}.

Now, we close this section by enlisting  the dark energy parametrizations that we wish to study in this work. We consider four well known DE parametrizations as follows. The first one is the Chevallier-Polarski-Linder model \cite{Chevallier:2000qy, Linder:2002et} having the following expression

\begin{eqnarray}\label{model-cpl}
w_x(z) = w_0 + w_a \frac{z}{1+z}
\end{eqnarray}
where $w_0$ is the present value of $w_x (z)$ and $w_a = dw_x (z)/dz$  at $z=0$, is another free parameter of this model.

As a second model, we consider the Logarithmic parametrization introduced by G. Efstathiou \cite{Efstathiou:1999tm}
\begin{eqnarray}\label{model-log}
w_x (z) = w_0 + w_a \ln (1+z)
\end{eqnarray}
where $w_0$ and $w_a$ parameters have the same meanings as described for the CPL parametrization.

We then consider another DE parametrization widely known as the Jassal-Bagla-Padmanabhan (JBP) parametrization \cite{Jassal:2005qc}

\begin{eqnarray}\label{model-jbp}
w_x (z) = w_0 + w_a \frac{z}{(1+z)^2}
\end{eqnarray}
and here too, $w_0$ and $w_a$ parameters have the same meanings as described for the above two models, namely CPL and Logarithmic.

Finally, we end up with  the Barboza-Alcaniz parametrization
\cite{Barboza:2008rh}
\begin{eqnarray}\label{model-ba}
w_x (z) = w_0 + w_a \frac{z (1+z)}{1+z^2}
\end{eqnarray}
where $w_0$, $w_a$ have the same meanings as described above for other DE parametrizations.

\section{Method of simulating GW data and its use}
\label{sec-gw}

In this section we shall describe the method for
simulating the Gravitational Waves Standard Sirens (GWSS) data, each data point of which consists of $(z, d_{{L}} (z), \sigma_{d_{L}})$ of a GW source, where $d_{L} (z)$ is the luminosity distance at the redshift $z$ and $\sigma_{d_{L}}$ is the associated error with $d_{L} (z)$. The constraining ability of this catalogue, together with other astronomical datasets, is further investigated in various cosmological models, for instance \cite{Zhao:2010sz,Cai:2016sby,Wang:2018lun}. The simulation of GW data is model dependent, thus one needs to choose the fiducial values of model parameters. In this paper, each set of parameters used in the GW simulation is decided by other observational data under a specific cosmological model, and after that the aforementioned GW data as well as the real data from different observational sources are combined to constrain the same model. This procedure has been followed in section \ref{sec-results}.

The initial step to generate the GWSS data is performed by simulating the redshift distribution of the sources.
In this paper we assume the redshifts of all observed GW sources are available. Practically, this is achieved by employing techniques such as identifying the Electromagnetic counterparts. Our interest is focused on GW events originate from 2 types of binary systems: the binary system of a Black Hole (BH) and a Neutron Star (NS) identified as BHNS as well as binary neutron star, named as BNS.

Following some earlier works in this direction \cite{Zhao:2010sz,Cai:2016sby,Wang:2018lun}, the redshift distribution of the observable sources is given by

\begin{equation}
P(z)\propto \frac{4\pi d_C^2(z)R(z)}{H(z)(1+z)},
\label{equa:pz}
\end{equation}
where $d_C(z)$ represents the comoving distance at the redshift $z$;
$R(z)$ is the merger rate of binary system (BHNS or BNS) with the fitting form
\cite{Cai:2016sby,Schneider:2000sg,Cutler:2009qv}

\begin{equation}
R(z)=\begin{cases}
1+2z, & z\leq 1, \\
\frac{3}{4}(5-z), & 1<z<5, \\
0, & z\geq 5.
\end{cases}
\label{equa:rz}
\end{equation}

Based on the prediction of the Advanced LIGO-Virgo network, the detailed configuration of our simulation is as follows. The ratio between observed BHNS and BNS events is set to be $0.03$, which makes BNS the overwhelming majority of GW sources. By roughly considering the mass distribution of the astrophysical objects NS and BH, we perform random sampling of their masses from uniform distributions $U(M_{\odot}, 2 M_{\odot})$ and $U(3 M_{\odot}, 10 M_{\odot})$ respectively, with $M_{\odot}$ being one solar mass.
For more details, we refer to \cite{Cai:2016sby,Wang:2018lun}.

Thus, according to the redshift and mass distribution described above, the catalogue of the GWSS data can be easily obtained through the introduction of the fiducial model which could be any well motivated cosmological model.
Now, for the spatially flat Universe, technically, one could find the expression for $H(z)$ for the concerned cosmological model and consequently, the
luminosity distance $d_L (z)$ of the GW sources can now be calculated through the relation

\begin{equation}
d_L (z) = {(1 + z)}\int_0^z {\frac{{dz'}}{{H(z')}}}.
\label{equa:dl}
\end{equation}
Hence, the mean luminosity distances of all the GW sources can be generated using Eq.~(\ref{equa:dl}).
That means, the $d_{L} (z)$ vs. $z$ relation can be obtained for every GW event for the concerned cosmological model which as mentioned could be any well motivated cosmological model. Although in some earlier works, $\Lambda$CDM has been considered to the fiducial model, in a similar fashion, instead of the $\Lambda$CDM model, one may fix some other dark energy models to generate the simulated GW data, since there is no such strict rule to select the $\Lambda$CDM model as the fiducial one. In this work we have not fixed $\Lambda$CDM as the fiducial model which is usually done (for instance, see \cite{Zhao:2010sz,Cai:2016sby}), rather we have considered the dynamical DE models as the fiducial models. We shall describe about it later in more detail.

Now, while measuring the luminosity distance of the GW source, certainly, one needs to calculate the associated error which we denote by $\sigma_{d_{L}}$.
In order to calculate this error, one needs the expression of GW signal, i.e., the strain of GW interferometers. Note that, since the GW amplitude relies on $d_{L} (z)$, one can extract the information regarding $d_{L} (z)$ once other parameters (e.g. masses of the binary system, etc.) are evaluated from the waveform.
This is the reason why the GW events are often referred to as the standard sirens, analogous to the Supernovae Type Ia standard candles.
As a consequence, the error of GW detection (given in terms of GW SNR) is passed to $\sigma_{d_{L} (z)}$ via Fisher matrix.

In the following we enter into the main part of this section where we describe the strain of GW interferometers.

Considering the transverse-traceless (TT) gauge, the strain $h(t)$ in the GW interferometers can be given by  \cite{Cai:2016sby,Wang:2018lun}
\begin{equation*}
h(t)=F_+(\theta, \phi, \psi)h_+(t)+F_\times(\theta, \phi, \psi)h_\times(t),
\end{equation*}
where $F_{+}$ and $F_{\times}$ are the beam pattern functions of the Einstein Telescope; $\psi$ is the polarization angle; the angles $\theta$, $\phi$ effectively describe the location of the GW source with respect to the GW detector (here Einstein Telescope); $h_{+} = h_{xx} = -h_{-yy}$, $h_{\times} = h_{xy} = h_{yx}$ (two independent components of the GW's tensor $h_{\mu \nu}$ in the transverse-traceless (TT) gauge), see the details here \cite{Cai:2016sby}. Now, one can write down the antenna pattern functions of the Einstein Telescope as \cite{Zhao:2010sz,Cai:2016sby,Wang:2018lun}

\begin{align}
F_+^{(1)}(\theta, \phi, \psi)=&~~\frac{{\sqrt 3 }}{2}\Bigl[\frac{1}{2}(1 + {\cos ^2}(\theta ))\cos (2\phi )\cos (2\psi ) \nonumber\\
                              &~~- \cos (\theta )\sin (2\phi )\sin (2\psi )\Bigr],\nonumber\\
F_\times^{(1)}(\theta, \phi, \psi)=&~~\frac{{\sqrt 3 }}{2}\Bigl[\frac{1}{2}(1 + {\cos ^2}(\theta ))\cos (2\phi )\sin (2\psi ) \nonumber\\
                              &~~+ \cos (\theta )\sin (2\phi )\cos (2\psi )\Bigr].\nonumber
\end{align}

For the remaining two interferometers, their antenna pattern functions can be derived using above equations for $F_+^{(1)}(\theta, \phi, \psi)$ and $F_\times^{(1)}(\theta, \phi, \psi)$ and substituting $\phi$ by $\phi+ 120^\circ$ or $\phi+240^\circ$, because the three interferometers form an equilateral triangle, and hence, they make $60^\circ$ with each other.

Then, we follow the works of \cite{Zhao:2010sz,Li:2013lza} to derive the Fourier transform $\mathcal{H}(f)$ of the time domain waveform $h(t)$ considering the stationary phase approximation that leads to, 
$\mathcal{H}(f)=\mathcal{A}f^{-7/6}\exp[i(2\pi ft_0-\pi/4+2\psi(f/2)-\varphi_{(2.0)})],$
where $\mathcal{A}$ is the Fourier amplitude having the following expression
\begin{align}
\mathcal{A}=&~~\frac{1}{d_L}\sqrt{F_+^2(1+\cos^2(\omega))^2+4F_\times^2\cos^2(\omega)}\nonumber\\
            &~\times \sqrt{5\pi/96}\pi^{-7/6}\mathcal{M}_c^{5/6},\nonumber
\end{align}
in which $\mathcal{M}_c$ is dubbed as the ``chirp mass'' which is related to the total mass $M$ of the coalescing binary system ($M = m_1+m_2$; here the component masses are $m_1$, $m_2$) and the symmetric mass ratio (SMR) $\eta=m_1 m_2/M^2$, by the relation $\mathcal{M}_c =M \eta^{3/5}$.
Note that the masses here are actually the observed masses, which is related to the intrinsic masses as $M_{\rm obs}=(1+z)M_{\rm int}$, exhibiting an enhancement of a factor $(1 + z)$.

Furthermore, in the expression for $\mathcal{A}$,
the symbol $\omega$ denotes the angle of inclination of the binary's orbital angular momentum with the line of sight. Since the short gamma ray bursts (SGRBs) are usually expected to be strongly beamed, the coincidence observations of SGRBs suggest that the binaries should be aligned in such a way so that  $\omega \simeq 0$ with its maximal inclination about $\omega=20^\circ$.
Here, we make a comment that averaging the Fisher matrix over the inclination (i.e., $\omega$) and the polarization (i.e., $\psi$) under the constraint $\omega <90^\circ$ is almost (roughly) the same as setting $\omega=0$, considered in the simulation of \cite{Li:2013lza}.
Thus, during the simulation of the GW sources, one can safely consider $\omega=0$. However, during the estimation of the practical uncertainty of $d_L$, the uncertainty of inclination should be considered positively.

When the waveform of GW are known, one can calculate the signal-to-noise ratio (SNR). The SNR plays a very crucial role in detection of the GW event, because a GW detection is confirmed if the combined SNR of at least $8$ is found in the Einstein Telescope \cite{ET,Sathyaprakash:2012jk} (see also ~\cite{Cai:2016sby,Zhao:2010sz,Cai:2017aea,Yang:2017bkv} for more details in this direction). In general, the combined SNR for the network employing three independent interferometers (just like in the Einstein Telescope) is: $\rho=\sqrt{\sum\limits_{i=1}^{3}(\rho^{(i)})^2},$
where $\rho^{(i)}=\sqrt{\left\langle \mathcal{H}^{(i)},\mathcal{H}^{(i)}\right\rangle}$, and the inner product inside the square root follows \cite{Zhao:2010sz,Cai:2016sby,Wang:2018lun}
\begin{equation}
\left\langle{a,b}\right\rangle=4\int_{f_{\rm lower}}^{f_{\rm upper}}\frac{\tilde a(f)\tilde b^\ast(f)+\tilde a^\ast(f)\tilde b(f)}{2}\frac{df}{S_h(f)},
\label{euqa:product}
\end{equation}
where the sign ``$\sim$'' placed over the symbols
denotes their Fourier transformations and $S_h(f)$ is the one-side noise power spectral density which for this article is taken to be the same as in \cite{Zhao:2010sz}.

Now, the instrumental error (following Fisher matrix approach) on $d_{L}$ can be estimated through the relation
\begin{align}\label{error}
\sigma_{d_L}^{\rm inst}\simeq \sqrt{\left\langle\frac{\partial \mathcal H}{\partial d_L},\frac{\partial \mathcal H}{\partial d_L}\right\rangle^{-1}}.
\end{align}
Assuming that $d_{L}$ is independent of other parameters, 
and using the relation
$\mathcal H \propto d_L^{-1}$, from (\ref{error}), one can deduce that,
$\sigma_{d_L}^{\rm inst}\simeq d_L/\rho$. Now, 
when we estimate the uncertainty of the mesurement $d_L$, we should take into account the inclination $\omega$. At the same time we must consider the correlation between $d_L$ and $\omega$. While taking into account such correlation, the maximal effect of the inclination on the SNR which is a factor of $2$ (between $\omega=0$ and $\omega=90^\circ$) is considered.  Now, in order to provide with an estimation of the ability of the GWSS to constrain the cosmological parameters, we double the estimation of the error imposed on the luminosity distance that goes as \cite{Li:2013lza}:  $\sigma_{d_L}^{\rm inst}\simeq \frac{2d_L}{\rho}$. Moreover, 
under the short-wave approximation, GW are lensed in the same way as electromagnetic waves during propagation, resulting in an additional weak lensing error which is modeled as $\sigma_{d_L}^{\rm lens}$ = $0.05z d_L$ in \cite{Cai:2016sby}. Consequently, the combined error is
$\sigma_{d_L} = \sqrt{(\sigma_{d_L}^{\rm inst})^2+(\sigma_{d_L}^{\rm lens})^2}$, where
the errors $\sigma_{d_L}^{\rm inst}$ and  $\sigma_{d_L}^{\rm lens}$, are already defined above.

Thus, following the method described above, 
one is now able to generate the future GWSS dataset consisting of
($z$, $d_{L} (z)$, $\sigma_{d_{L}(z)}$). As argued in \cite{Cai:2016sby},
the constraining ability of Planck on cosmological parameters can only be reached with at least 1000 GW events, corresponding to 10 years of observation by the Einstein Telescope.therefore, data of 1000 GW events are mocked in this work.

Finally, we come to the last part of this section where we describe the approach to
use the simulated GW data. The analysis with GW data is similar to the standard cosmological probes.  For the GW standard siren measurements with $N$ simulated data points, the  $\chi^2$ function is given by
\begin{align}
\chi_{\rm GW}^2=\sum\limits_{i=1}^{N}\left[\frac{\bar{d}_L^i-d_L(\bar{z}_i;\vec{\Theta})}{\bar{\sigma}_{d_L}^i}\right]^2,
\label{equa:chi2}
\end{align}
where $\bar{z}_i$, $\bar{d}_L^i$, and $\bar{\sigma}_{d_L}^i$ are respectively the $i$-th redshift, luminosity distance at this redshift, and the error of the luminosity distance of the simulated GW data for this particular redshift. Here, $\vec{\Theta}$ represents the set of cosmological parameters that we need to constrain.

We conclude this section with the following remark which we believe to be important in the context of simulating GW data. We notice that different models of GW sources have been proposed in earlier investigations, e.g. \cite{Cusin:2018rsq, Dvorkin:2016okx, Chen:2018rzo}, and the types or distributions of GW sources may vary from model to model. While, for the purpose of this paper, what really matters is their impact on the observables, especially the error of luminosity distance. A detailed investigation regarding the merger of astrophysical binary systems is reported in Ref. \cite{Chen:2018rzo}, where the exact expressions for NS/BH merger rates are derived based on the physical process of star formation, and the current abundances of binary systems are normalized by the constraint from observational data \cite{Abbott:2017vtc,TheLIGOScientific:2017qsa}. For the sake of cross-check, we have also conducted a simulation following their approach and assume the same form of lensing error (this part is not considered in \cite{Chen:2018rzo}). It turns out that the resulting ${\delta d_{\rm L}}$ of these two methods are quite close, as is shown in Fig. \ref{error}. And this is partly because the contribution of gravitational lensing (green line of Fig. \ref{error}) takes majority in the whole error budget (yellow and blue lines of of Fig. \ref{error}), that means, $\sigma_{d_{\rm L}}^{\rm lens} > \sigma_{d_{\rm L}}^{\rm inst}$. Thus, the validity of our approach is confirmed. 

\begin{figure}
	\includegraphics[width=0.9\linewidth]{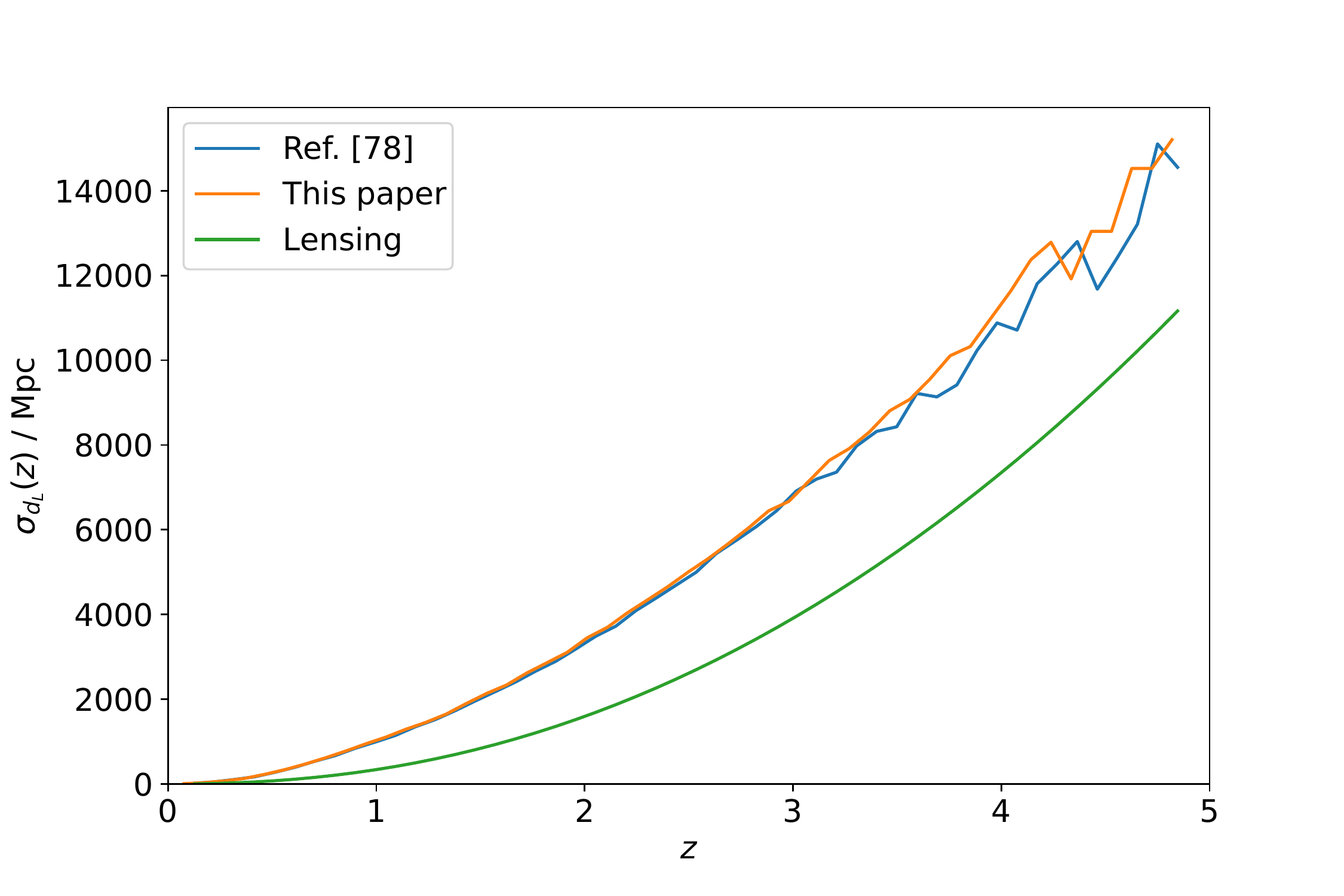}
	\caption{\textit{Error of luminosity distance based on two sets of simulations. It is clear from the figure that the results of these two methods are quite close. In the plot we also show the contribution of gravitational lensing, which takes majority in the whole error budget, i.e. $\sigma_{d_{\rm L}}^{\rm lens} > \sigma_{d_{\rm L}}^{\rm inst}$. }}
	\label{error}
\end{figure}

\begin{figure*}%[h]
\includegraphics[width=0.45\textwidth]{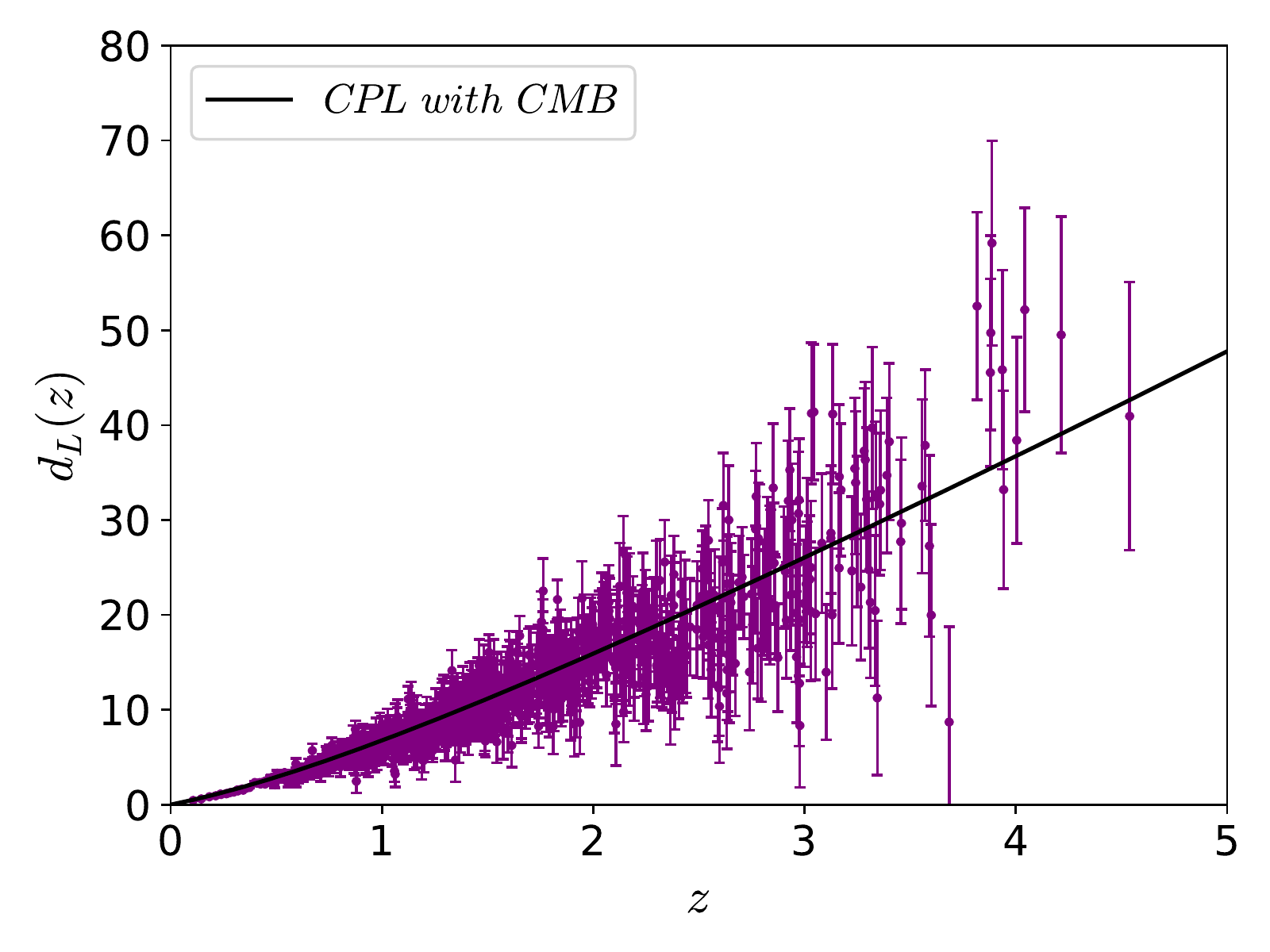}
\includegraphics[width=0.45\textwidth]{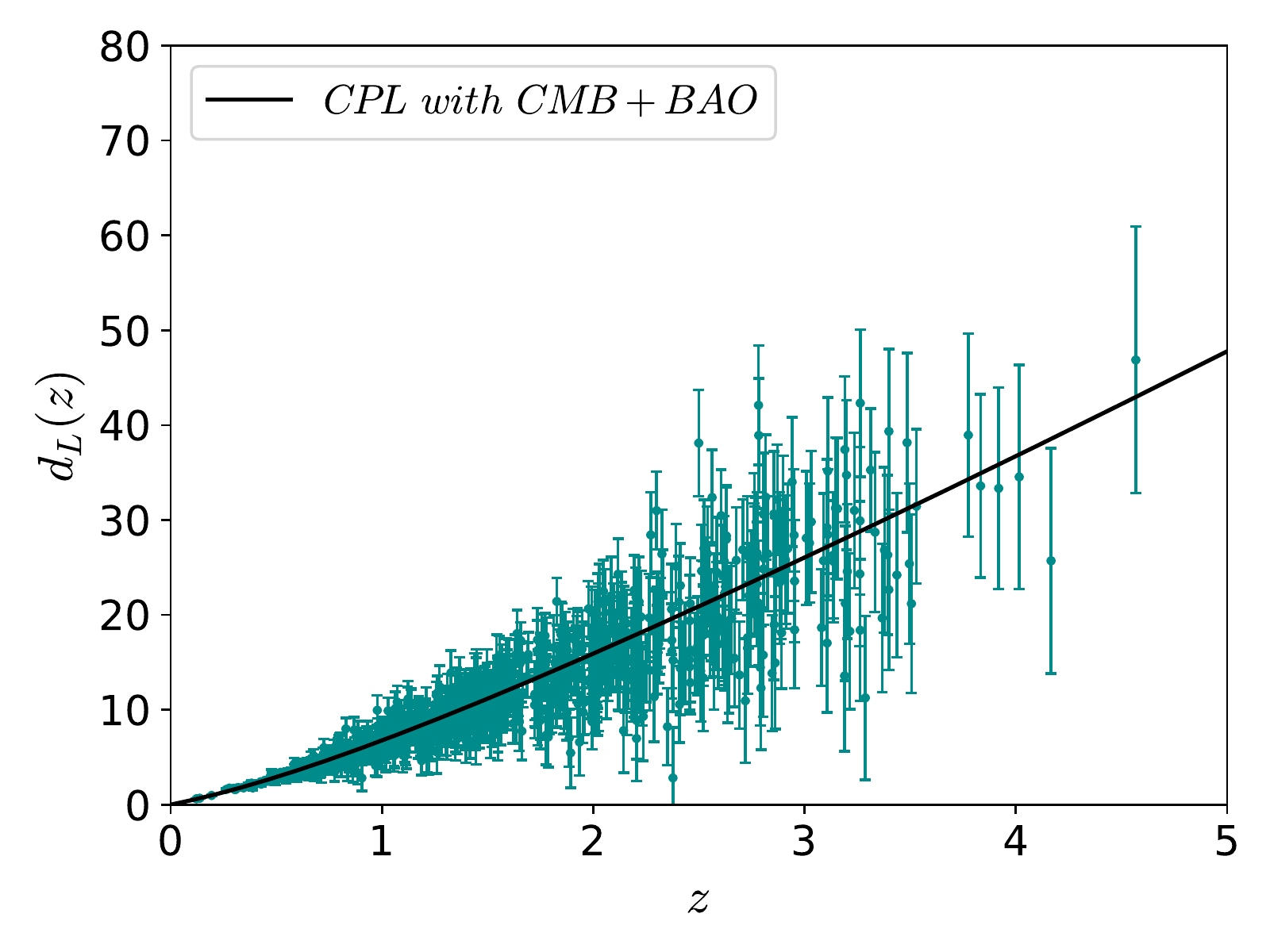}
\includegraphics[width=0.45\textwidth]{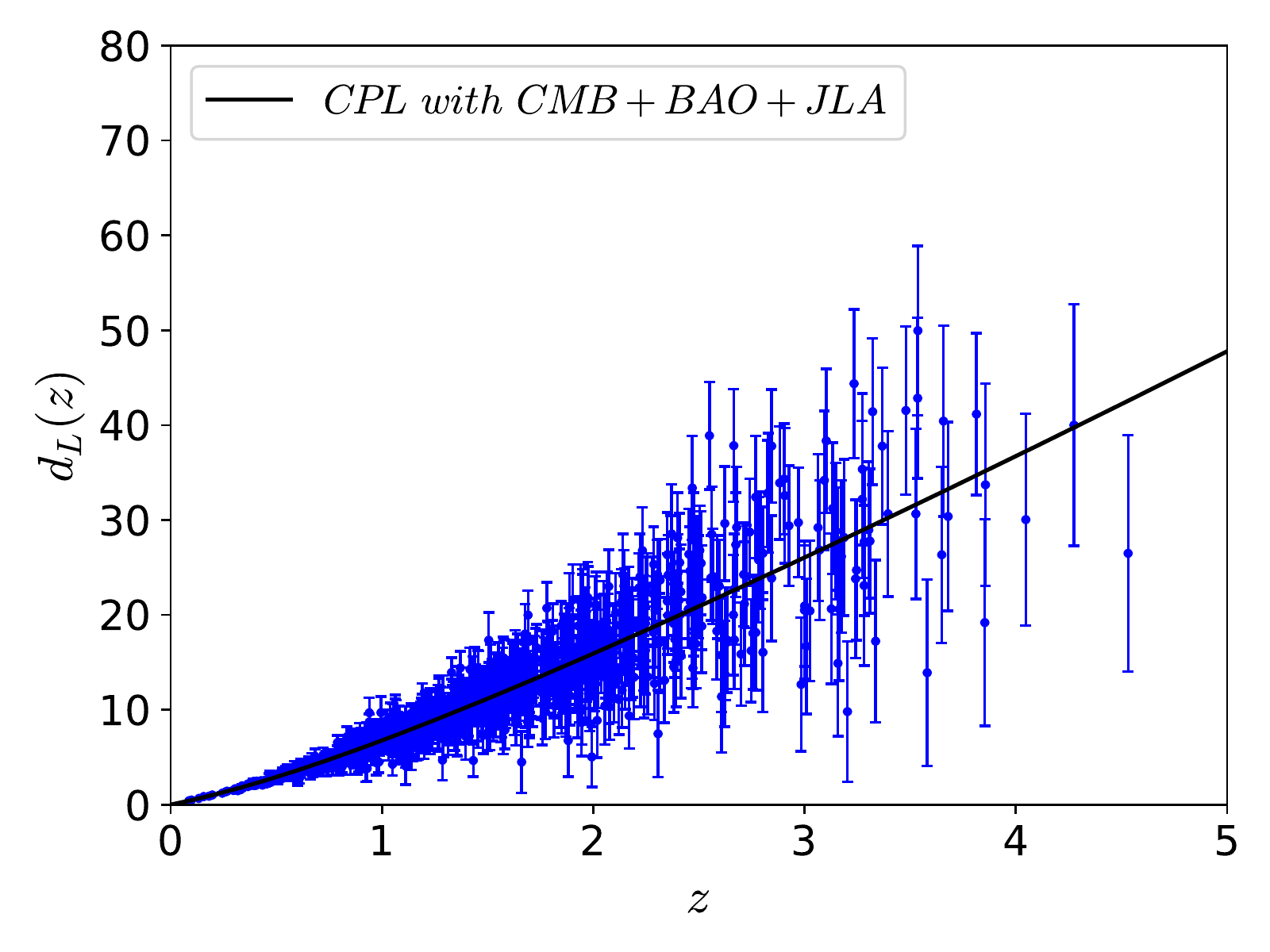}
\includegraphics[width=0.45\textwidth]{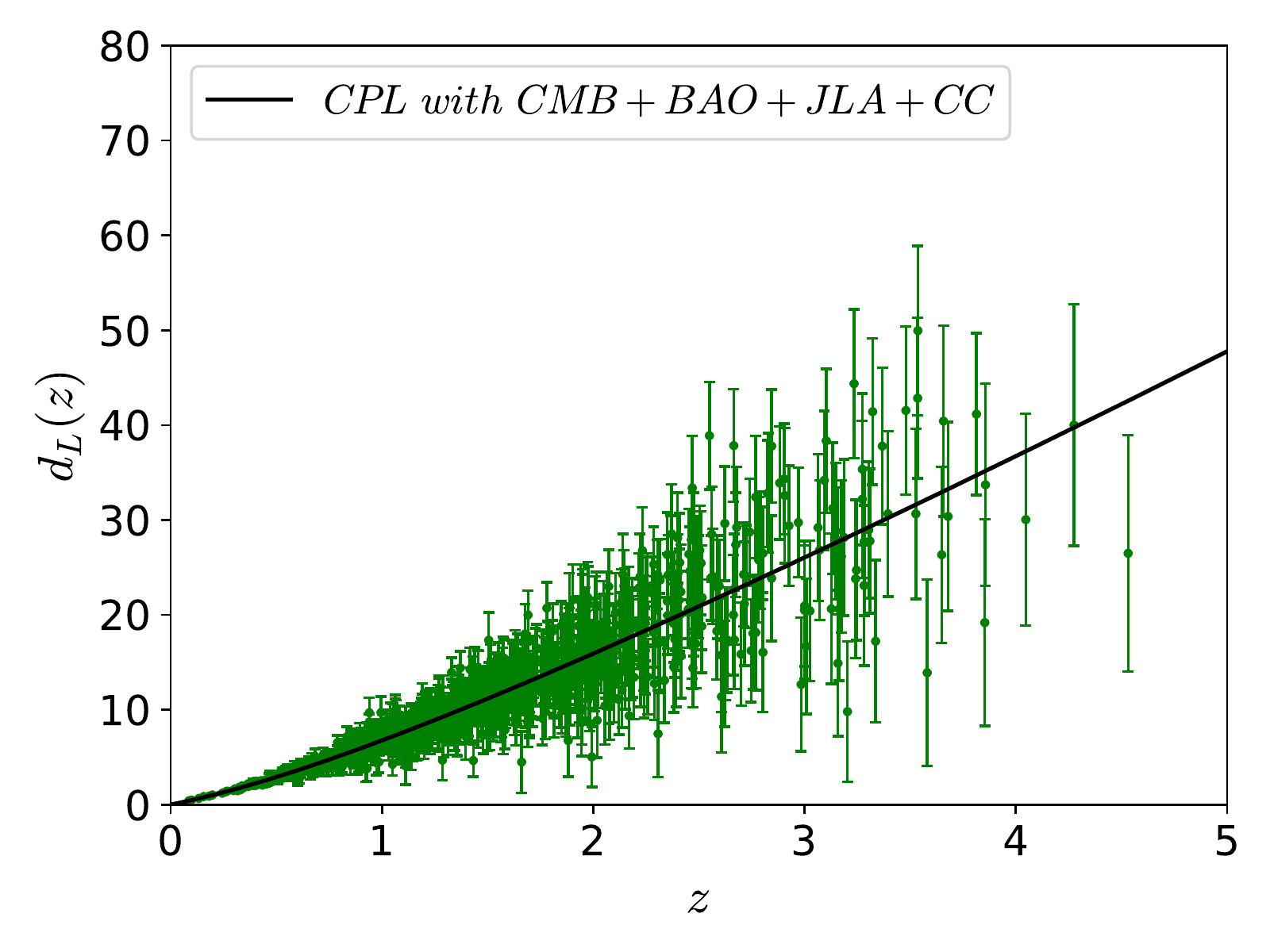}
\caption{\textit{For the fiducial CPL model, we first constrain the cosmological parameters using the datasets CMB, CMB+BAO, CMB+BAO+JLA and CMB+BAO+JLA+CC and then we use the best-fit values of the parameters for ``each dataset'' to generate the corresponding GW catalogue. Following this, in each panel we show $d_L (z)$ vs $z$ catalogue with the corresponding error bars for 1000 simulated GW events. The upper left and upper right panels respectively present the catalogue ($z$, $d_L (z)$) with the corresponding error bars for 1000 simulated events derived using the CMB alone and CMB+BAO dataset. The lower left and lower right panels respectively present the catalogue ($z$, $d_L (z)$) with the corresponding error bars for 1000 simulated events derived using the CMB+BAO+JLA and CMB+BAO+JLA+CC datasets.}}
\label{fig-dL-cpl}
\end{figure*}

\section{Standard Cosmological Probes and the total likelihood analysis including GW}
\label{sec-data}

Here we summarize the standard observational data used to analyze the models. In the following we outline a brief desctiption for each dataset.

\begin{enumerate}

\item \textit{Cosmic Microwave Background (CMB) data:} The cosmic microwave background radiation is an important cosmological data to analyze the dark energy models. In particular, we use the Planck 2015 measurements \cite{Adam:2015rua, Aghanim:2015xee}
 that include the high- and low- $\ell$TT likelihoods in the mutiple range
$ 2 \leq \ell \leq 2508$ as well as the high- and low- $\ell$ polarization likelihoods. The entire data set is identified as Planck TTTEEE+lowTEB.

\item \textit{Baryon Acoustic Oscillations (BAO) data:} In
this work we employ four distinct BAO data measured by
different observational surveys. Precisely, we take the
(i) 6dF Galaxy Survey (6dFGS) measurement at
$z_{\emph{\emph{eff}}}=0.106$
\cite{Beutler:2011hx}, (ii) the Main Galaxy Sample of Data Release 7 of Sloan
Digital Sky Survey (SDSS-MGS) at $z_{\emph{\emph{eff}}}=0.15$
\cite{Ross:2014qpa}, (iii) the CMASS sample from the latest Data
Release 12 (DR12) of the Baryon Oscillation Spectroscopic Survey (BOSS) at
$z_{\mathrm{eff}}=0.57$   and finally (iv) the LOWZ sample from BOSS DR12 at $z_{\mathrm{eff}}=0.32$ \cite{Gil-Marin:2015nqa}.

\item \textit{Supernovae Type Ia:} The joint light curve sample \cite{Betoule:2014frx}
from Supernovae Type Ia (SNIa) data scattered in the redshift region $z \in [0.01, 1.30]$ have been considered. The total number of SNIa in this region is 740.

\item \textit{Cosmic Chronometers:} We also add the Hubble parameter measurements from the cosmic chronometers. The cosmic chronometers are the most massive and passively evolving galaxies. The measurements of the Hubble parameters from the cosmic chronometers are promising to estimate the cosmological parameters, see \cite{Moresco:2016mzx}. The total number of Hubble data points we consider in this analysis is thirty distributed in the redshift region $0 < z < 2$.

\end{enumerate}

Now in order to extract the observational constraints on the proposed dynamical DE parametrizations for several combinations of the cosmological datasets, we use an efficient package, namely the  Markov Chain Monte Carlo package \textit{cosmomc} \cite{Lewis:2002ah, Lewis:2013hha} which is equipped with the well known
convergence statistic by Gelman-Rubin \cite{Gelman-Rubin}. The \textit{cosmomc} package also includes the support for Planck 2015 Likelihood Code \cite{Aghanim:2015xee}. One can avail this code from the website \url{http://cosmologist.info/cosmomc/.}, and it is freely available. 
The parameter space that we shall constrain in this work is as follows: 
\begin{align}
\mathcal{P} \equiv\Bigl\{\Omega_{b}h^2, \Omega_{c}h^2, 100\theta_{MC}, \tau, n_{s}, log[10^{10}A_{s}], w_0, w_a \Bigr\},\nonumber
\end{align}
where $\Omega_{b} h^2$, $\Omega_{c}h^2$ are respectively the physical density for baryons and cold dark matter; $\theta_{MC}$ is the ratio of sound horizon to the angular diameter distance; $\tau$ refers to the reionization optical depth; $n_{s}$ is the scalar spectral index; $A_s$ is the amplitude of the primordial scalar power spectrum; $w_0$, $w_a$ are the key parameters of all the DE parametrizations. In Table \ref{priors} we describe the flat priors on the cosmological parameters used during the analysis of the models.

\begin{table}[ht]
\begin{center}
\begin{tabular}{|c|c|}
\hline
Parameter                    & Prior\\
\hline
$\Omega_{\rm b} h^2$         & $[0.005,0.1]$\\
$\Omega_{\rm c} h^2$                           & $[0.01, 0.99]$\\
$\tau$                       & $[0.01,0.8]$\\
$n_s$                        & $[0.5, 1.5]$\\
$\log[10^{10}A_{s}]$         & $[2.4,4]$\\
$100\theta_{MC}$             & $[0.5,10]$\\
$w_0$                        & $[-2, 0]$\\
$w_a$                        & $[-3, 3]$\\
\hline
\end{tabular}
\end{center}
\caption{The flat priors on various cosmological parameters used for constraining the dynamical dark energy models. }
\label{priors}
\end{table}

\section{Results and analysis}
\label{sec-results}

Let us now summarize the main observational results extracted from the dynamical DE models (\ref{model-cpl}), (\ref{model-log}), (\ref{model-jbp}) and (\ref{model-ba})
after the inclusion of the simulated GW data. In the following we describe the results for each model in detail.

\begingroup
\squeezetable
\begin{center}
\begin{table*}
\begin{tabular}{ccccccccc}
\hline\hline
Parameters & CMB & CMB+BAO & CMB+BAO+JLA & CMB+BAO+JLA+CC &\\ \hline

$\Omega_c h^2$ & $    0.1190_{-    0.0014-    0.0027}^{+    0.0014+    0.0027}$ & $    0.1191_{-    0.0013-    0.0027}^{+    0.0014+    0.0026}$ & $    0.1191_{-    0.0013-    0.0026}^{+    0.0013+    0.0025}$ & $    0.1190_{-    0.0013-    0.0025}^{+    0.0013+    0.0024}$ & \\

$\Omega_b h^2$ & $    0.02228_{-    0.00016-    0.00031}^{+    0.00015+    0.00031}$ & $    0.02226_{-    0.00015-    0.00029}^{+    0.00015+    0.00029}$ & $    0.02226_{-    0.00014-    0.00030}^{+    0.00014+    0.00030}$ & $    0.02228_{-    0.00016-    0.00029}^{+    0.00014+    0.00030}$ & \\

$100\theta_{MC}$ & $    1.04081_{-    0.00032-    0.00064}^{+    0.00032+    0.00062}$
& $    1.04078_{-    0.00032-    0.00064}^{+    0.00033+    0.00063}$ &  $    1.04079_{-    0.00032-    0.00063}^{+    0.00032+    0.00063}$ & $    1.04081_{-    0.00032-    0.00063}^{+    0.00033+    0.00063}$ & \\

$\tau$ & $    0.075_{-    0.017-    0.034}^{+    0.018+    0.034}$ & $    0.078_{-    0.017-    0.034}^{+    0.017+    0.034}$ & $    0.080_{-    0.017-    0.034}^{+    0.017+    0.034}$ & $ 0.081_{-    0.017-    0.034}^{+    0.017+    0.033}$ & \\

$n_s$ & $    0.9667_{-    0.0044-    0.0087}^{+    0.0044+    0.0089}$  & $    0.9665_{-    0.0044-    0.0084}^{+    0.0044+    0.0091}$ & $    0.9666_{-    0.0044-    0.0089}^{+    0.0045+    0.0088}$ &  $ 0.9665_{-    0.0043-    0.0082}^{+    0.0043+    0.0085}$ & \\

${\rm{ln}}(10^{10} A_s)$ & $    3.083_{-    0.034-    0.068}^{+    0.035+    0.066}$ & $    3.090_{-    0.033-    0.066}^{+    0.034+    0.066}$ & $    3.092_{-    0.033-    0.067}^{+    0.033+    0.066}$ & $    3.094_{-    0.033-    0.065}^{+    0.033+    0.066}$ & \\

$w_0$ & $   -1.218_{-    0.597-    0.782}^{+    0.302+    0.856}$ & $   -0.524_{-    0.236-    0.514}^{+    0.374+    0.524}$  & $   -0.909_{-    0.123-    0.201}^{+    0.095+    0.216}$ &  $   -0.909_{-    0.116-    0.209}^{+    0.099+    0.213}$ & \\

$w_a$ &   $< -0.446 < 0.526$ & $   -1.403_{-    1.021-    1.466}^{+    0.731+    1.570}$ & $   -0.409_{-    0.277-    0.777}^{+    0.517+    0.689}$ &  $   -0.399_{-    0.297-    0.724}^{+    0.423+    0.676}$ & \\

$\Omega_{m0}$ & $    0.218_{-    0.081-    0.097}^{+    0.028+    0.146}$ & $    0.344_{-    0.026-    0.054}^{+    0.032+    0.051}$ & $    0.308_{-    0.011-    0.019}^{+    0.009+    0.020}$ &  $    0.308_{-    0.011-    0.019}^{+    0.010+    0.020}$ & \\

$\sigma_8$ & $    0.960_{-    0.065-    0.185}^{+    0.118+    0.152}$ & $    0.803_{-    0.030-    0.051}^{+    0.024+    0.053}$ & $    0.835_{-    0.017-    0.035}^{+    0.018+    0.035}$ &  $    0.835_{-    0.017-    0.033}^{+    0.017+    0.034}$ & \\

$H_0$ & $   83.06_{-    7.98-   21.61}^{+   15.10+   18.40}$ & $   64.36_{-    3.23-    4.67}^{+    2.05+    5.26}$ &  $   67.94_{-    1.08-    2.05}^{+    1.09+    2.10}$ &  $   67.92_{-    1.09-    2.10}^{+    1.09+    2.14}$ & \\

\hline
\hline
\end{tabular}
\begin{tabular}{ccccccccc}

Parameters & CMB+GW & CMB+BAO+GW & CMB+BAO+JLA+GW & CMB+BAO+JLA+CC+GW & \\ \hline

$\Omega_c h^2$ & $    0.1186_{-    0.0012-    0.0024}^{+    0.0012+    0.0024}$ & $    0.1188_{-    0.0013-    0.0025}^{+    0.0013+    0.0025}$ & $    0.1189_{-    0.0012-    0.0023}^{+    0.0012+    0.0024}$ & $    0.1188_{-    0.0013-    0.0025}^{+    0.0013 +    0.0025}$\\

$\Omega_b h^2$ & $    0.02233_{-    0.00014-    0.00027}^{+    0.00014+    0.00028}$ & $    0.02231_{-    0.00015-    0.00030}^{+    0.00015+    0.00028}$ & $    0.02226_{-    0.00016-    0.00030}^{+    0.00015+    0.00030}$  & $    0.02231_{-    0.00015-    0.00029}^{+    0.00015+    0.00029}$\\

$100\theta_{MC}$ & $    1.04088_{-    0.00030-    0.00062}^{+    0.00031+    0.00060}$ & $    1.04088_{-    0.00032-    0.00063}^{+    0.00032+    0.00062}$ & $    1.04079_{-    0.00032-    0.00063}^{+    0.00032+    0.00061}$ & $    1.04088_{-    0.00030-    0.00061}^{+    0.00031+    0.00061}$\\

$\tau$ & $    0.079_{-    0.017-    0.033}^{+    0.017+    0.033}$ & $    0.081_{-    0.017-    0.033}^{+    0.017+    0.034}$ & $    0.081_{-    0.017-    0.0315}^{+    0.017+    0.034}$ & $    0.082_{-    0.017-    0.034}^{+    0.018+    0.034}$ \\

$n_s$ & $    0.9677_{-    0.0042-    0.0081}^{+    0.0041+    0.0082}$ & $    0.9675_{-    0.0043-    0.0087}^{+    0.0043+    0.0086}$ & $    0.9670_{-    0.0041-    0.0082}^{+    0.0041+    0.0078}$ & $    0.9675_{-    0.0042-    0.0086}^{+    0.0043+    0.0086}$ \\

${\rm{ln}}(10^{10} A_s)$ & $    3.089_{-    0.033-    0.067}^{+    0.034+    0.064}$ & $    3.093_{-    0.033-    0.064}^{+    0.034+    0.065}$ & $    3.094_{-    0.033-    0.062}^{+    0.033+    0.067}$ & $    3.096_{-    0.034-    0.066}^{+    0.035+    0.066}$ \\

$w_0$ & $   -1.168_{-    0.212-    0.361}^{+    0.180+    0.385}$ & $   -0.465_{-    0.200-    0.360}^{+    0.189+    0.359}$ & $   -0.904_{-    0.080-    0.144}^{+    0.070+    0.155}$ & $   -0.902_{-    0.062-    0.124}^{+    0.064+    0.124}$ \\

$w_a$ & $   -1.081_{-    0.640-    1.558}^{+    0.842+    1.303}$ & $   -1.523_{-    0.562-    1.160}^{+    0.642+    1.071}$ & $   -0.256_{-    0.227-    0.523}^{+    0.263+    0.549}$  & $   -0.373_{-    0.226-    0.500}^{+    0.263+    0.451}$ \\

$\Omega_{m0}$ & $    0.218_{-    0.010-    0.019}^{+    0.010+    0.020}$ & $    0.349_{-    0.016-    0.031}^{+    0.017+    0.031}$ & $    0.318_{-    0.006-    0.012}^{+    0.006+    0.012}$ & $    0.309_{-    0.004-    0.009}^{+    0.004+    0.009}$ \\

$\sigma_8$ & $    0.945_{-    0.022-    0.040}^{+    0.020+    0.043}$ & $    0.797_{-    0.019-    0.033}^{+    0.017+    0.036}$ & $    0.822_{-    0.015-    0.027}^{+    0.015+    0.029}$ & $    0.831_{-    0.015-    0.029}^{+    0.015+    0.029}$ \\

$H_0$ & $   80.75_{-    1.92-    3.37}^{+    1.71+    3.68}$ & $   63.77_{-    1.52-    2.77}^{+    1.37+    2.80}$ & $   66.98_{-    0.55-    1.10}^{+    0.55+    1.12}$ & $   67.72_{-    0.35-    0.71}^{+    0.36+    0.71}$ \\

\hline\hline
\end{tabular}
\caption{68\% and 95\% CL constraints on the Chevallier-Polarski-Linder parametrization (\ref{model-cpl}) using various combinations of the observational data with and without the GW data. The upper panel represents the constraints on the model without the GW data while in the lower panel we present the corresponding constraints using the GW data. For the CMB only case the upper limits of the $w_a$ parameter at 68\% and 95\% CL are reported.  
Here, $\Omega_{m0}$ is the present value of $\Omega_m = \Omega_b +\Omega_c$ and $H_0$ is in the units of km s$^{-1}$Mpc$^{-1}$.}
\label{tab:results-cpl}
\end{table*}
\end{center}
\endgroup
\begin{figure*}
\includegraphics[width=0.45\textwidth]{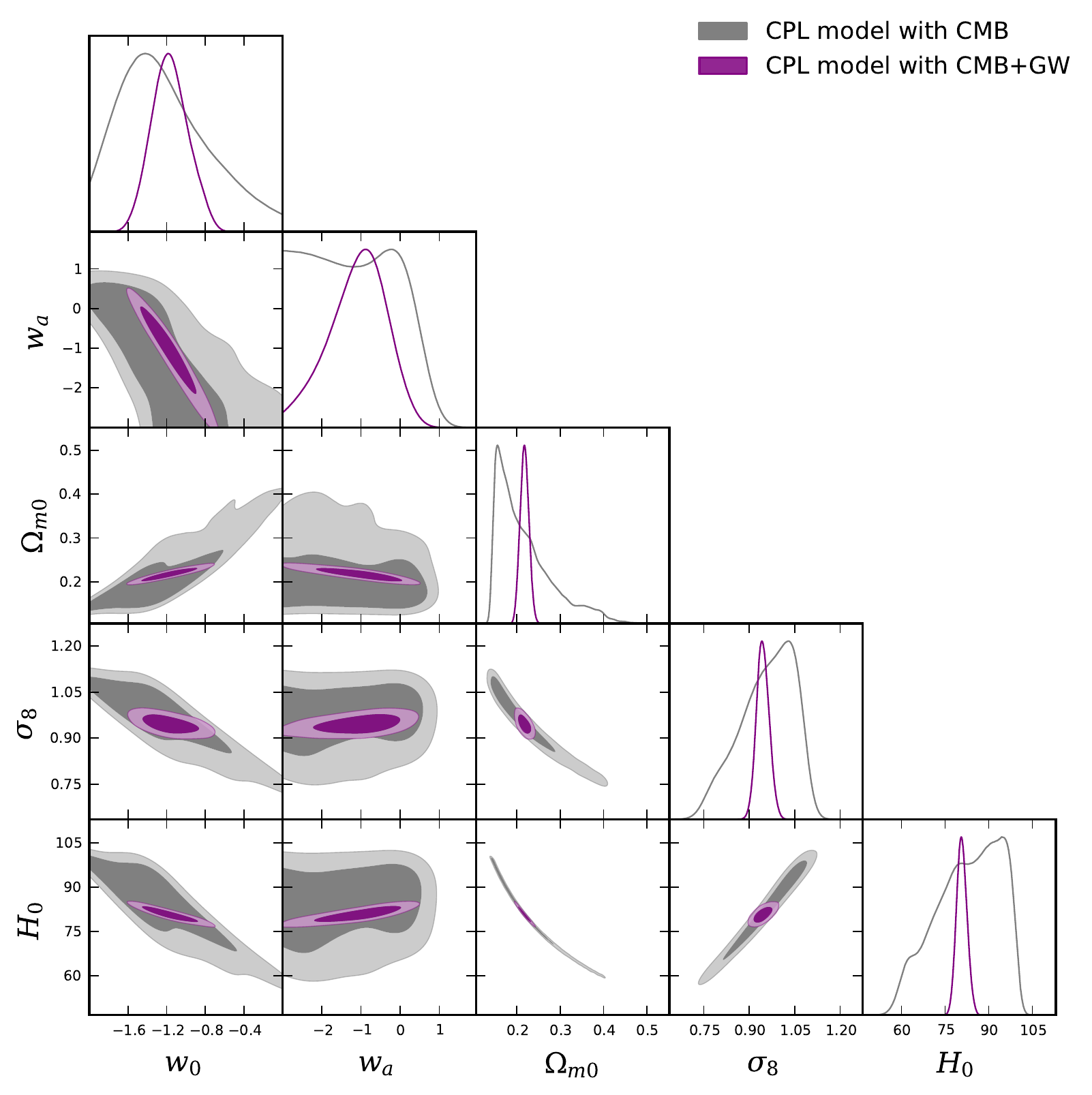}
\includegraphics[width=0.45\textwidth]{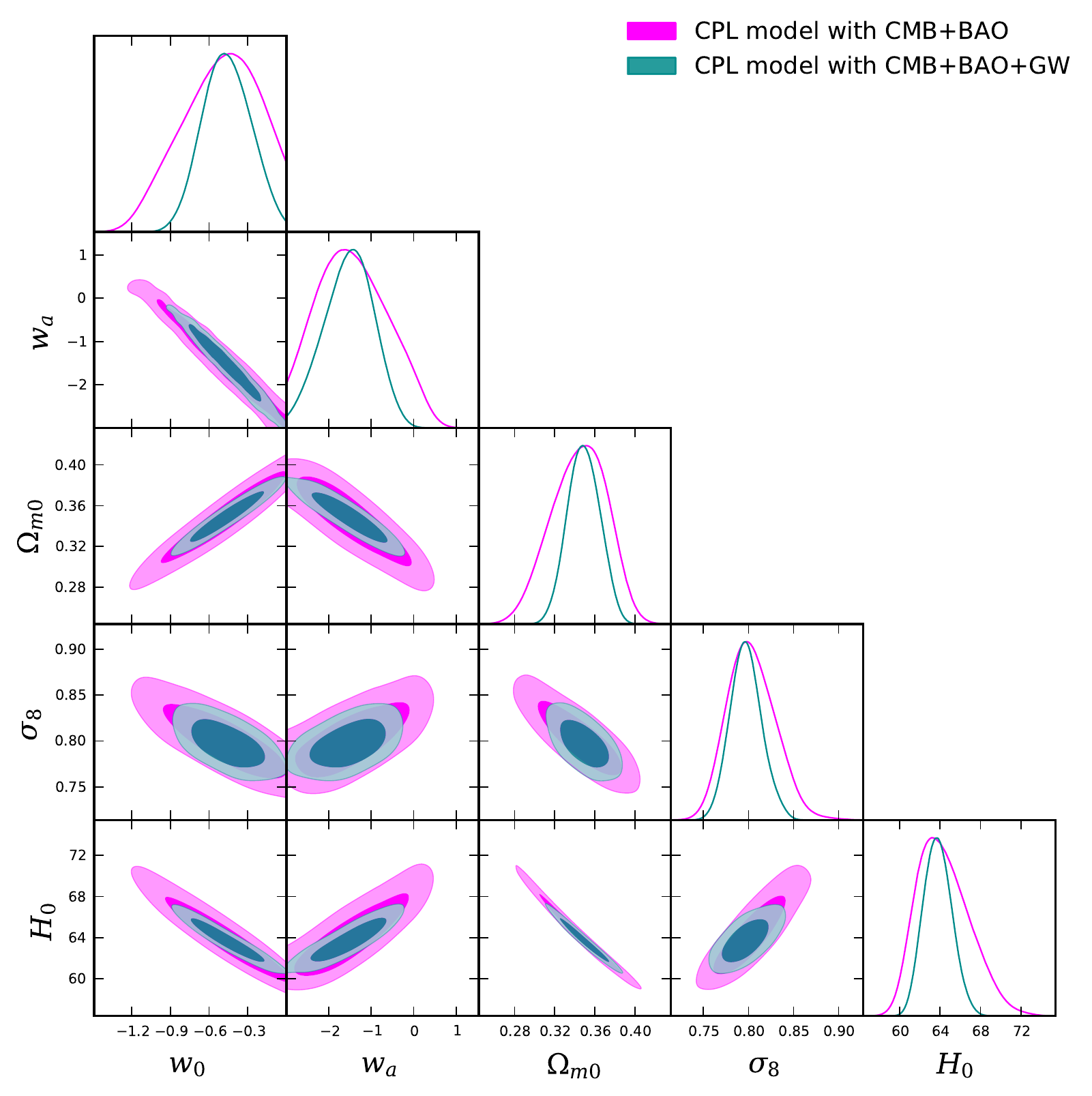}
\includegraphics[width=0.45\textwidth]{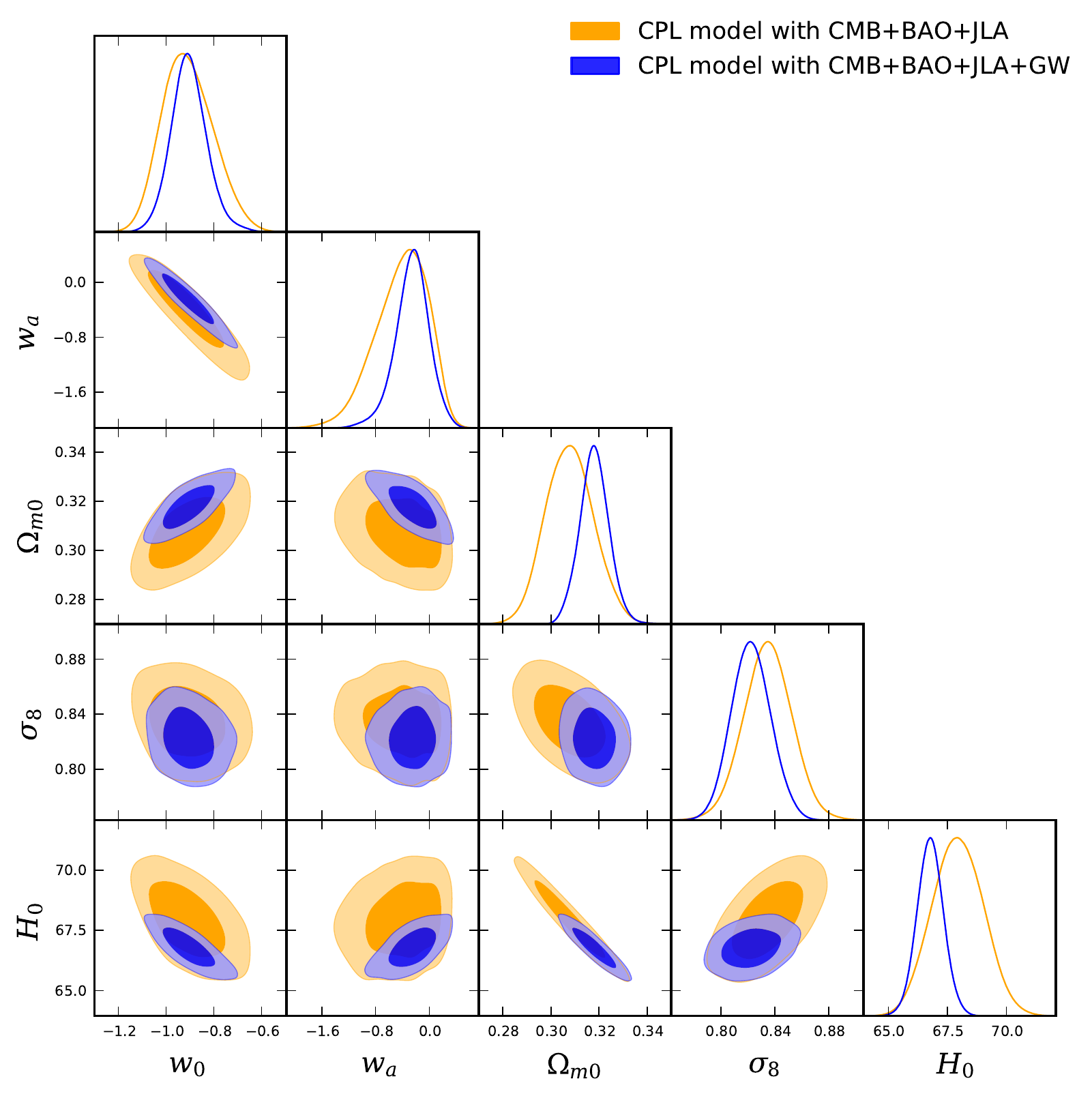}
\includegraphics[width=0.45\textwidth]{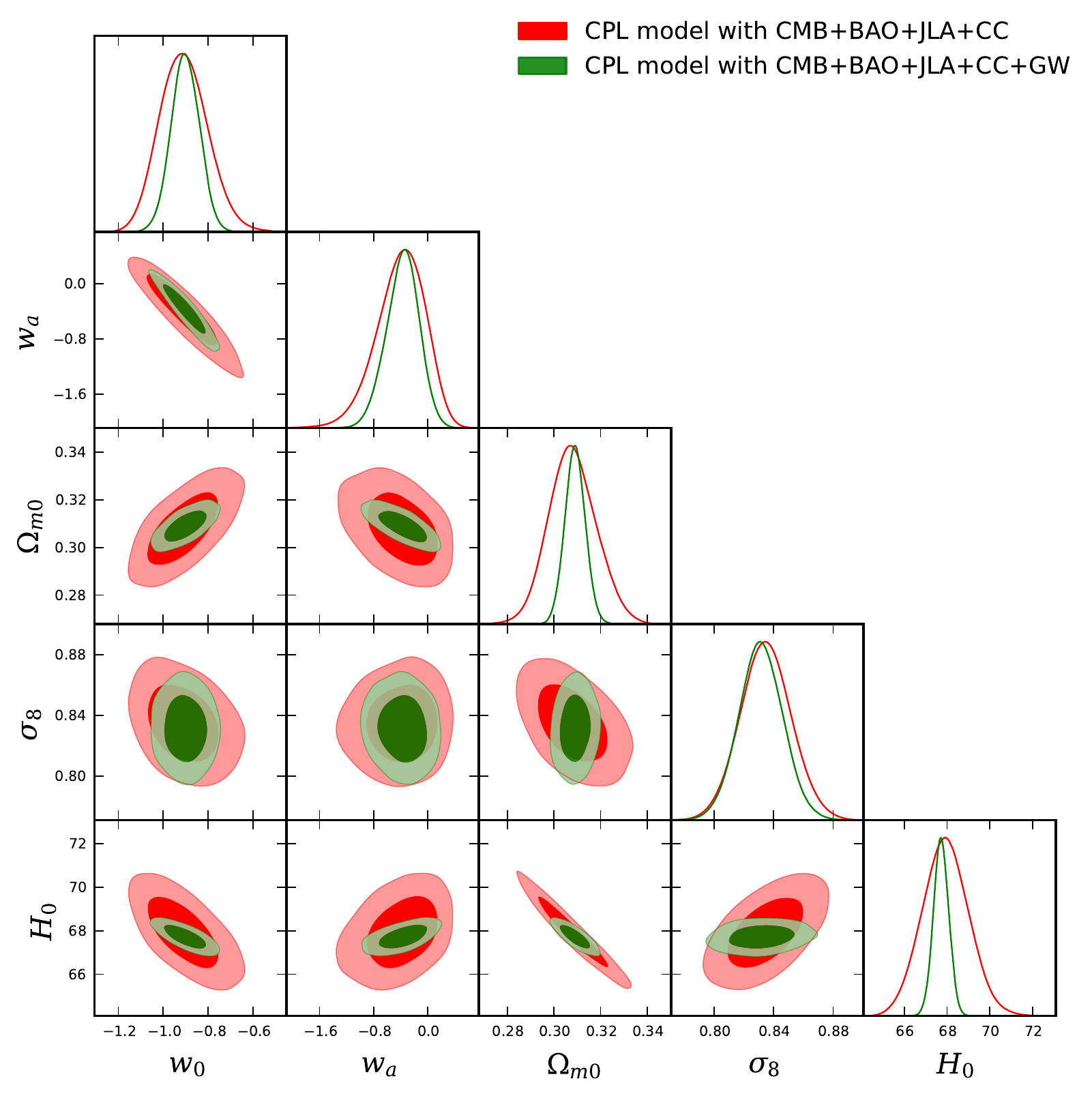}
\caption{\textit{68\% and 95\% CL contour plots for various combinations of some selected parameters of the CPL model (\ref{model-cpl}) using different  observational data in presence (absence) of the GW data.}}
\label{fig-contour-cpl}
\end{figure*}
\begin{figure*}
\includegraphics[width=0.36\textwidth]{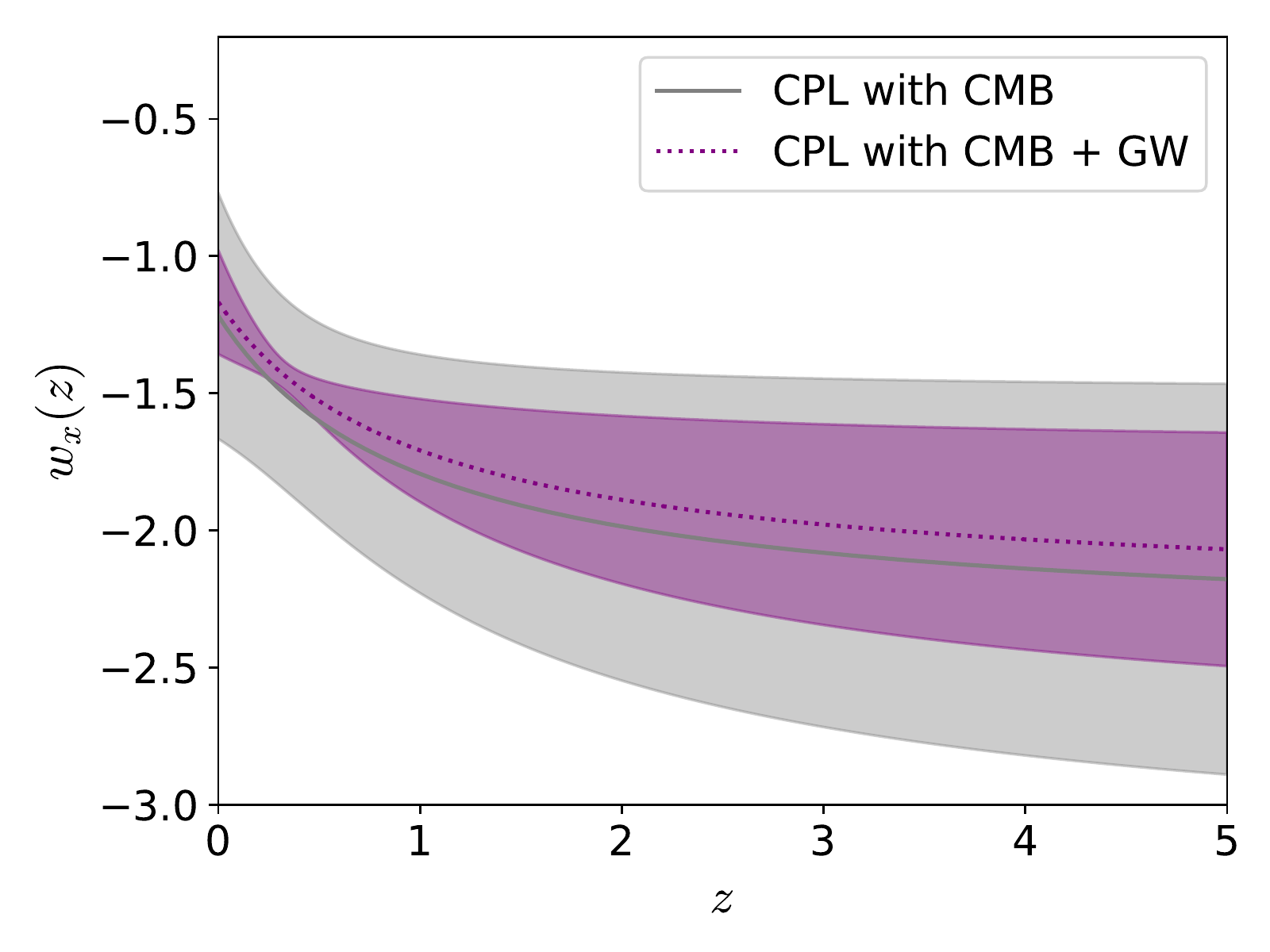}
\includegraphics[width=0.36\textwidth]{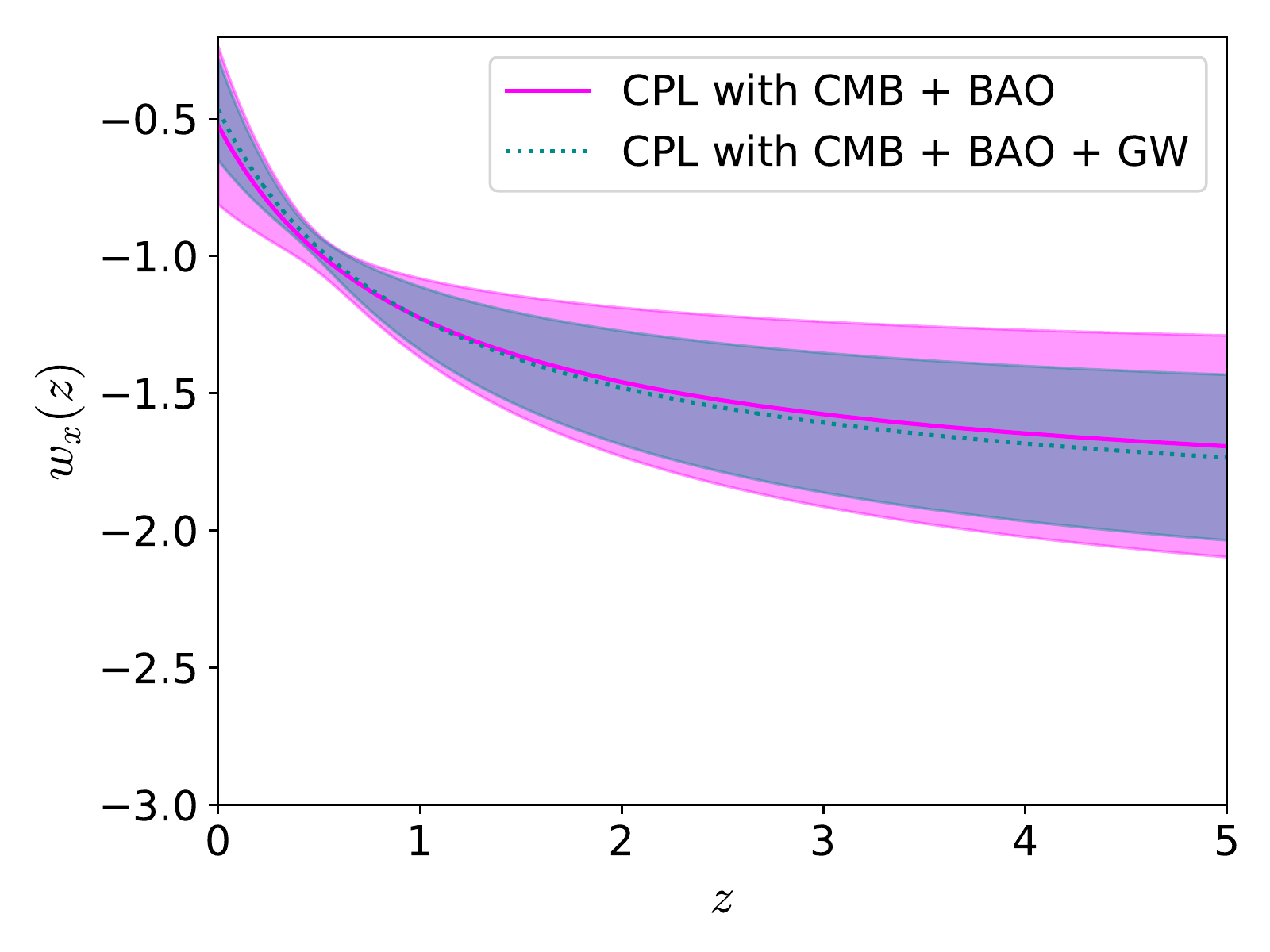}
\includegraphics[width=0.36\textwidth]{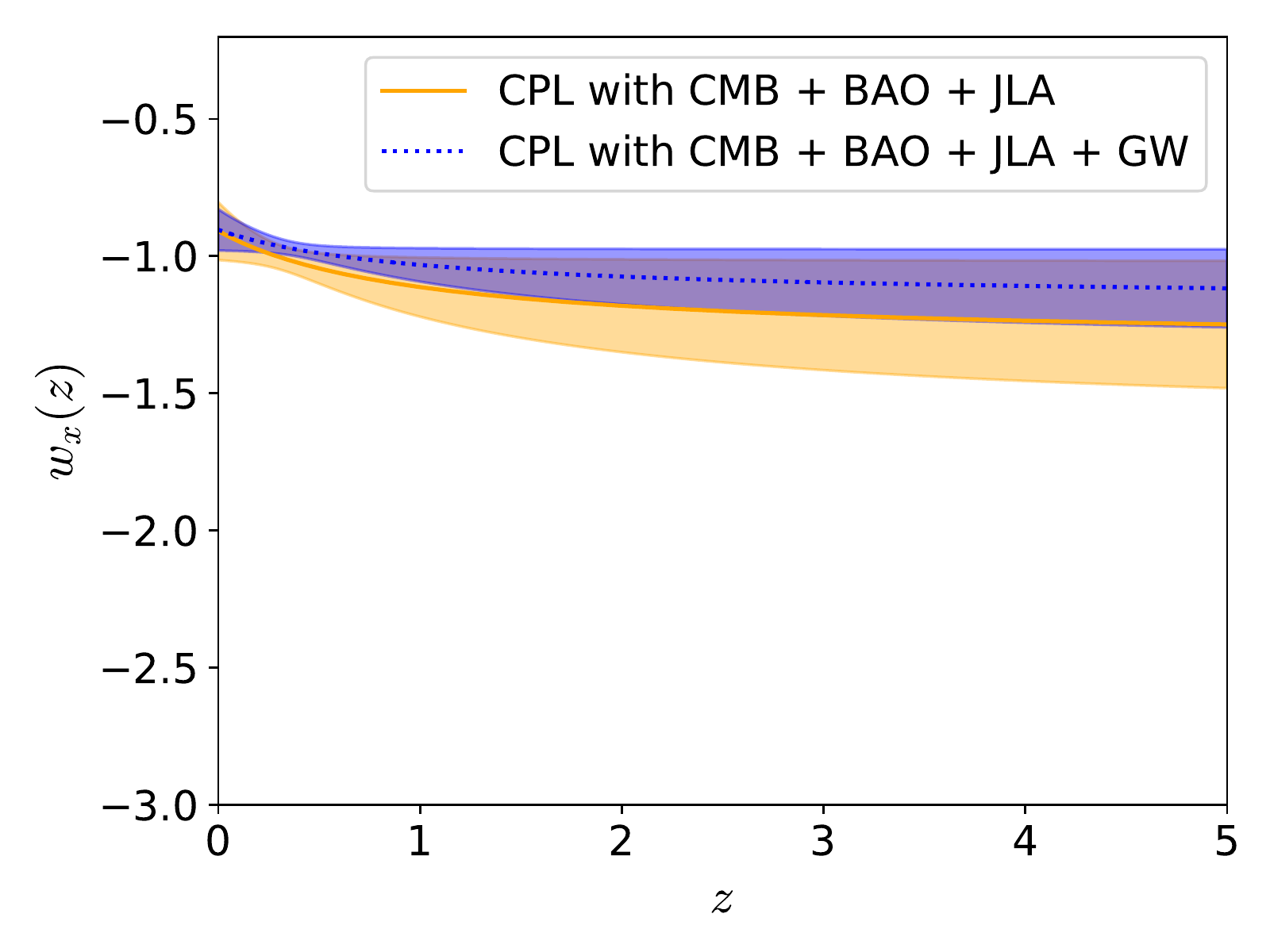}
\includegraphics[width=0.36\textwidth]{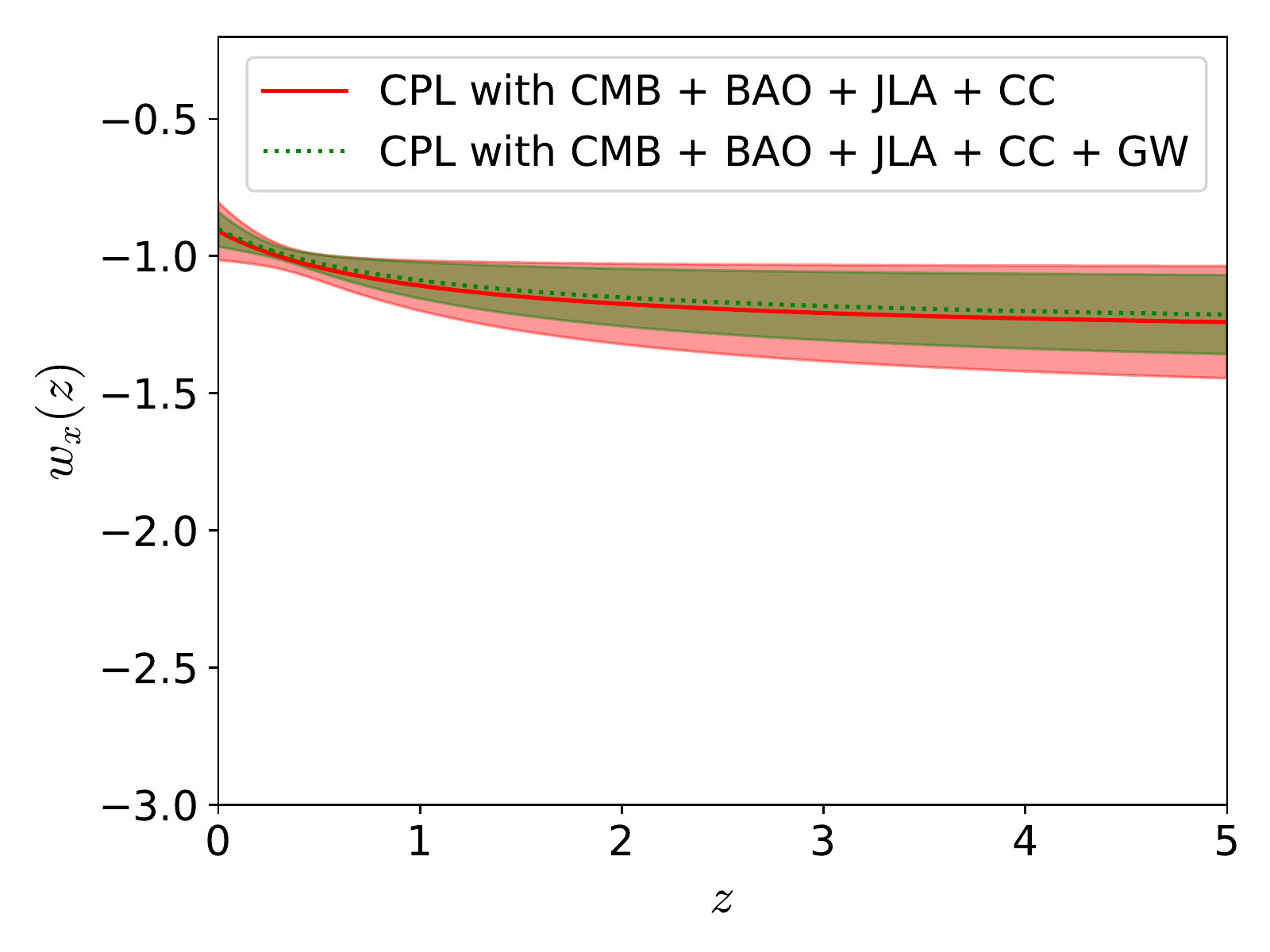}
\caption{\textit{The evolution of the dark energy equation of state for the CPL parametrziation is shown for different datasets taking the mean values of the key parameters $w_0$ and $w_a$ from the corresponding analysis with and without the GW data. The solid curves stand for the evolution of $w_x (z)$ for the standard cosmological probes while the dotted curves stand for the dataset in presence of the GW data. The shaded regions show the 68\% CL constraints on these two parameters.} }
\label{fig-w-cpl}
\end{figure*}
\begin{figure*}
\includegraphics[width=0.35\textwidth]{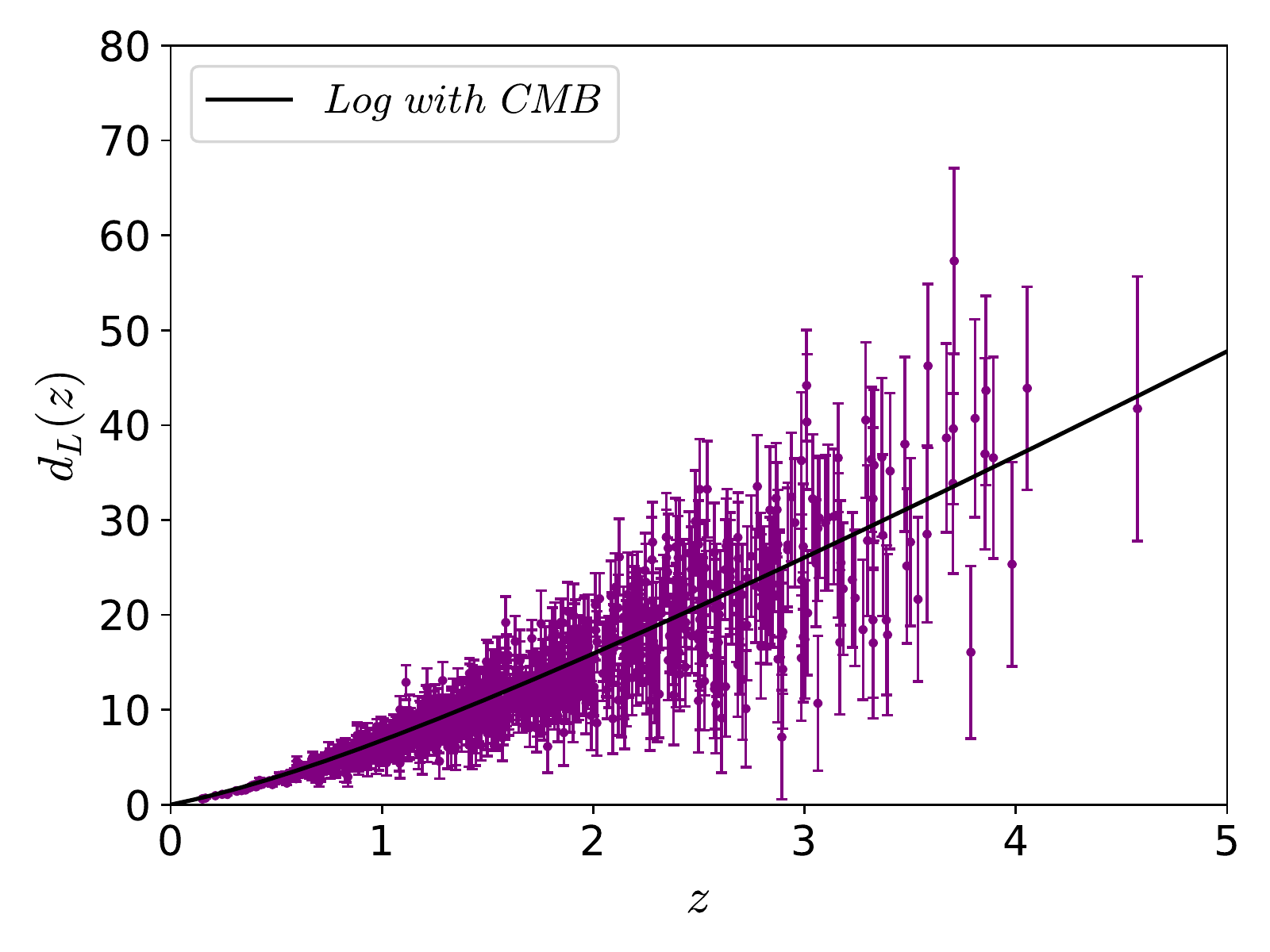}
\includegraphics[width=0.35\textwidth]{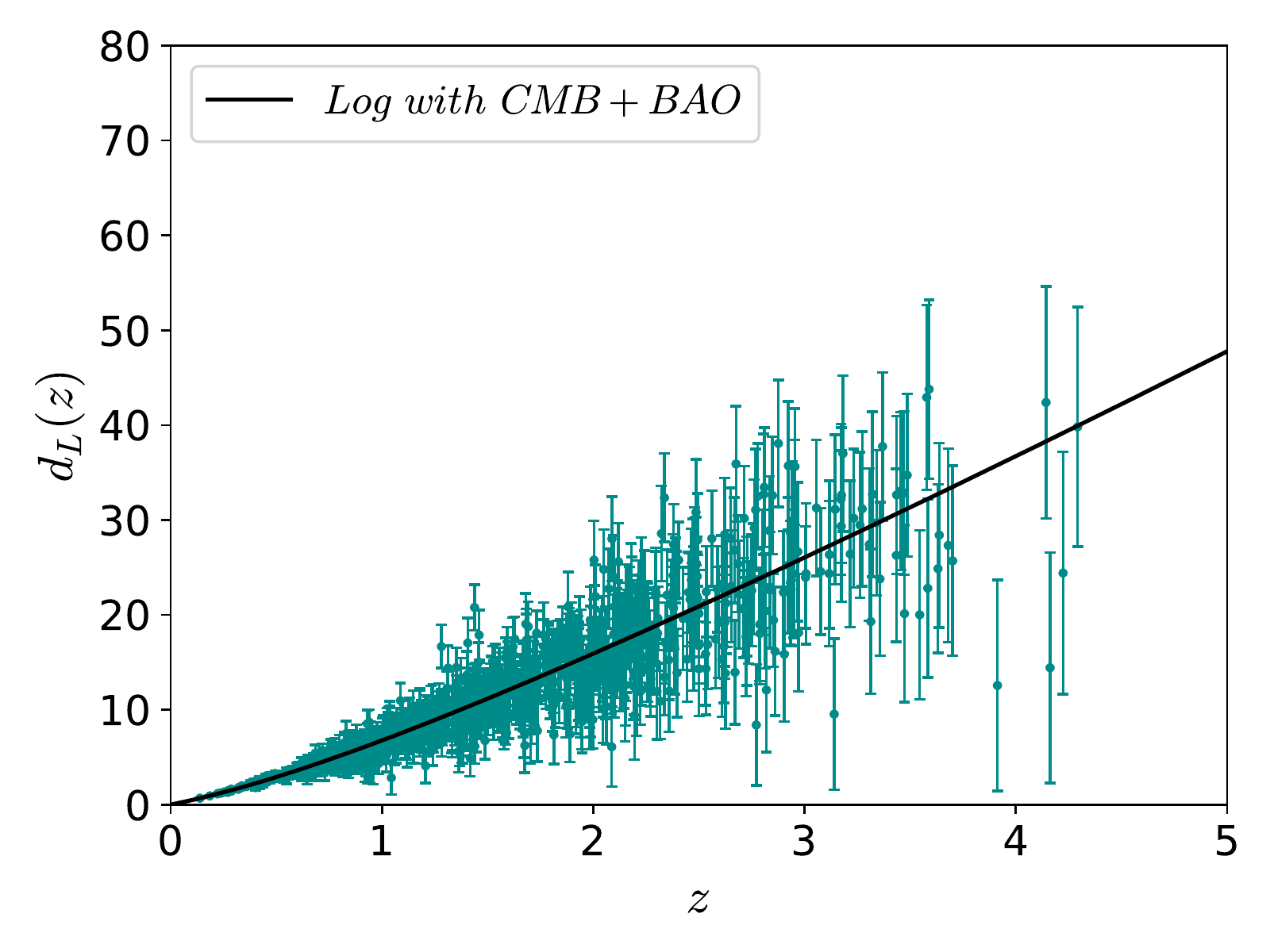}
\includegraphics[width=0.35\textwidth]{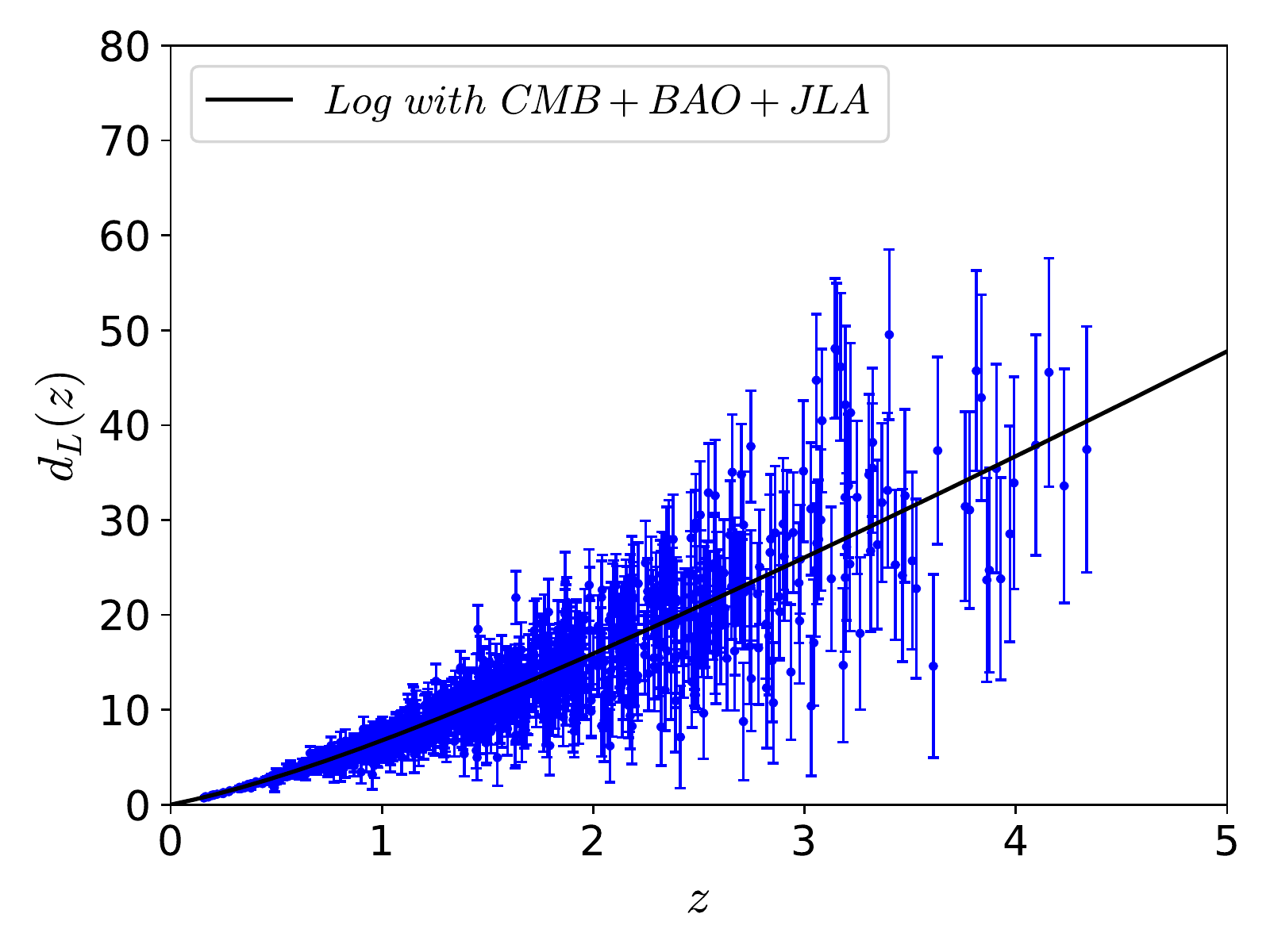}
\includegraphics[width=0.35\textwidth]{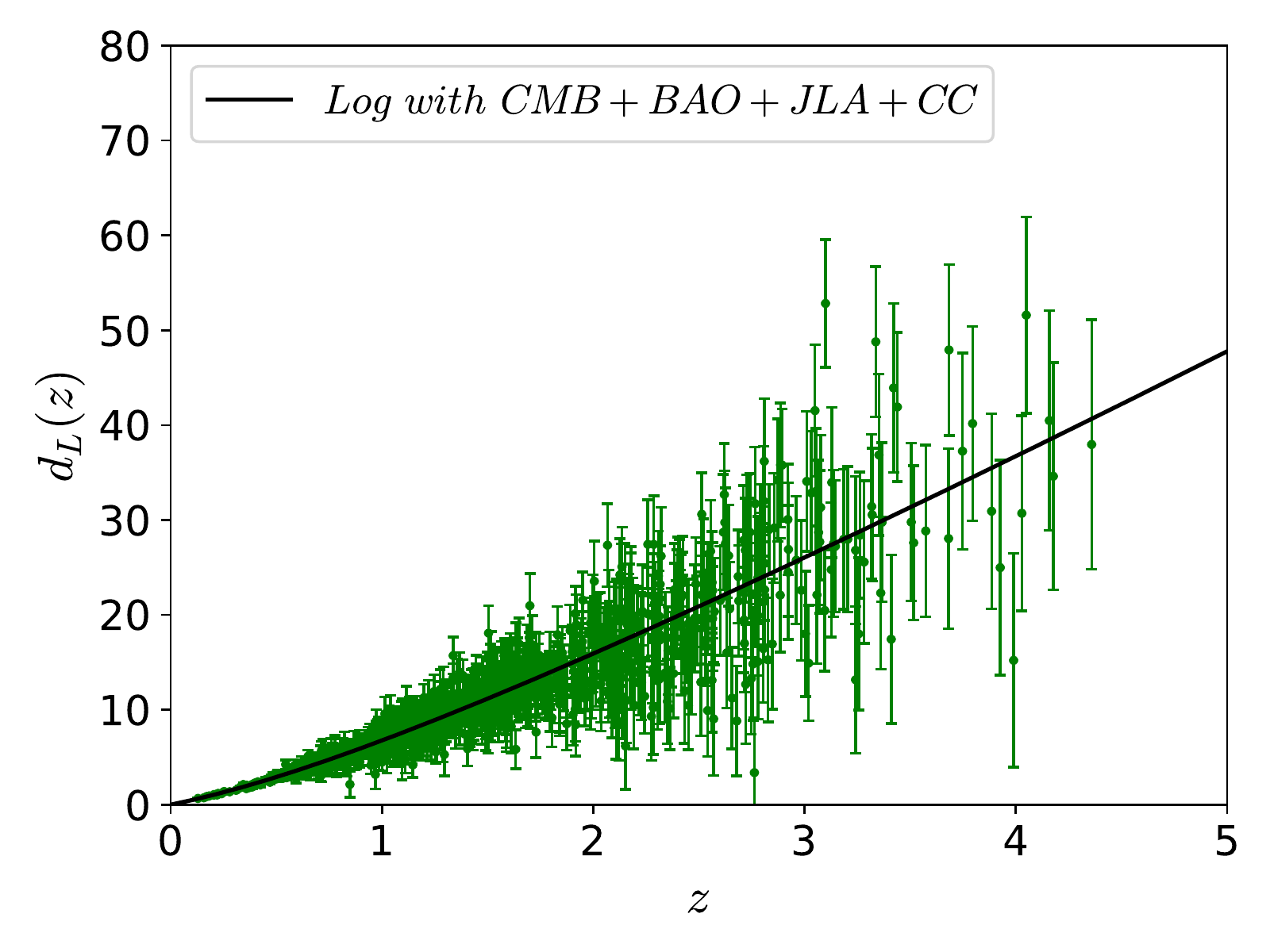}
\caption{\textit{For the fiducial logarithmic model, we first constrain the cosmological parameters using the datasets CMB, CMB+BAO, CMB+BAO+JLA and CMB+BAO+JLA+CC and then we use the best-fit values of the parameters for ``each dataset'' to generate the corresponding GW catalogue. Following this, in each panel we show $d_L (z)$ vs $z$ catalogue with the corresponding error bars for 1000 simulated GW events. The upper left and upper right panels respectively present the catalogue ($z$, $d_L (z)$) with the corresponding error bars for 1000 simulated GW events derived using the CMB alone and CMB+BAO dataset. The lower left and lower right panels respectively present the catalogue ($z$, $d_L (z)$) with the corresponding error bars for 1000 simulated  GW events derived using the CMB+BAO+JLA and CMB+BAO+JLA+CC datasets. }}
\label{fig-dL-log}
\end{figure*}

\subsection{CPL parametrization}
\label{results-cpl}

First of all, we have constrained the CPL parametrization of eqn. (\ref{model-cpl}) using the standard cosmological probes, such as CMB, BAO, JLA and CC (summarized in the upper half of the Table \ref{tab:results-cpl}), and then using the best-fit values of the model parameters of this model, we have generated the GW catalogue comprising 1000 simulated GW events. So, here we have considered CPL as the fiducial model. In Fig. \ref{fig-dL-cpl}, we have shown the luminsity distance $d_L (z) $ versus $z$ graphics for the 1000 simulated GW events. Now, incorporating the simulated GW events with the standard cosmological probes, we have constrained the CPL parametrization. The summary of the observational constraints on the CPL model after the inclusion of the simulated GW data are shown in the lower half of Table \ref{tab:results-cpl}.

In Fig. \ref{fig-contour-cpl} we present the comparisons between the constraining results of the datasets before and after the inclusion of the GW data to the standard cosmological probes mentioned above, where in particular, we show the
one-dimensional (1D) marginalized posterior distributions for some selected parameters of the model as well as the two-dimensional (2D) contour plots between several combinations of the model parameters of this parametrization. Specifically, the upper left panel of Fig. \ref{fig-contour-cpl} presents the comparisons between the datasets CMB and CMB+GW; the upper right panel of Fig. \ref{fig-contour-cpl} is for CMB+BAO and CMB+BAO+GW; the lower left panel of Fig. \ref{fig-contour-cpl} is for CMB+BAO+JLA and CMB+BAO+JLA+GW; finally the lower right panel of Fig. \ref{fig-contour-cpl} is for CMB+BAO+JLA+CC and CMB+BAO+JLA+CC+GW. In the following we describe the effects of GW on the model parameters corresponding to different observational datasets.

In the second column of Table \ref{tab:results-cpl}, we present the observational constraints on the model parameters for the datasets CMB and CMB+GW. One can clearly notice that the inclusion of GW to CMB is effective to reduce the error bars on some of the parameter space of this model, see the top left panel of Fig. \ref{fig-contour-cpl} for a better view on the parameter space. In particular, one can note the significant improvement in the estimations of the Hubble constant as follows, $H_0 = 83.06_{-7.98}^{+   15.10}$ (68\% CL, CMB), $H_0 = 80.75_{-1.92}^{+ 1.71}$ (68\% CL, CMB+GW). We note that due to the inclusion of GW to CMB, the error bars on $H_0$ are reduced by several factors. 
We also note that the matter density parameter at present, $\Omega_{m0}$, for CMB alone is constrained to be small compared to the Planck's estimation \cite{Ade:2015xua} and the inclusion of GW to CMB again improves the parameter space, but slightly (see the lower half of Table \ref{tab:results-cpl}). However, significant improvement is found in the estimation of $\sigma_8$ where one can notice that the inclusion of GW to CMB reduces the error bars by several factors.  Concerning the two key parameters of this model, namely, $w_0$ and $w_a$, the effects of GW to CMB are quite evident. The addition of GW to CMB significantly improves the parameter space by reducing the error bars: $w_0 = -1.218_{- 0.597}^{+ 0.302} $ (68\% CL, CMB) and $w_0= -1.168_{-    0.212}^{+    0.180}$ (68\% CL, CMB+GW). Although the deviation in the mean value of $w_0$, defined by $|\Delta w_0| = |w_0 ({\rm CMB}) - w_0 ({\rm CMB+GW})| = 0.05$, is very small, but the effective nature of  GW is visible through its constraining power in terms of the reduction of the error bars on $w_0$. Overall, the inclusion of GW to CMB shifts  $w_0$ towards $-1$ boundary, although its phantom nature is still allowed within 68\% CL. 
The constraints on $w_a$ for CMB alone is not stringent (the upper linit is $w_a < 0.526$ at 95\% CL),  but the inclusion of GW again reduces its error bars with $w_a =  -1.081_{- 0.640}^{+    0.842}$ (68\% CL, CMB+GW).  
In fact, the power of GW is  clear from both the 1D posterior distributions of some parameters  as well as the 2D contour plots shown in the top left panel of Fig. \ref{fig-contour-cpl}. From this figure (top left panel of Fig. \ref{fig-contour-cpl}), one can clearly understand that a significant improvement in the parameter space is due to the inclusion of GW to CMB.

We now present the cosmological constraints from CMB+BAO and CMB+BAO+GW. With these,
we could be able to see how GW data affect this particular combination.  The summary of the observational constraints are shown in the third column of Table \ref{tab:results-cpl} and the corresponding grapgical variations are shown in the top right panel of Fig. \ref{fig-contour-cpl}.
From the table, one can see that the inclusion of BAO to both CMB and CMB+GW, lowers $H_0$ returning similar mean values as follows, $H_0 = 64.36_{-    3.23}^{+    2.05}$ (68\%, CMB+BAO) and $H_0 = 63.77_{- 1.52}^{+ 1.37} $ (68\%, CMB+BAO+GW). The error bars on $H_0$ is reduced after the inclusion of GW data. One can also notice that for both the analyses,  $w_0$ allows very higher values and $w_a$ takes very lower values, exactly same as recently found in \cite{Scolnic:2017caz}. The interesting fact is that, after the inclusion of BAO to CMB, all the parameters are correlated with each other (see the top right panel of \ref{fig-contour-cpl}), and this remains true even after the inclusion of GW to the combined analysis CMB+BAO.  But, indeed, it is quite clear that the dataset CMB+BAO+GW provides better constraints than CMB+BAO.

We now discuss the cosmological constraints in presence of the JLA data to the previous datasets, that means, precisely we discuss the constraints from CMB+BAO+JLA and CMB+BAO+JLA+GW. The summary of the observational constraints is shown in the fourth column of Table \ref{tab:results-cpl} and the graphical distributions are shown in the bottom left panel of Fig. \ref{fig-contour-cpl}. From this analysis, it is again clear that the inclusion of GW data reduces the error bars on all the parameters.
In particular, one can see the 68\% CL constraints on the Hubble constant as, 
$H_0=    67.94_{- 1.08}^{+    1.09}$ (CMB+BAO+JLA), 
$H_0= 66.98_{-    0.55}^{+    0.55}$ (CMB+BAO+JLA+GW) which show that the inclusion of GW shifts $H_0$ towards its lower values and the error bars are reduced by a factor of $2$. Concerning the two key parameters of this model, that means, $w_0$ and $w_a$, we have some interesting observation. We see that
for both the combinations,  $w_0$ approaches near the `$-1$' border
with $w_0 = -0.909_{- 0.123}^{+    0.095}$ (68\% CL, CMB+BAO+JLA) and $w_0 = -0.904_{-    0.080}^{+    0.070}$ (68\% CL, CMB+BAO+JLA+GW). 
From the highest peak of the 1D posterior distributions of $w_0$ (see the bottom left panel of Fig. \ref{fig-contour-cpl}) for both the datasets, $w_0> -1$ is strongly supported while the tails of the posterior distributions of this parameter are lying from quintessence to the phantom regime due to the error bars on $w_0$.  The improvement in $w_a$ is also transparent:
$w_a =  -0.409_{-    0.277}^{+    0.517}$ (68\% CL, CMB+BAO+JLA) and $w_a = -0.256_{-    0.227}^{+ 0.263}$ (68\% CL, CMB+BAO+JLA+GW). So, from both the observational datasets, dynamical nature is allowed while one can also note that,
$w_a =0$ is also not excluded  in 68\% CL.  Finally, we mention the correlations between the parameters clearly shown in the bottom left panel of Fig. \ref{fig-contour-cpl}, where we see that such correlations are not affected by the GW data.   However, we mention that the inclusion of JLA decreases the correlation between some of the combinations of the parameters.  And in particular, we find that some of parameters are uncorrelated, for instance, we see that $\sigma_8$ seems to be uncorrelated with $w_0$ and $w_a$.

We now discuss the last two analyses for this model, namely with CMB+BAO+JLA+CC and its companion CMB+BAO+JLA+CC+GW. The summary of the observational constraints is shown in the last column of Table \ref{tab:results-cpl} and in the bottom right panel of Fig. \ref{fig-contour-cpl} we compare these datasets. From the analysis, we clearly notice that the inclusion of the GW data improves the parameters space in an effective way. In fact, the maximum effects are seen in $H_0$ and $\Omega_{m0}$ (see the 1D posterior distributions of these parameters as well).
In particular, one can look at the improvements of the Hubble parameter after the inclusion of GW data: $H_0 = 67.92_{-    1.09}^{+    1.09}$ (68\%, CMB+BAO+JLA+CC) and $H_0= 67.72_{-    0.35}^{+    0.36}$ (68\% CL, CMB+BAO+JLA+CC+GW). Furthermore, the estimations of other parameters can also be visualized in a similar fashion. Concerning the key parameters of this parametrization, namely, $(w_0, w_a)$, we observe significant changes on their constraints. Looking at the 68\% CL costraints on $w_0$ where  $w_0 = -0.909_{-  0.116}^{+    0.099}$ (CMB+BAO+JLA+CC) and $w_ 0 =  - 0.902_{-0.062}^{+    0.064}$ (CMB+BAO+JLA+CC+GW), one can see that after the inclusion of GW the at 68\% upper CL error bars  on $w_0$ are reduced by a factor of $2$. For the other parameter $w_a$: $w_a = -0.399_{- 0.297}^{+    0.423}$ (68\% CL, CMB+BAO+JLA+CC) and $w_a  =  -0.373_{-    0.226}^{+    0.263}$ (68\% CL, CMB+BAO+JLA+CC+GW), although the reduction of the error bars are not much significant compared to $w_0$, however, such improvements are clearly visualized. Moreover, looking at the constraints on $w_0$, one can also argue that for both the datasets, the dark energy equation of state at present exhibits its quintessential nature (i.e., $w_0 > -1$).  This feature is actually clear if one looks at the highest peaks of the 1D posterior distributions of $w_0$ in Fig. \ref{fig-contour-cpl} (see the bottom right panel of this figure). Additionally, we find that for the final combination, that means for CMB+BAO+JLA+CC+GW, within 68\% CL, $w_a \neq 0$. It means that a dynamical character is allowed within this confidence level.   Concerning the correlations between the parameters, one may draw similar conclusions as found in previous two datasets, namely, CMB+BAO+JLA and CMB+BAO+JLA+GW.

Finally, using the mean values of $(w_0, w_a)$ from all the datasets, in Fig. \ref{fig-w-cpl} we have shown the qualitative evolution of the dark energy equation of state $w_x (z)$ for this model. The solid lines in each plot stand for the $w_x (z)$ curve for the usual cosmological probe and the dotted lines depict the evolution of $w_x (z)$
in presence of the GW data. In each plot the shaded regions (with similar colors to the corresponding curves) present the 68\% regions for the parameters $w_0, w_a$ corresponding to each dataset (with or without the GW data). From this figure (i.e., Fig. \ref{fig-w-cpl}) one can see the addition of GW to the standard cosmological data certainly improves the parameter space. The maximum effects of GW are visible with the CMB alone data.

\begingroup
\squeezetable
\begin{center}
\begin{table*}
\begin{tabular}{ccccccccccccccccc}
\hline\hline
Parameters & CMB & CMB+BAO & CMB+BAO+JLA & CMB+BAO+JLA+CC &\\ \hline

$\Omega_c h^2$ & $    0.1190_{-    0.0014-    0.0027}^{+    0.0014+    0.0028}$ & $    0.1193_{-    0.0014-    0.0026}^{+    0.0013+    0.0026}$ & $    0.1193_{-    0.0014-    0.0025}^{+    0.0013+    0.0025}$ & $    0.1192_{-    0.0013-    0.0025}^{+    0.0012+    0.0026}$  \\

$\Omega_b h^2$ & $    0.02229_{-    0.00016-    0.00031}^{+    0.00016+    0.00031}$ & $    0.02225_{-    0.00015-    0.00030}^{+    0.00015+    0.00031}$ & $    0.02226_{-    0.00015-    0.00029}^{+    0.00015+    0.00030}$ &  $    0.02226_{-    0.00015-    0.00030}^{+    0.00015+    0.00030}$  \\

$100\theta_{MC}$ & $    1.04081_{-    0.00032-    0.00066}^{+    0.00033+    0.00065}$ & $    1.04075_{-    0.00032-    0.00065}^{+    0.00032+    0.00065}$ & $    1.04075_{-    0.00033-    0.00065}^{+    0.00032+    0.00062}$ & $    1.04078_{-    0.00032-    0.00065}^{+    0.00031+    0.00059}$  \\

$\tau$ & $    0.074_{-    0.017-    0.034}^{+    0.017+    0.034}$  & $    0.076_{-    0.017-    0.034}^{+    0.018+    0.033}$ & $    0.078_{-    0.017-    0.033}^{+    0.017+    0.034}$ & $    0.079_{-    0.017-    0.034}^{+    0.018+    0.033}$  \\

$n_s$ &  $    0.9668_{-    0.0045-    0.0090}^{+    0.0045+    0.0087}$ &  $    0.9659_{-    0.0045-    0.0087}^{+    0.0045+    0.0089}$  & $    0.9661_{-    0.0044-    0.0085}^{+    0.0043+    0.0088}$ & $    0.9663_{-    0.0042-    0.0085}^{+    0.0043+    0.0087}$  \\

${\rm{ln}}(10^{10} A_s)$ & $    3.081_{-    0.034-    0.067}^{+    0.034+    0.067}$ & $    3.085_{-    0.034-    0.068}^{+    0.035+    0.065}$ & $    3.089_{-    0.0335-    0.0671}^{+    0.0334+    0.0646}$  & $    3.091_{-    0.034-    0.066}^{+    0.034+    0.065}$  \\

$w_0$ & $-1.058^{+0.354 +0.865}_{-0. 550-0.759}$ & $   -0.429_{-    0.223-    0.386}^{+    0.265+    0.429}$ & $   -0.895_{-    0.098-    0.169}^{+    0.084+    0.177}$ & $   -0.894_{-    0.097-    0.158}^{+    0.072+    0.166}$  \\

$w_a$ & $   -1.579_{-    1.421-    1.421}^{+    1.579+    1.579}$ & $   -1.301_{-    0.570-    0.967}^{+    0.549+    0.979}$ & $   -0.365_{-    0.083-    0.450}^{+    0.365+    0.365}$ & $   -0.352_{-    0.137-    0.416}^{+    0.293+    0.352}$  \\

$\Omega_{m0}$ & $    0.219_{-    0.082-    0.097}^{+    0.030+    0.136}$ & $    0.356_{-    0.024-    0.047}^{+    0.026+    0.043}$ & $    0.308_{-    0.011-    0.021}^{+    0.010+    0.021}$ & $    0.309_{-    0.010-    0.019}^{+    0.010+    0.020}$  \\

$\sigma_8$ & $    0.959_{-    0.067-    0.176}^{+    0.122+    0.152}$ & $    0.795_{-    0.026-    0.044}^{+    0.023+    0.048}$  & $    0.835_{-    0.018-    0.034}^{+    0.017+    0.034}$ & $    0.835_{-    0.017-    0.035}^{+    0.017+    0.033}$  \\

$H_0$ & $   82.78_{-    8.34-   20.63}^{+   15.48+   18.54}$ & $   63.30_{-    2.52-    4.02}^{+    1.87+    4.32}$ & $   67.93_{-    1.19-    2.18}^{+    1.11+    2.22}$ & $   67.84_{-    1.14-    2.01}^{+    1.05+    2.13}$ &
\\

\hline
\end{tabular}
\begin{tabular}{cccccccccccc}
Parameters & CMB+GW & CMB+BAO+GW & CMB+BAO+JLA+GW & CMB+BAO+JLA+CC+GW  \\ \hline

$\Omega_c h^2$ & $    0.1179_{-    0.0012-    0.0023}^{+    0.0012+    0.0023}$ & $    0.1192_{-    0.0013-    0.0027}^{+    0.0013+    0.0026}$ & $    0.1194_{-    0.0013-    0.0025}^{+    0.0012+    0.0026}$ & $    0.1192_{-    0.0012-    0.0023}^{+    0.0012+    0.0024}$\\

$\Omega_b h^2$ & $    0.02241_{-    0.00014-    0.00027}^{+    0.00013+    0.00029}$ & $    0.02228_{-    0.00015-    0.00028}^{+    0.00015+    0.00030}$ & $    0.02227_{-    0.00015-    0.00030}^{+    0.00015+    0.00030}$ & $    0.02224_{-    0.00014-    0.00027}^{+    0.00015+    0.00028}$\\

$100\theta_{MC}$ & $    1.04101_{-    0.00031-    0.00060}^{+    0.00032+    0.00060}$ & $    1.04079_{-    0.00032-    0.00063}^{+    0.00033+    0.00063}$ & $    1.04077_{-    0.00031-    0.00061}^{+    0.00031+    0.00062}$ & $    1.04076_{-    0.00031-    0.00061}^{+    0.00032+    0.00061}$\\

$\tau$ & $    0.082_{-    0.017-    0.034}^{+    0.017+    0.034}$ & $    0.078_{-    0.018-    0.034}^{+    0.017+    0.034}$ & $    0.078_{-    0.017-    0.033}^{+    0.017+    0.033}$ & $    0.079_{-    0.017-    0.034}^{+    0.017+    0.034}$\\

$n_s$ & $    0.9698_{-    0.004-    0.008}^{+    0.004+    0.009}$ & $    0.9667_{-    0.0045-    0.0089}^{+    0.0045+    0.0089}$ & $    0.9658_{-    0.0043-    0.0083}^{+    0.0043+    0.0086}$ & $    0.9660_{-    0.0041-    0.0081}^{+    0.0041+    0.0081}$\\

${\rm{ln}}(10^{10} A_s)$ & $    3.095_{-    0.034-    0.067}^{+    0.034+    0.066}$ &
$    3.090_{-    0.034-    0.067}^{+    0.034+    0.066}$ & $    3.089_{-    0.033-    0.064}^{+    0.033+    0.064}$ & $    3.090_{-    0.033-    0.067}^{+    0.033+    0.066}$\\

$w_0$ & $   -1.056_{-    0.196-    0.344}^{+    0.179+    0.356}$ & $   -0.607_{-    0.186-    0.336}^{+    0.172+    0.348}$ & $   -0.919_{-    0.085-    0.138}^{+    0.071+    0.152}$ & $   -0.902_{-    0.078-    0.126}^{+    0.057+    0.139}$\\

$w_a$ & $   -1.500_{-    0.571-    1.275}^{+    0.718+    1.133}$ & $   -0.955_{-    0.388-    0.848}^{+    0.543+    0.897}$ & $   -0.399_{-    0.172-    0.391}^{+    0.251+    0.388}$ & $   -0.252_{-    0.089-    0.309}^{+    0.220+    0.252}$\\

$\Omega_{m0}$ & $    0.207_{-    0.009-    0.016}^{+    0.009+    0.017}$ & $    0.335_{-    0.015-    0.030}^{+    0.015+    0.031}$ & $    0.300_{-    0.007-    0.013}^{+    0.007+    0.014}$ & $    0.316_{-    0.007-    0.013}^{+    0.006+    0.014}$ \\

$\sigma_8$ & $    0.959_{-    0.020-    0.039}^{+    0.020+    0.040}$ & $    0.812_{-    0.017-    0.033}^{+    0.017+    0.034}$ & $    0.845_{-    0.017-    0.030}^{+    0.015+    0.032}$ & $    0.827_{-    0.016-    0.030}^{+    0.016+    0.031}$\\

$H_0$ & $   82.57_{-    1.65-    3.10}^{+    1.66+    3.25}$ & $   65.19_{-    1.48-    2.65}^{+    1.31+    2.83}$ & $   68.88_{-    0.76-    1.43}^{+    0.73+    1.47}$ & $   67.11_{-    0.69-    1.37}^{+    0.68+    1.37}$\\

\hline\hline
\end{tabular}
\caption{68\% and 95\% CL constraints on the Logarithmic parametrization (\ref{model-log}) using various combinations of the observational data with and without the GW data. The upper panel represents the constraints without the GW data while in the lower panel we present the corresponding constraints using the GW data.
Here, $\Omega_{m0}$ is the present value of $\Omega_m = \Omega_b +\Omega_c$ and $H_0$ is in the units of km s$^{-1}$Mpc$^{-1}$.}
\label{tab:results-log}
\end{table*}
\end{center}
\endgroup
\begin{figure*}
\includegraphics[width=0.45\textwidth]{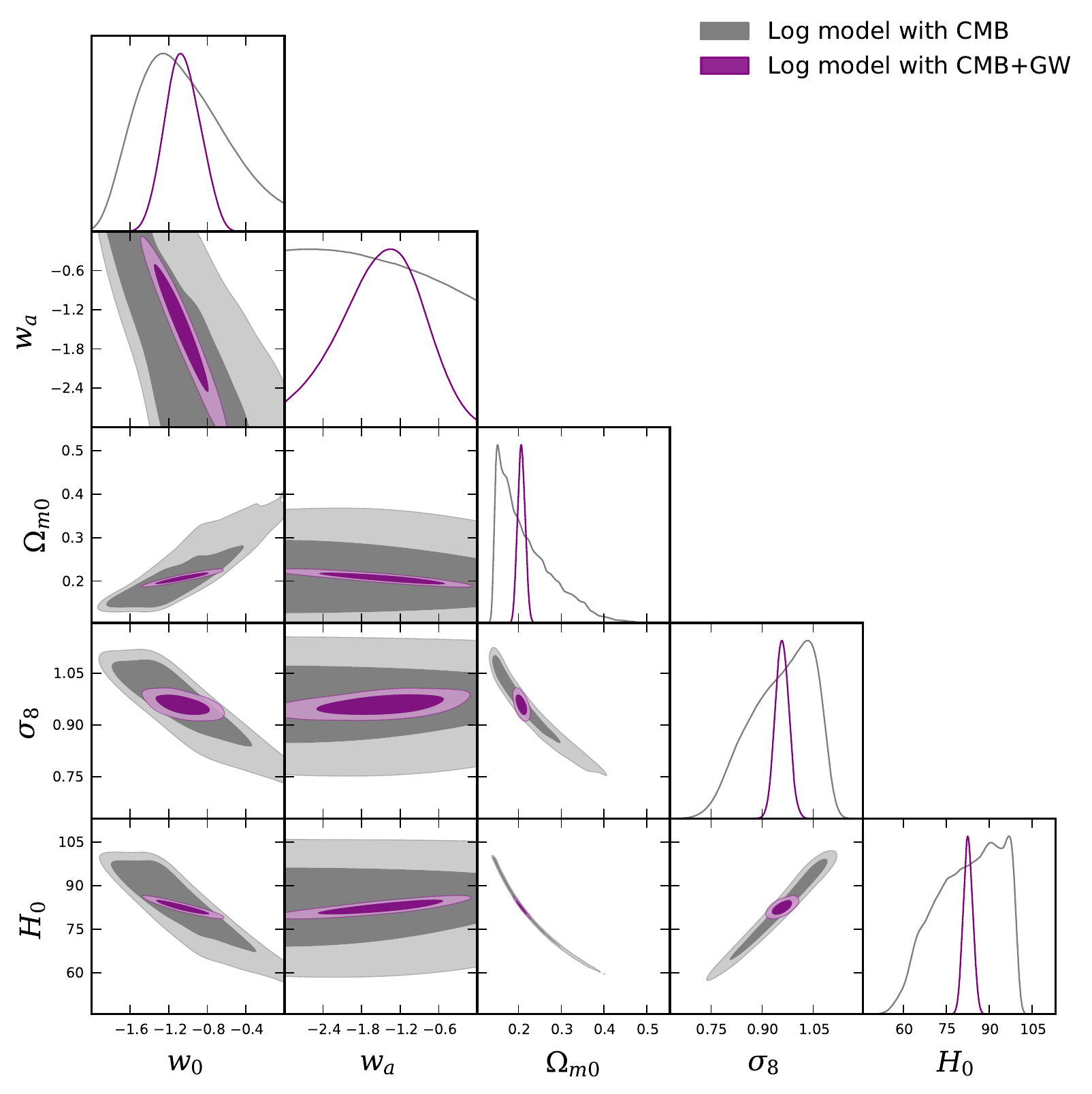}
\includegraphics[width=0.45\textwidth]{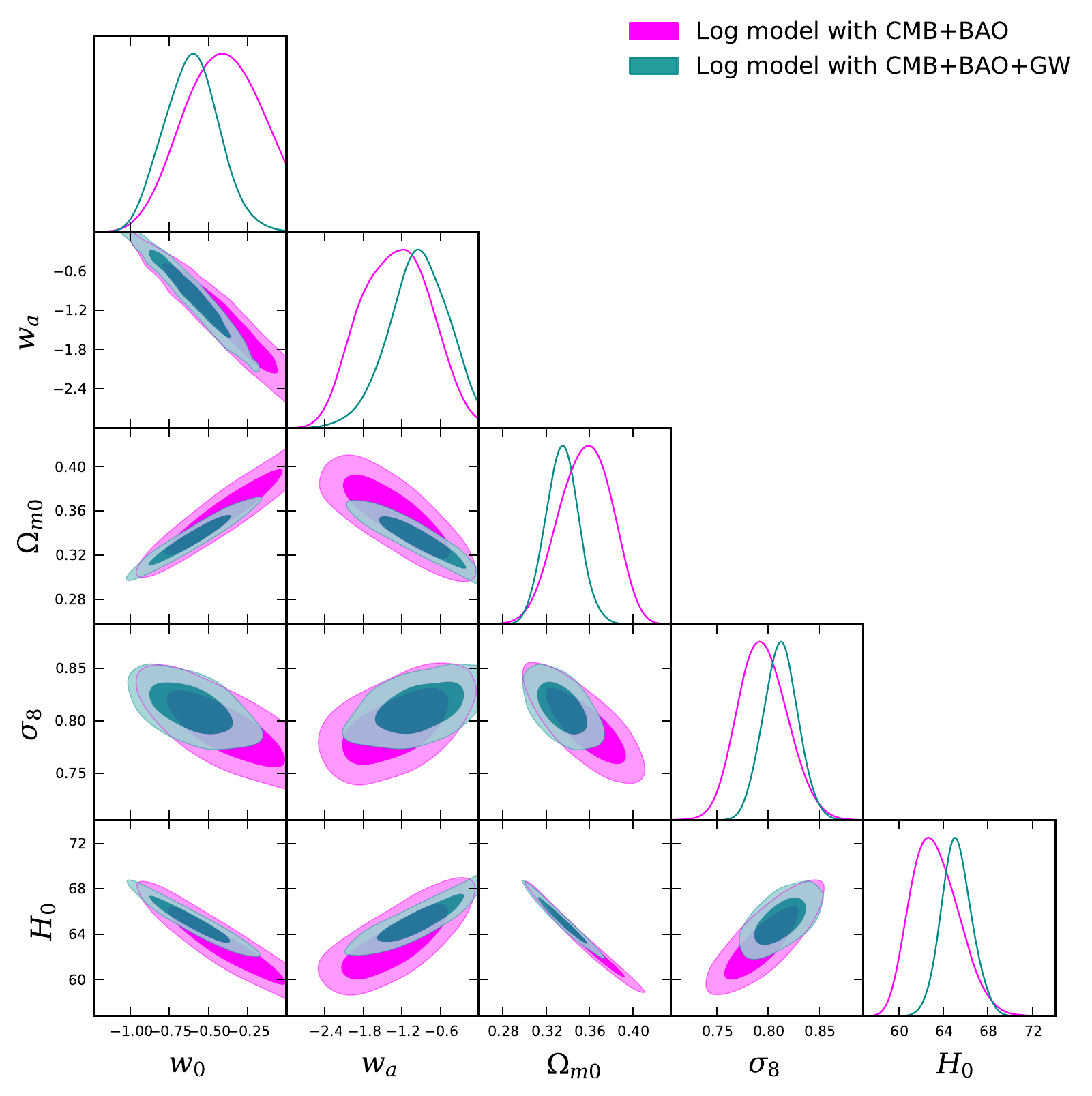}
\includegraphics[width=0.45\textwidth]{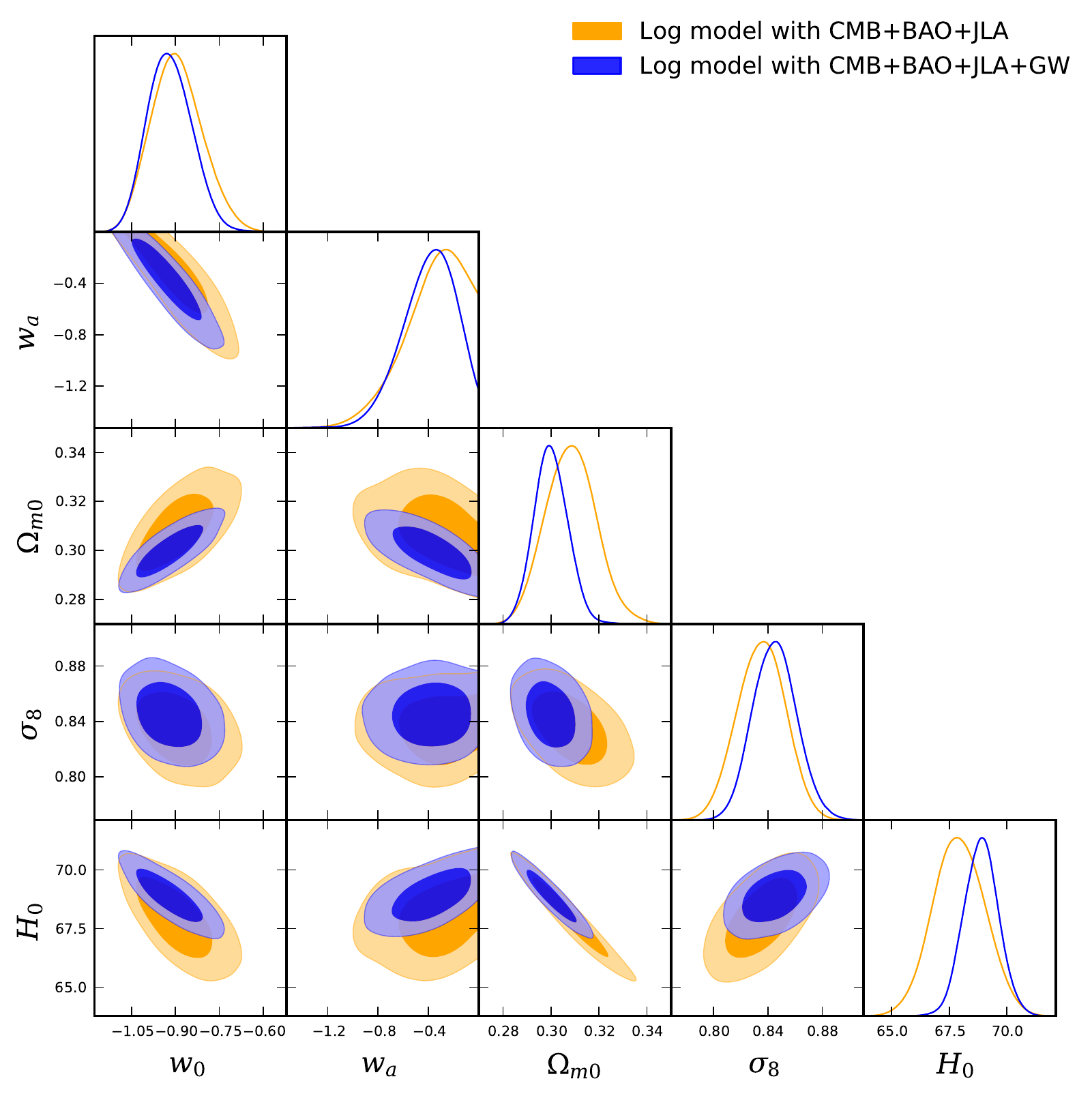}
\includegraphics[width=0.45\textwidth]{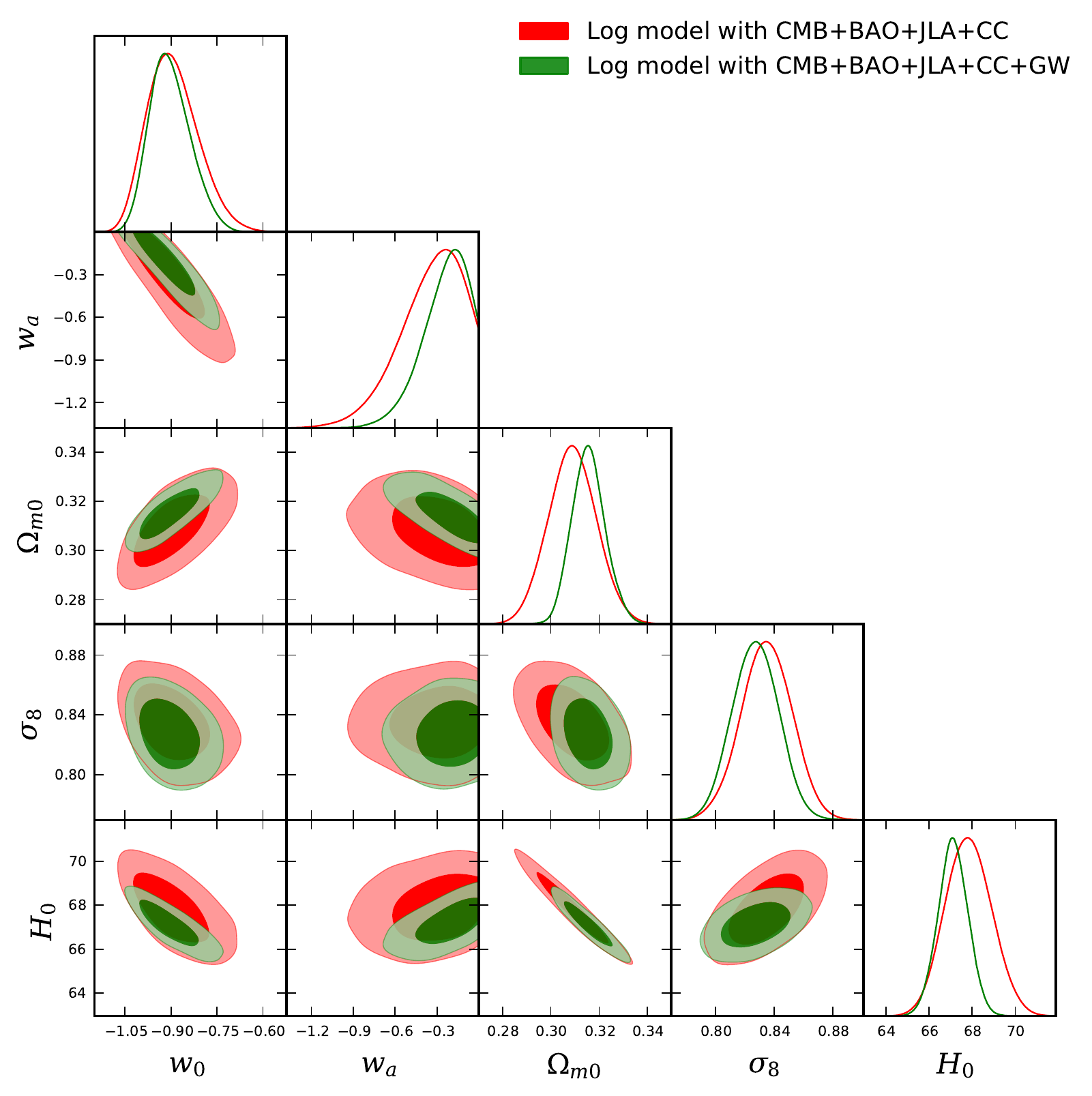}
\caption{\textit{In this figure we have shown the 68\% and 95\% confidence-level contour plots for various combinations of some selected parameters of the Logarithmic parametrization (\ref{model-log}) using different  observational data in presence (absence) of the GW data. }}
\label{fig-contour-log}
\end{figure*}
\begin{figure*}
\includegraphics[width=0.36\textwidth]{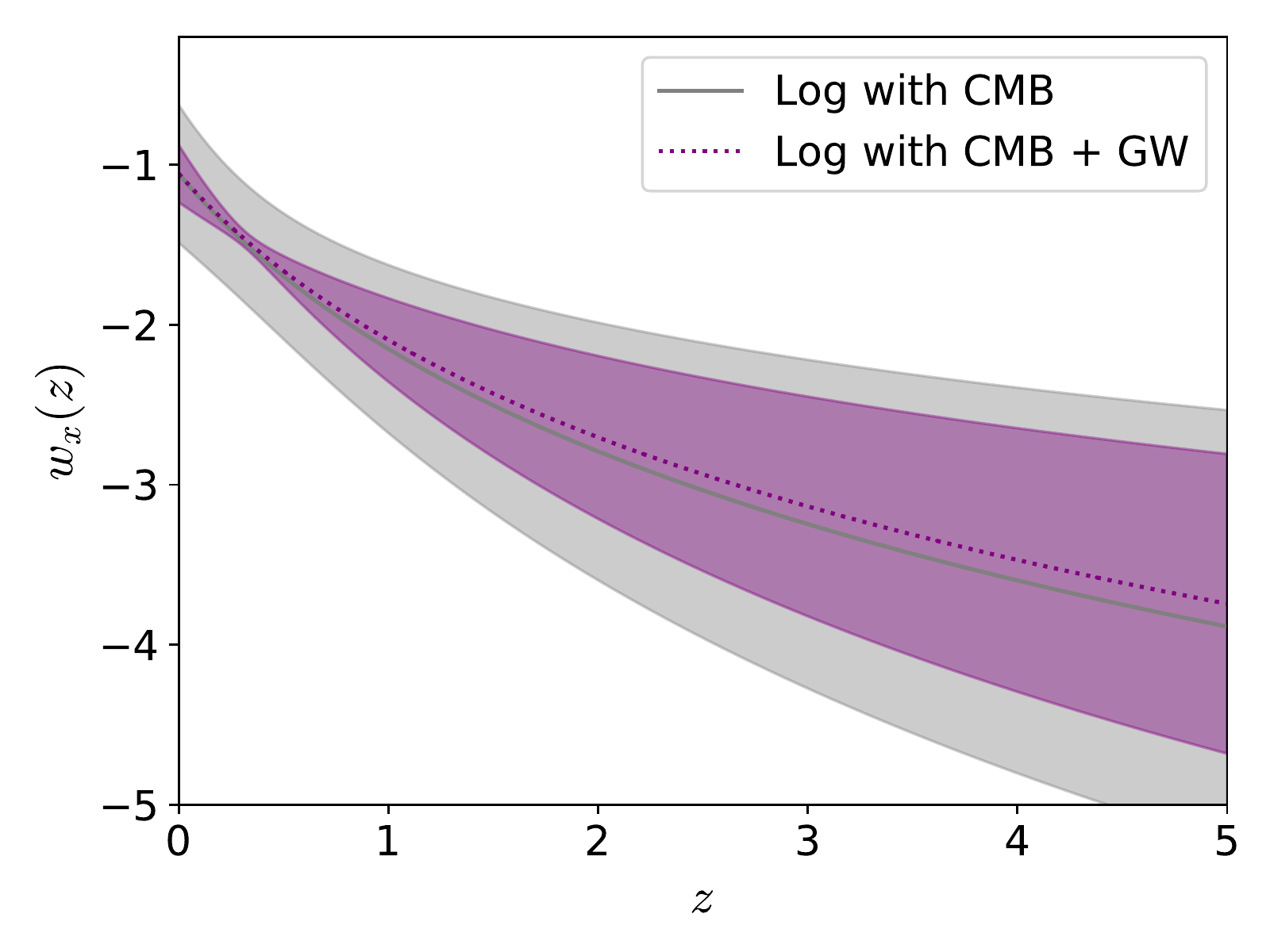}
\includegraphics[width=0.36\textwidth]{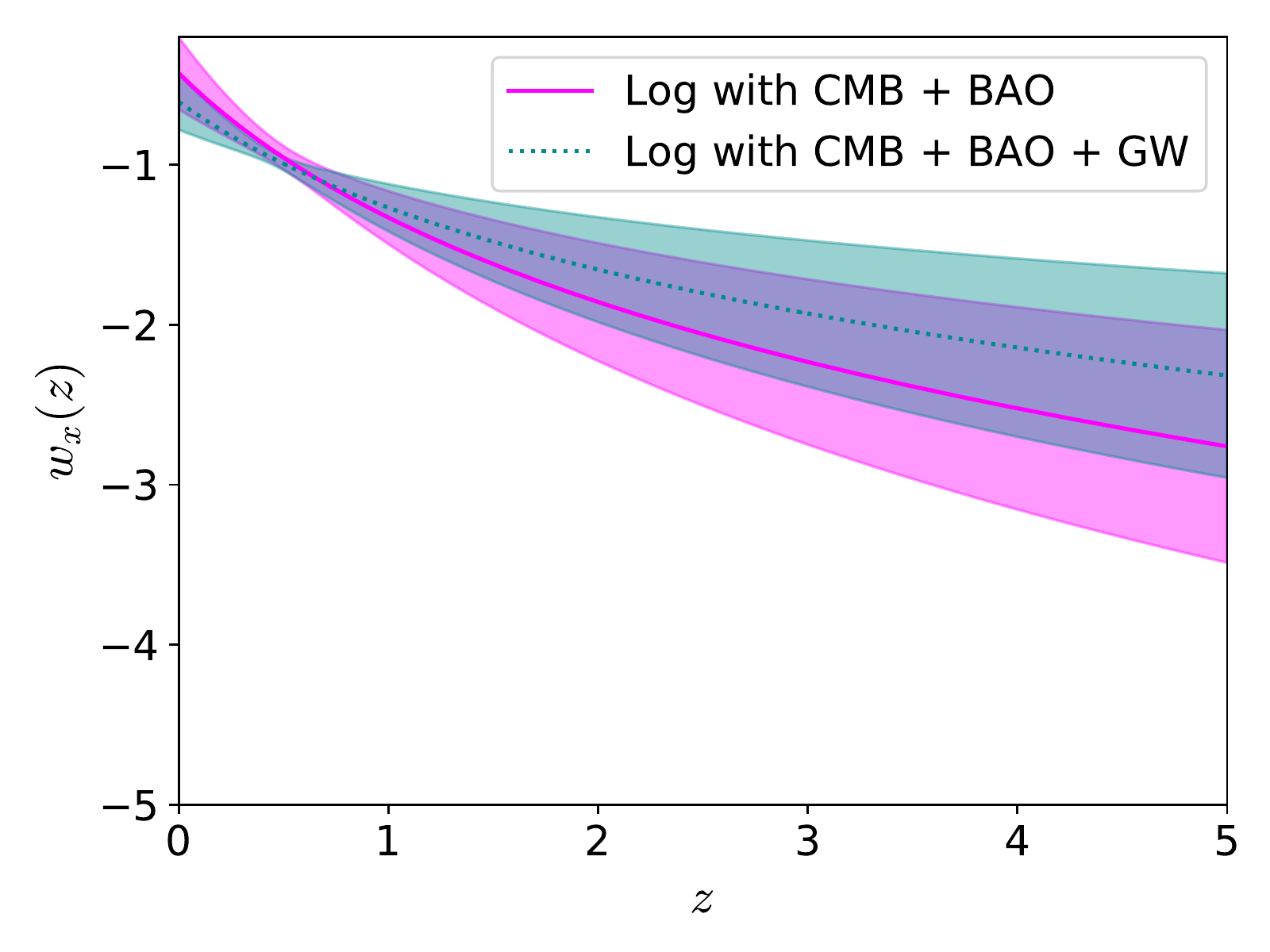}
\includegraphics[width=0.36\textwidth]{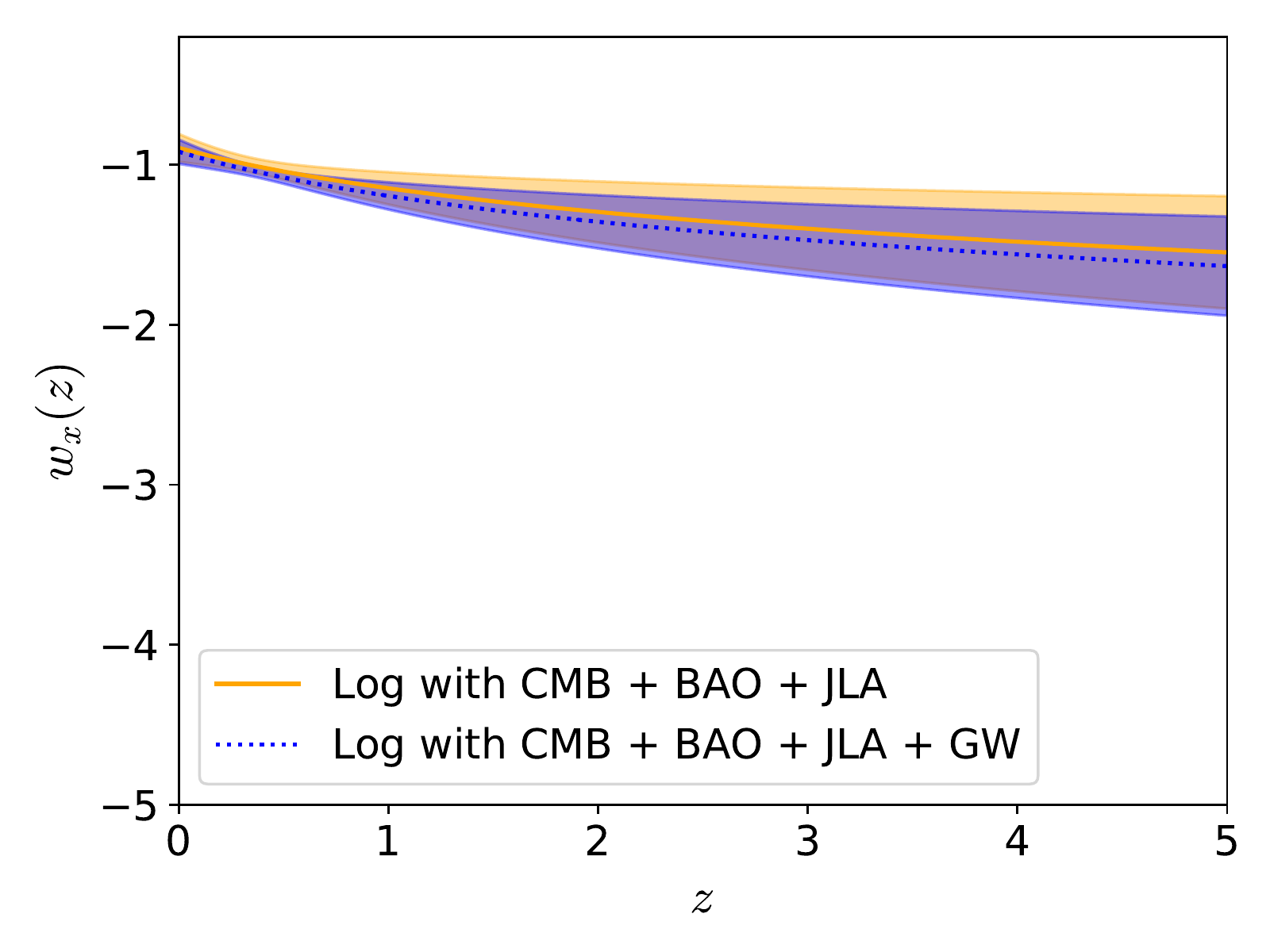}
\includegraphics[width=0.36\textwidth]{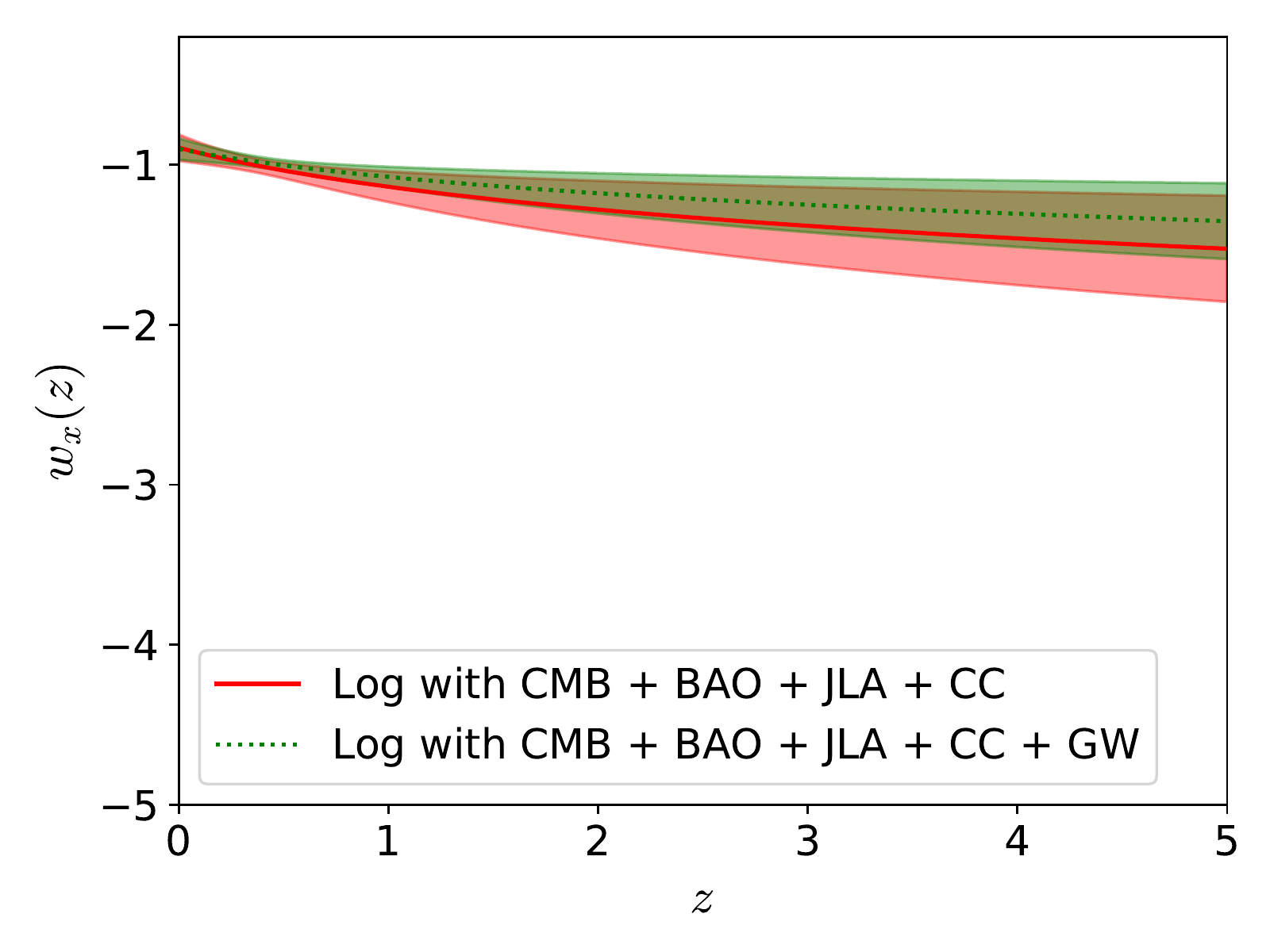}
\caption{\textit{The evolution of the dark energy equation of state for the Logarithmic parametrization has been shown for different datasets taking the mean values of the key parameters $w_0$ and $w_a$ from the analyses with and without the GW data. The solid curves stand for the evolution of $w_x (z)$ for the standard cosmological probes while the dotted curves for the dataset in presence of the GW data. The shaded regions show the 68\% CL constraints on these two parameters.}}
\label{fig-w-log}
\end{figure*}

\subsection{Logarithmic parametrization}
\label{results-log}

In a similar fashion, we  constrain the Logarithmic parametrization (\ref{model-log}) using the standard cosmological probes, such as CMB, BAO, JLA and CC (summarized in the upper half of the Table \ref{tab:results-log}), and then using the best-fit values of the model parameters, we have generated the GW catalogue comprising 1000 simulated GW events. In Fig. \ref{fig-dL-log}, we have shown the relation $d_L (z) $ vs $z$ for the 1000 simulated GW events. Now, using the simulated GW events with the standard cosmological probes, we have constrained this parametrization. The summary of the observational constraints on the CPL model after the inclusion of the simulated GW data are shown in the lower half of Table \ref{tab:results-log}.

In Fig. \ref{fig-contour-log} we show the 1D marginalized posterior distributions for some specific parameters of this model as well as the 2D contour plots considering several combinations of the model parameters.
From a first look at the upper and lower halves of Table
\ref{tab:results-log}, one could clearly see that
the inclusion of GW data to the standard cosmological probes significantly improves the model parameters of this parametrization, a similar observation already found in  CPL parametrization. Let us now describe how GW works with different observational datasets presented here.

We begin the analyses with CMB data alone and CMB+GW. The results of both the analyses are summarized in the second column of Table \ref{tab:results-log}. In the top left panel of Fig. \ref{fig-contour-log}, we compare the constaints on the model parameters from the datasets from which one can clearly see that the inclusion of GW data significantly reduces the error bars on the model parameters. 
In particular, one can notice that both the datasets (CMB and CMB+GW) return very high values of the Hubble constant with similar mean values while the error bars on $H_0$ are reduced significantly after the inclusion of GW to CMB. As one can see that, $H_0 = 82.78_{- 8.34}^{+   15.48}$ (68\% CL, CMB)  and $H_0 = 82.57_{- 1.65}^{+ 1.66}$ (68\% CL, CMB+GW).   It shows that the inclusion of GW reduces the error bars by a factor more than $5$. In fact, for 68\% upper CL errors, this reduction is very very high.  The dark energy equation of state at present, i.e., $w_0$ is constrained to be very close to the cosmological constant boundary `$w_0 = -1$' from both the datasets. One can see that, at 68\% CL CMB data alone constrain, $w_0 =-1.058_{-    0.550}^{+    0.354}$ while for CMB+GW, $w_0 = -1.056_{- 0.196}^{+    0.179}$ (68\% CL). This clearly shows that the addition of GW to CMB significantly reduces the error bars on $w_0$, almost by a factor (not less than) of $2$ and thus reflects the constraining power of GW. Concerning the another free parameter of the model, $w_a$, we find that  CMB alone cannot constrain it while the inclusion of GW could constrain it well
with $w_a  = -1.500_{-    0.571}^{+    0.718}$ (68\% CL, CMB+GW). So, this clearly reflects the constraining power of GW. 

We note that the constraints on $w_a$ are not so stringent due to high error bars. Furthermore, the correlations between the parameters (see the top left panel of Fig. \ref{fig-contour-log}) follow a similar trend as seen for the same datasets with CPL model (i.e., top left panel of Fig. \ref{fig-contour-cpl}). Overall, the constraining power of GW is quite clear from the results.

When BAO is added to CMB (see the third column of Table \ref{tab:results-log} summarizing the results), we find that $H_0$ is significantly lowered with small error bars compared to the constraints from CMB giving, $H_0 = 63.30_{-2.52}^{+    1.87}$ (68\% CL, CMB+BAO), and when the GW is added to CMB+BAO, the error bars are further decreased, but the mean value of $H_0$ slightly increases with, $H_0 =  65.19_{- 1.48}^{+    1.31}$ (68\% CL, CMB+BAO+GW).  The constraints on $\Omega_{m0}$ are significantly high for both the datasets ($\Omega_{m0} = 0.356_{-0.024}^{+    0.026}$ at 68\% CL for CMB+BAO, and $\Omega_{m0} = 0.335_{- 0.015}^{+    0.015}$ at 68\% CL for CMB+BAO+GW), similar to what we found for the CPL parametrization with the same datasets (we refer to the third colum of Table \ref{tab:results-cpl} for comparisons).  Concerning the key parameters $w_0, w_a$ of the model, we see that $w_0$ is very far from $w_0 = -1$ and $w_a$ is very high (considering its magnitude). In Fig. \ref{fig-contour-log} we have compared the constraints between the datasets from which we can see that the parameters are correlated with each other. This result has already been found for the CPL parametrization with the same datasets (compare the top right panels of Fig. \ref{fig-contour-cpl} and \ref{fig-contour-log}).

We now discuss the next analyses with JLA. In particular, we focus on the constraints from CMB+BAO+JLA and CMB+BAO+JLA+GW. The results are summarized in the third column of Table \ref{tab:results-log} and in the bottom left panel of Fig. \ref{fig-contour-log} we compare the constraints from the datasets. We see that the inclusion of JLA improves the constraints from CMB+BAO, that means, the constraints from CMB+BAO+JLA are more stringent than CMB+BAO and the inclusion of GW gives more finest constraints on the parameters.  Specifically, looking at the constraints on the Hubble constant given by, $H_0 = 67.93_{-    1.19}^{+    1.11} $ (68\% CL, CMB+BAO+JLA) and 
$H_0 = 68.88_{-    0.76}^{+    0.73}$ (68\% CL, CMB+BAO+JLA+GW), one can see that the inclusion of GW shifts $H_0$ towards its higher values with lower error bars. 
Concerning the dark energy equation of state at present, $w_0$, we see that for CMB+BAO+JLA dataset, $w_0 = -0.895_{-    0.098}^{+    0.084}$ (68\% CL) and $w_ 0 = -0.919_{-    0.085}^{+    0.071}$ (68\% CL, CMB+BAO+JLA+GW). It shows that the inclusion of GW shifts the dark energy equaton of state towards the cosmological constant boundary with some improvements in the error bars. More precisely, looking at the 1D posterior distributions for $w_0$ shown in the bottom left panel of Fig. \ref{fig-contour-log}, one can see that the highest peaks of $w_0$ are quintessential. 
We also find that the constraints on $w_a$ are significantly lowered (considering its magnitude) compared to the previous two datasets, namley, CMB (and its companion CMB+GW) and CMB+BAO (and its companion CMB+BAO+GW). In particular, the estimations are, $w_a =  -0.365_{-    0.083}^{+    0.365}$ (68\% CL, CMB+BAO+JLA) and $w_a =  -0.399_{-    0.172}^{+    0.251}$ (68\% CL, CMB+BAO+JLA+GW).
Now, finally, looking the the lower left panel of Fig. \ref{fig-contour-log}  one can say that the correlations between the parameters $\sigma_8$ and $w_a$ seems to be absent while the correlations (either positive or negative) with others are still existing after the inclusion of JLA. We would also like to remark that such correlations are not affected by the GW data.

We finish the observational analyses after the inclusion of the Hubble parameter measurements from CC to the previous dataset CMB+BAO+JLA. The results are summarized in the last cloumn of Table \ref{tab:results-log} and the bottom right panel of Fig. \ref{fig-contour-log} corresponds to the comparisons between the datasets. We find that almost all parameters are constrained in a similar way to CMB+BAO+JLA except the key parameters $w_0$, $w_a$ where we have some different observations. We find here for both the datasets, $w_0> -1$ strictly at 68\% CL, this is different from the previous analyses where for CMB+BAO+JLA+GW, $w_0 < -1$ was allowed at 68\% CL. But of course the highest peaks of the 1D posterior distributions of $w_0$ for both CMB+BAO+JLA+CC and its companion CMB+BAO+JLA+CC+GW are in favour of a quintessential dark energy at present. 
The parameter $w_a$ becomes more stringent than its estimation from CMB+BAO+JLA reducing error bars: $w_a = -0.352_{- 0.137}^{+    0.293}$ (68\% CL, CMB+BAO+JLA+CC) and $ w_a = -0.252_{-0.089}^{+    0.220}$ (68\% CL, CMB+BAO+JLA+CC+GW) which show that the standard cosmological probe (i.e., CMB+BAO+JLA+CC) allow $w_a = 0$, in 68\% CL, but the inclusion of GW changes this conclusion favoring the dynamical DE for this parametrization within 68\% CL.

Last but not least, in Fig. \ref{fig-w-log}, we have shown the evolution of the dark energy state for this parametrization using the mean values of $w_0$ and $w_a$ for the observational datasets employed in the work. The solid lines in each plot stand for the $w_x (z)$ curve using the usual cosmological probe and the dotted lines represent the evolution of $w_x (z)$ in presence of the GW data. In each plot the shaded regions (with similar colors to the corresponding curves) present the 68\% regions for the parameters $w_0, w_a$ corresponding to each dataset (with or without the GW data).
This figure gives a qualitative nature of this dark energy equation of state in a nutshell. As one can see that the inclusion of GW to the standard cosmological data certainly improves the parameter space. The maximum effects of GW are visible with the CMB alone data.

\begingroup
\squeezetable
\begin{center}
\begin{table*}
\begin{tabular}{cccccccccccccccc}
\hline\hline
Parameters & CMB & CMB+BAO & CMB+BAO+JLA & CMB+BAO+JLA+CC  \\ \hline

$\Omega_c h^2$ & $    0.1191_{-    0.0014-    0.0028}^{+    0.0014+    0.0029}$  & $    0.1187_{-    0.0013-    0.0024}^{+    0.0013+    0.0025}$ & $    0.1188_{-    0.0013-    0.0026}^{+    0.0013+    0.0025}$ & $    0.1189_{-    0.0012-    0.0025}^{+    0.0013+    0.0025}$  \\

$\Omega_b h^2$ & $    0.02228_{-    0.00016-    0.00031}^{+    0.00015+    0.00031}$ & $    0.02229_{-    0.00015-    0.00029}^{+    0.00015+    0.00030}$  & $    0.02229_{-    0.00014-    0.00028}^{+    0.00014+    0.00029}$ & $    0.02229_{-    0.00014-    0.00030}^{+    0.00014+    0.00030}$  \\

$100\theta_{MC}$ & $    1.04080_{-    0.00033-    0.00066}^{+    0.00035+    0.00064}$
& $    1.04084_{-    0.00032-    0.00061}^{+    0.00031+    0.00063}$  & $    1.04084_{-    0.00030-    0.00061}^{+    0.00031+    0.00063}$ & $    1.04083_{-    0.00030-    0.00059}^{+    0.00031+    0.00063}$  \\

$\tau$ & $    0.076_{-    0.018-    0.035}^{+    0.018+    0.034}$ & $    0.083_{-    0.017-    0.033}^{+    0.017+    0.033}$ & $    0.081_{-    0.017-    0.035}^{+    0.017+    0.034}$ & $    0.081_{-    0.017-    0.033}^{+    0.018+    0.033}$  \\

$n_s$ & $    0.9664_{-    0.0046-    0.0091}^{+    0.0045+    0.0090}$ & $    0.9676_{-    0.0044-    0.0087}^{+    0.0044+    0.0085}$ & $    0.9673_{-    0.0043-    0.0083}^{+    0.0043+    0.0084}$ & $    0.9671_{-    0.0044-    0.0084}^{+    0.0044+    0.0089}$  \\

${\rm{ln}}(10^{10} A_s)$ & $    3.084_{-    0.035-    0.068}^{+    0.036+    0.066}$ & $    3.097_{-    0.034-    0.066}^{+    0.034+    0.064}$ & $    3.093_{-    0.034-    0.068}^{+    0.034+    0.067}$ & $    3.095_{-    0.033-    0.065}^{+    0.034+    0.066}$  \\

$w_0$ & $   -1.423_{-    0.491-    0.577}^{+    0.220+    0.674}$  & $   -0.692_{-    0.144-    0.423}^{+    0.279+    0.346}$ & $   -0.932_{-    0.177-    0.255}^{+    0.115+    0.293}$ & $   -0.893_{-    0.148-    0.247}^{+    0.120 +    0.268}$  \\

$w_a$ & $  < 3$ & $ < 0.214 $ & $   -0.508_{-    0.622-    1.734}^{+    1.017+    1.424}$ & $   -0.737_{-    0.689-    1.514}^{+    0.839+    1.446}$  \\

$\Omega_{m0}$ & $    0.210_{-    0.069-    0.085}^{+    0.027+    0.115}$ & $    0.323_{-    0.017-    0.036}^{+    0.021+    0.034}$ & $    0.306_{-    0.010-    0.019}^{+    0.010+    0.021}$  & $    0.307_{-    0.010-    0.018}^{+    0.010+    0.019}$  \\

$\sigma_8$ & $    0.967_{-    0.062-    0.161}^{+    0.106+    0.141}$  & $    0.819_{-    0.023-    0.045}^{+    0.022+    0.045}$ & $    0.832_{-    0.018-    0.035}^{+    0.018+    0.035}$ & $    0.833_{-    0.018-    0.035}^{+    0.018+    0.035}$  \\

$H_0$ & $   84.01_{-    7.82-   18.75}^{+   13.21+   17.11}$ & $   66.29_{-    2.26-    3.54}^{+    1.58+    3.80}$ & $   68.07_{-    1.09-    2.15}^{+    1.08+    2.15}$ & $   67.95_{-    1.04-    2.00}^{+    1.05+    2.09}$ \\

\hline
\end{tabular}
\begin{tabular}{cccccccccccc}
Parameters & CMB+GW & CMB+BAO+GW & CMB+BAO+JLA+GW & CMB+BAO+JLA+CC+GW & \\ \hline

$\Omega_c h^2$ & $    0.1182_{-    0.0012-    0.0023}^{+    0.0011+    0.0023}$ & $    0.1186_{-    0.0012-    0.0025}^{+    0.0012+    0.0024}$ & $    0.1190_{-    0.0012-    0.0024}^{+    0.0012+    0.0023}$ & $    0.1189_{-    0.0013-    0.0024}^{+    0.0013+    0.0024}$ \\

$\Omega_b h^2$ & $    0.02238_{-    0.00014-    0.00027}^{+    0.00013+    0.00027}$ & $    0.02232_{-    0.00016-    0.00028}^{+    0.00014+    0.00030}$ & $    0.02229_{-    0.00014-    0.00029}^{+    0.00014+    0.00028}$ & $    0.02225_{-    0.00014-    0.00029}^{+    0.00014+    0.00029}$ \\

$100\theta_{MC}$ & $    1.04095_{-    0.00031-    0.00060}^{+    0.00031+    0.00059}$ & $    1.04088_{-    0.00031-    0.00061}^{+    0.00031+    0.00060}$  & $    1.04081_{-    0.00031-    0.00059}^{+    0.00031+    0.00059}$ & $    1.04077_{-    0.00032-    0.00062}^{+    0.00032+    0.00061}$ \\

$\tau$ & $    0.080_{-    0.017-    0.033}^{+    0.017+    0.033}$ & $    0.083_{-    0.018-    0.034}^{+    0.018+    0.034}$  & $    0.081_{-    0.016-    0.034}^{+    0.018+    0.032}$ & $    0.082_{-    0.017-    0.033}^{+    0.017+    0.032}$ \\

$n_s$ & $    0.9689_{-    0.0042-    0.0081}^{+    0.0041+    0.0081}$ & $    0.9680_{-    0.0042-    0.0084}^{+    0.0042+    0.0081}$ & $    0.9667_{-    0.0042-    0.0082}^{+    0.0041+    0.0084}$ & $    0.9669_{-    0.0042-    0.0084}^{+    0.0041+    0.0087}$ \\

${\rm{ln}}(10^{10} A_s)$ & $    3.092_{-    0.033-    0.065}^{+    0.033+    0.065}$ & $    3.098_{-    0.035-    0.066}^{+    0.035+    0.065}$ & $    3.094_{-    0.032-    0.066}^{+    0.035+    0.062}$ & $    3.097_{-    0.033-    0.064}^{+    0.032+    0.064}$ \\

$w_0$ & $   -1.213_{-    0.097-    0.240}^{+    0.152+    0.218}$ & $   -0.672_{-    0.106-    0.370}^{+    0.234+    0.286}$ & $   -0.925_{-    0.131-    0.220}^{+    0.108+    0.225}$ & $   -0.982_{-    0.132-    0.193}^{+    0.080+    0.215}$ \\

$w_a$ & $ < -0.126$ & $ < -0.019 $ & $   -0.683_{-    0.549-    1.346}^{+    0.828+    1.234}$ & $   -0.029_{-    0.391-    1.141}^{+    0.755+    1.003}$ \\

$\Omega_{m0}$ & $    0.206_{-    0.003-    0.006}^{+    0.003+    0.006}$ & $    0.320_{-    0.009-    0.023}^{+    0.014+    0.020}$ & $    0.302_{-    0.007-    0.011}^{+    0.006+    0.012}$  & $    0.314_{-    0.008-    0.014}^{+    0.007+    0.015}$ \\

$\sigma_8$ & $    0.957_{-    0.016-    0.033}^{+    0.016+    0.032}$ & $    0.821_{-    0.017-    0.034}^{+    0.017+    0.036}$ & $    0.840_{-    0.014-    0.030}^{+    0.016+    0.029}$ & $    0.826_{-    0.015-    0.032}^{+    0.017+    0.030}$ \\

$H_0$ & $   82.73_{-    0.54-    0.97}^{+    0.49+    1.02}$ & $   66.54_{-    1.41-    2.08}^{+    0.89+    2.42}$ & $   68.60_{-    0.60-    1.19}^{+    0.61+    1.22}$ & $   67.19_{-    0.71-    1.50}^{+    0.79+    1.40}$ \\
\hline\hline
\end{tabular}
\caption{Observational constraints at 68\% and 95\% confidence levels for the JBP parametrization (\ref{model-jbp}) using various combinations of the observational data with and without the GW data.
The upper panel represents the constraints on the model without the GW data while in the lower panel we present the corresponding constraints using the GW data. For the $w_a$ parameter the sign `$<$' denotes that we report its 95\% CL constraint.
Here $\Omega_{m0}$ is the present value of $\Omega_m = \Omega_b +\Omega_c$ and $H_0$ is in the units of km s$^{-1}$Mpc$^{-1}$.  }
\label{tab:results-jbp}
\end{table*}
\end{center}
\endgroup
\begin{figure*}
\includegraphics[width=0.35\textwidth]{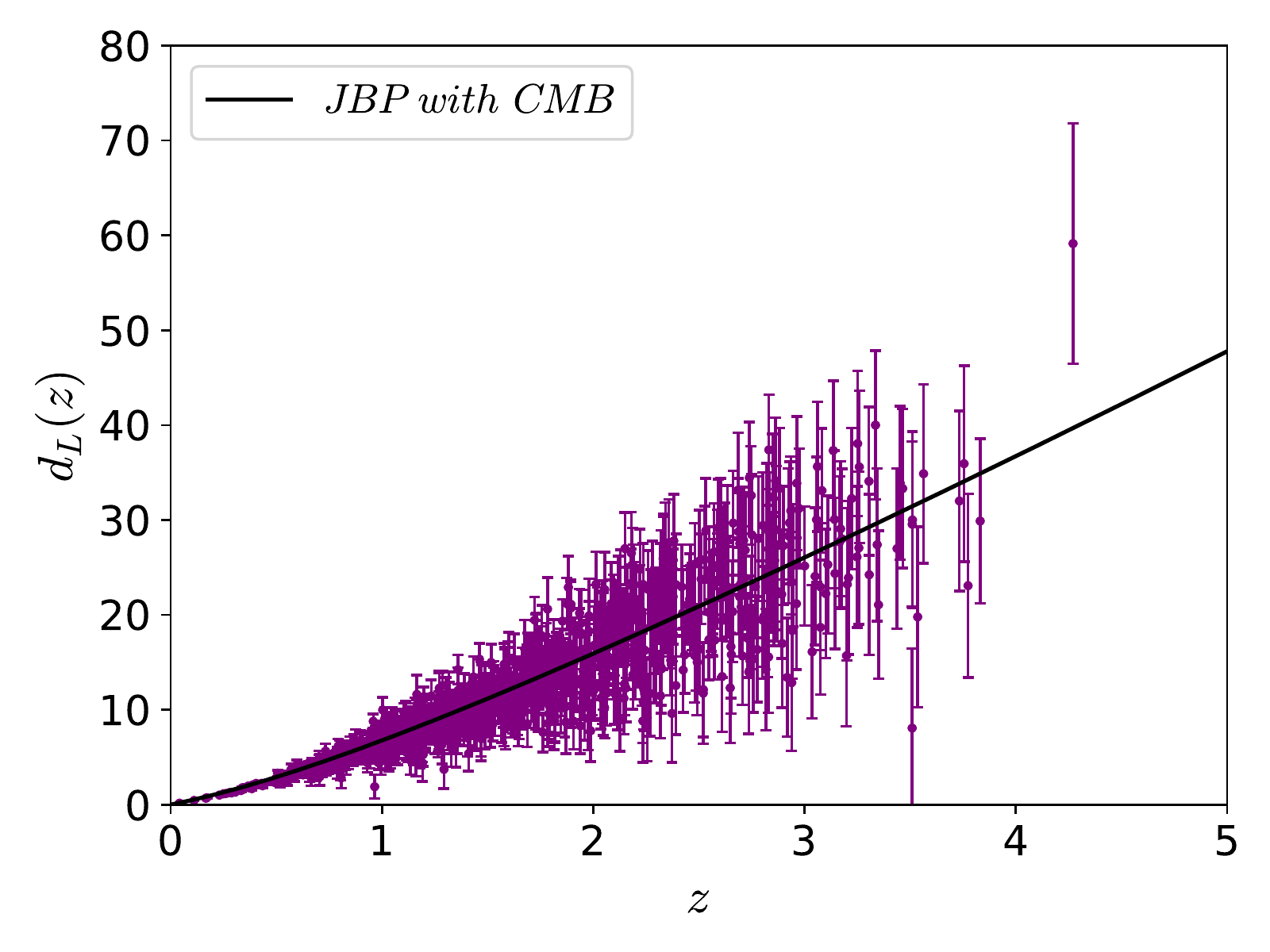}
\includegraphics[width=0.35\textwidth]{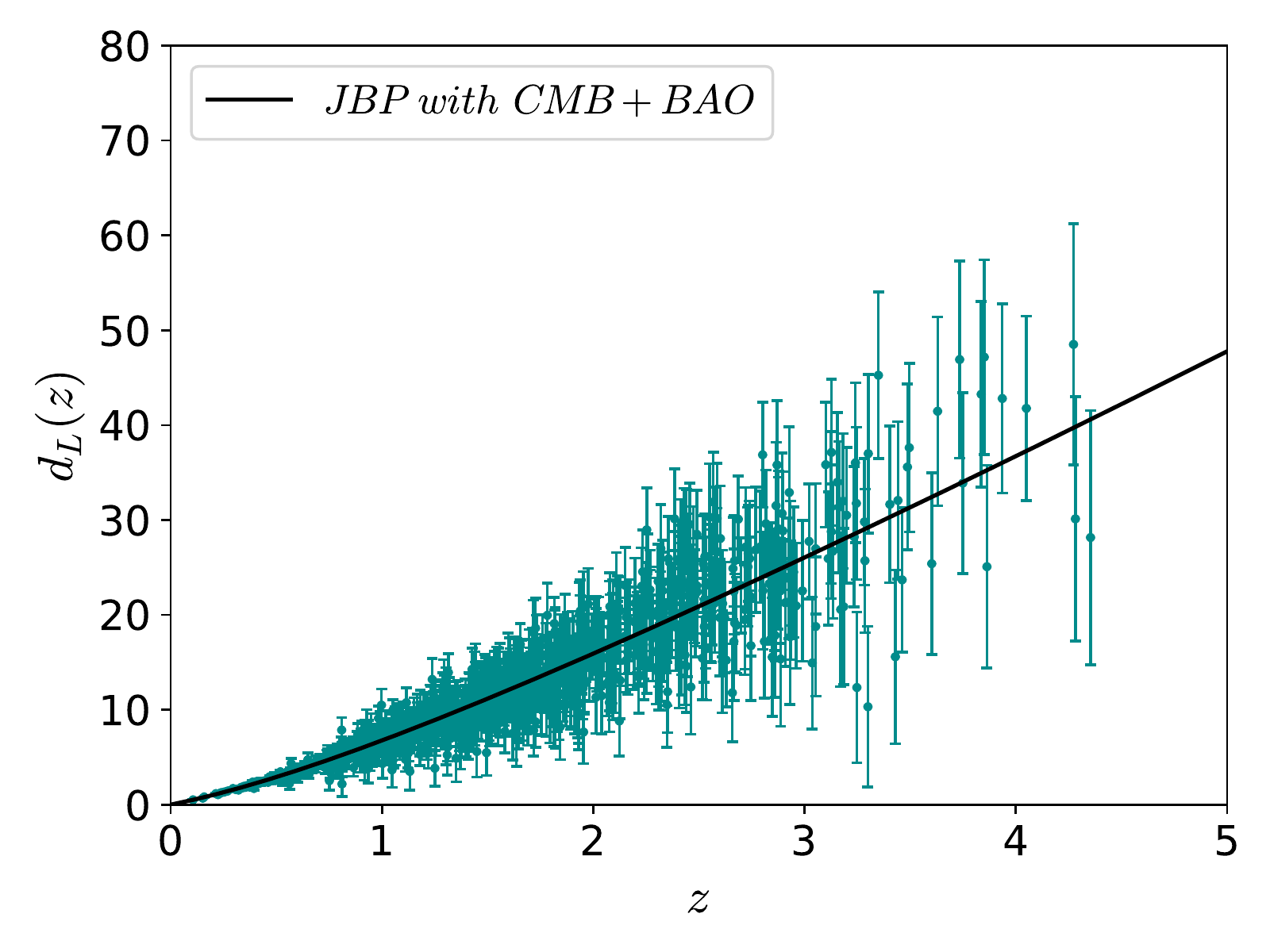}
\includegraphics[width=0.35\textwidth]{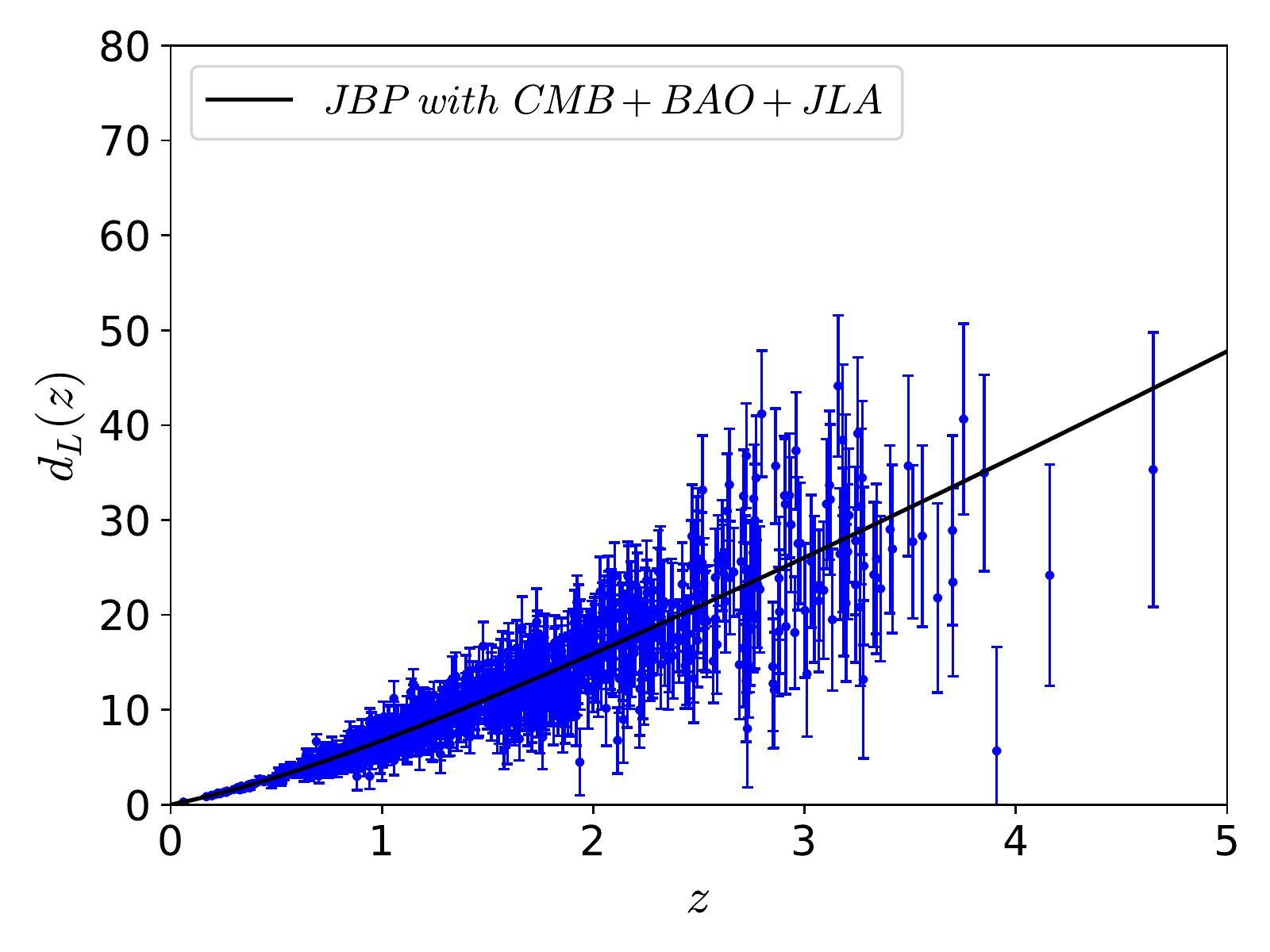}
\includegraphics[width=0.35\textwidth]{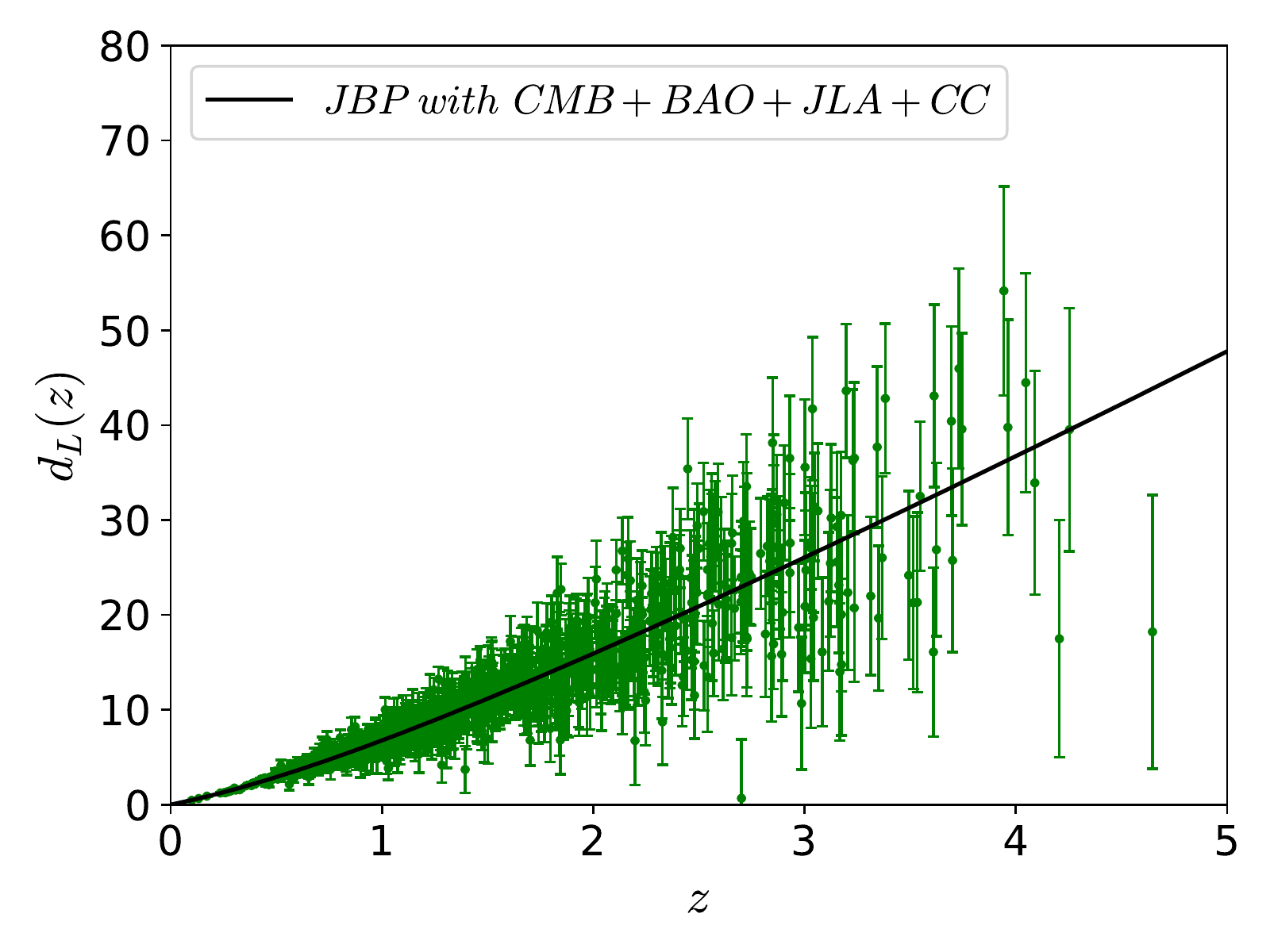}
\caption{\textit{For the fiducial JBP model, we first constrain the cosmological parameters using the datasets CMB, CMB+BAO, CMB+BAO+JLA and CMB+BAO+JLA+CC and then we use the best-fit values of the parameters for ``each dataset'' to generate the corresponding GW catalogue. Following this, in each panel we show $d_L (z)$ vs $z$ catalogue with the corresponding error bars for 1000 simulated GW events. The upper left and upper right panels respectively present the catalogue ($z$, $d_L (z)$) with the corresponding error bars for 1000 simulated GW events derived using the CMB alone and CMB+BAO dataset. The lower left and lower right panels respectively present the catalogue ($z$, $d_L (z)$) with the corresponding error bars for 1000 simulated GW events derived using the CMB+BAO+JLA and CMB+BAO+JLA+CC datasets.}}
\label{fig-dL-jbp}
\end{figure*}
\begin{figure*}
\includegraphics[width=0.45\textwidth]{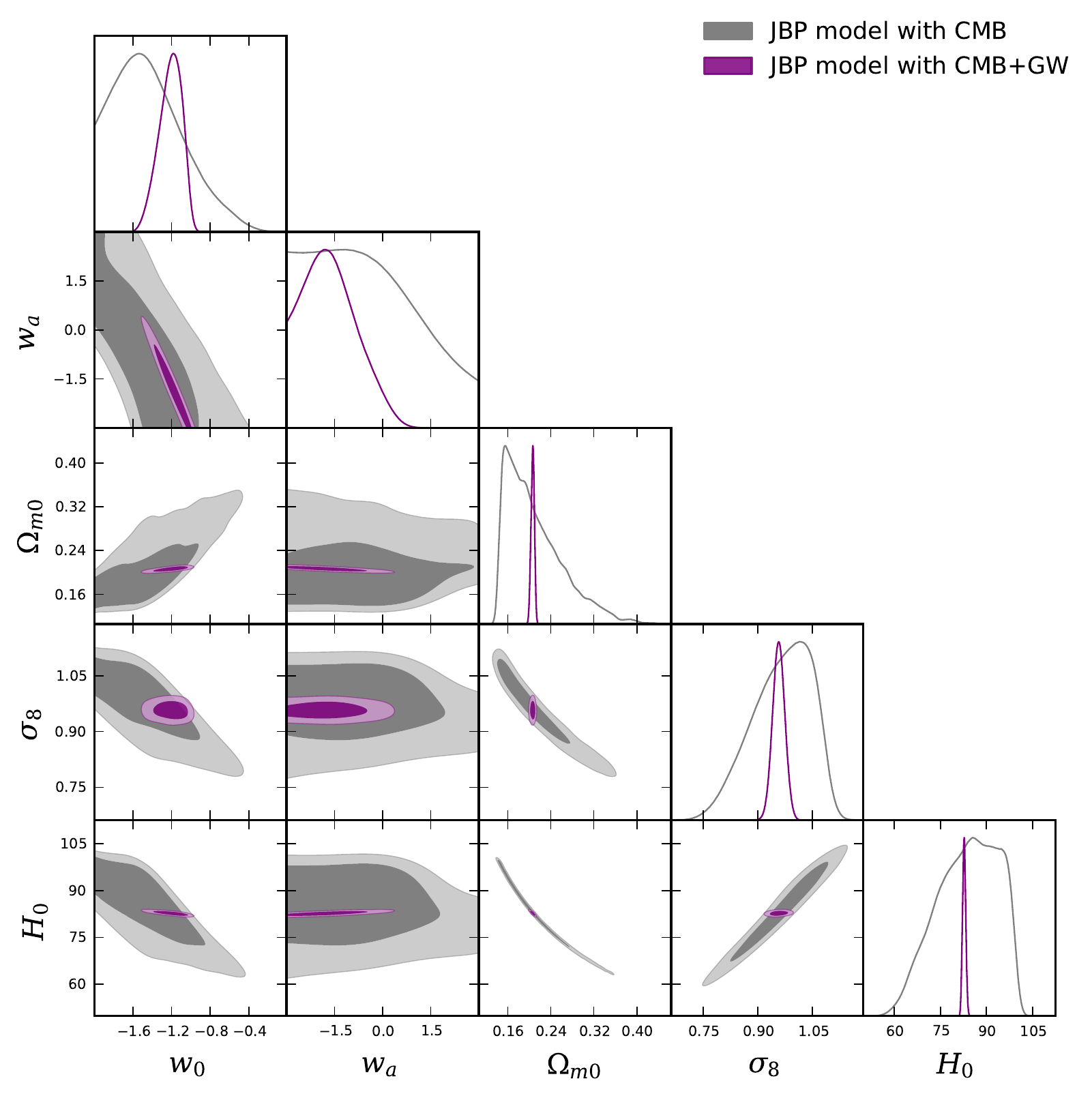}
\includegraphics[width=0.45\textwidth]{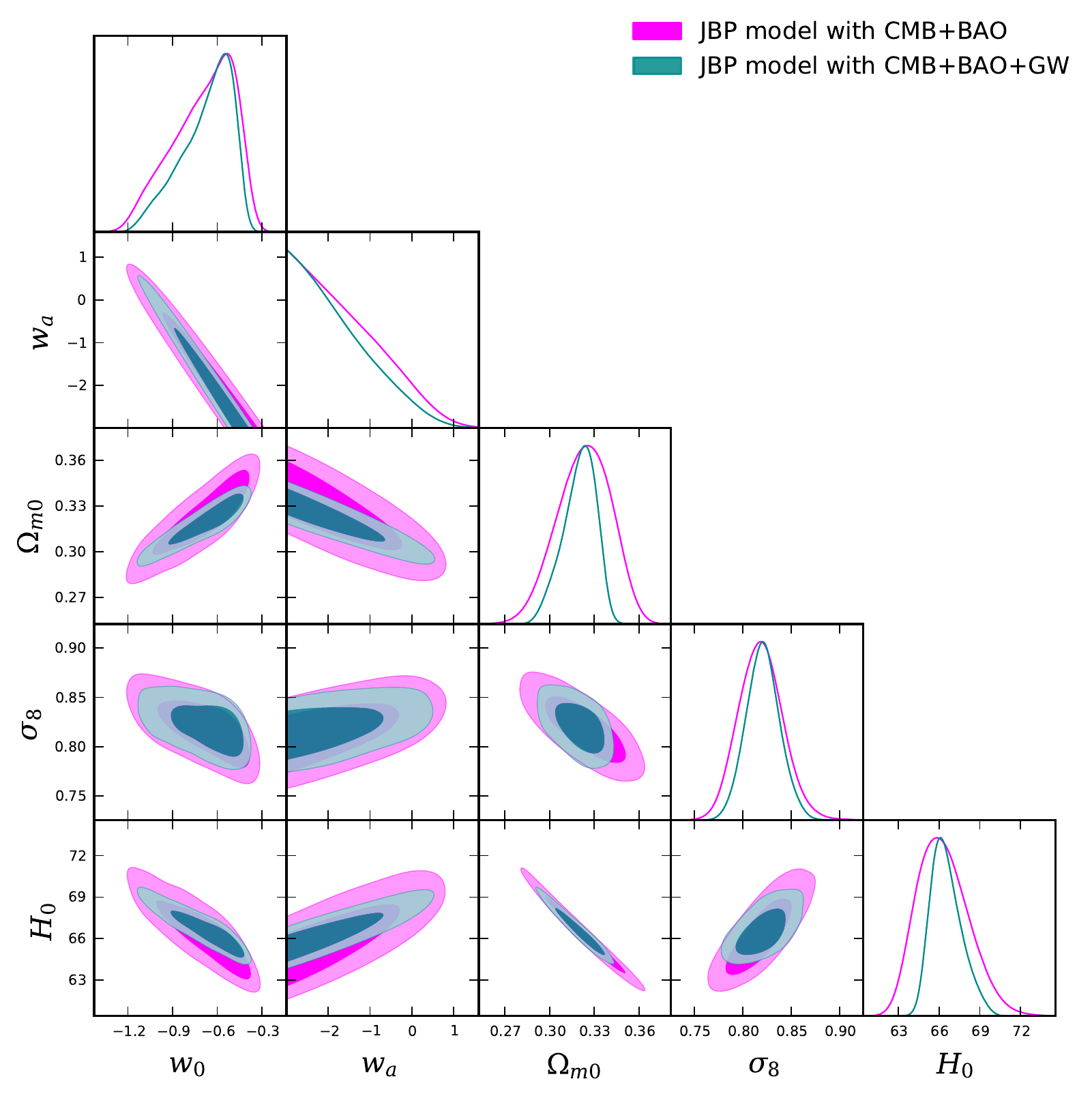}
\includegraphics[width=0.45\textwidth]{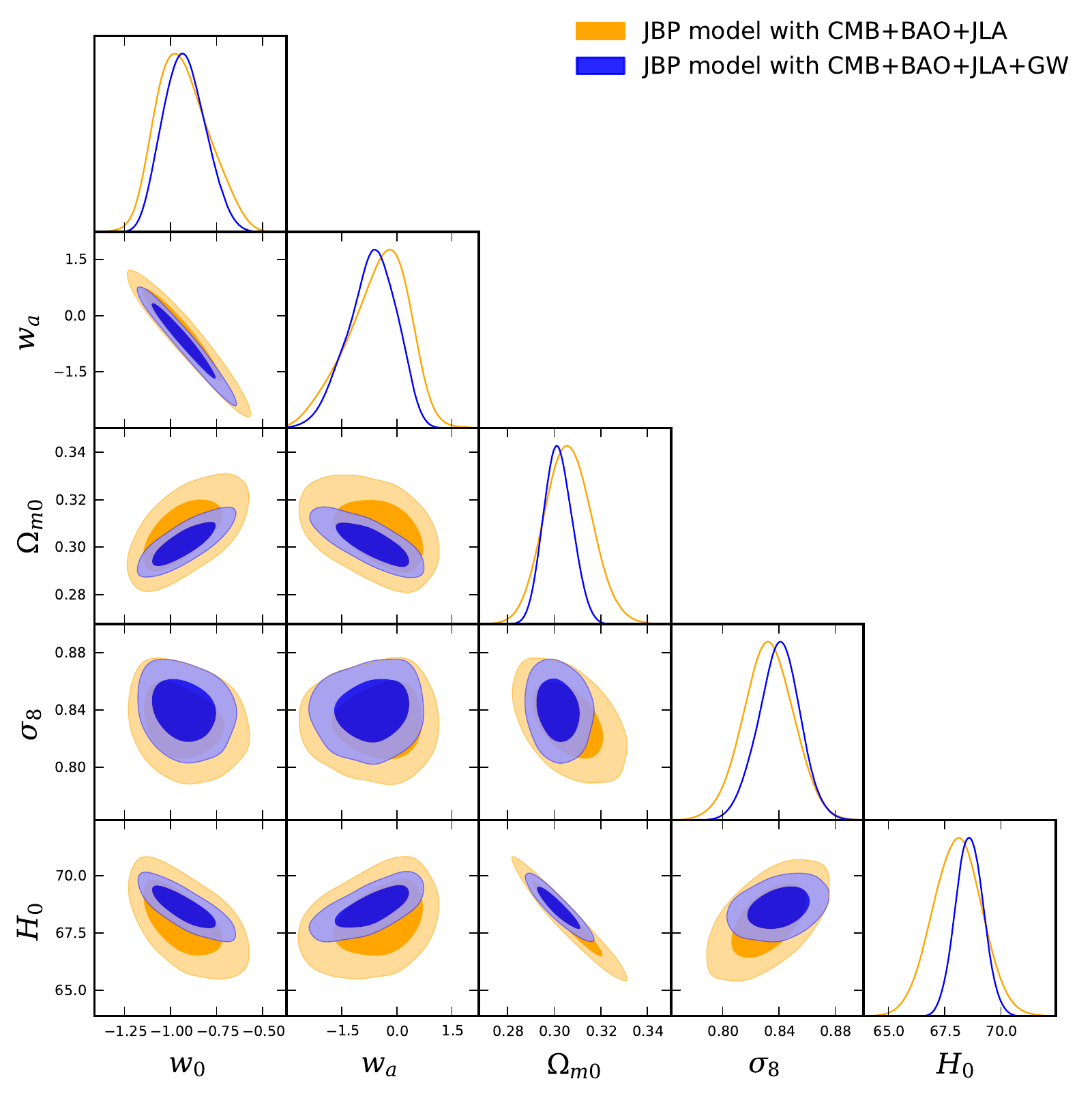}
\includegraphics[width=0.45\textwidth]{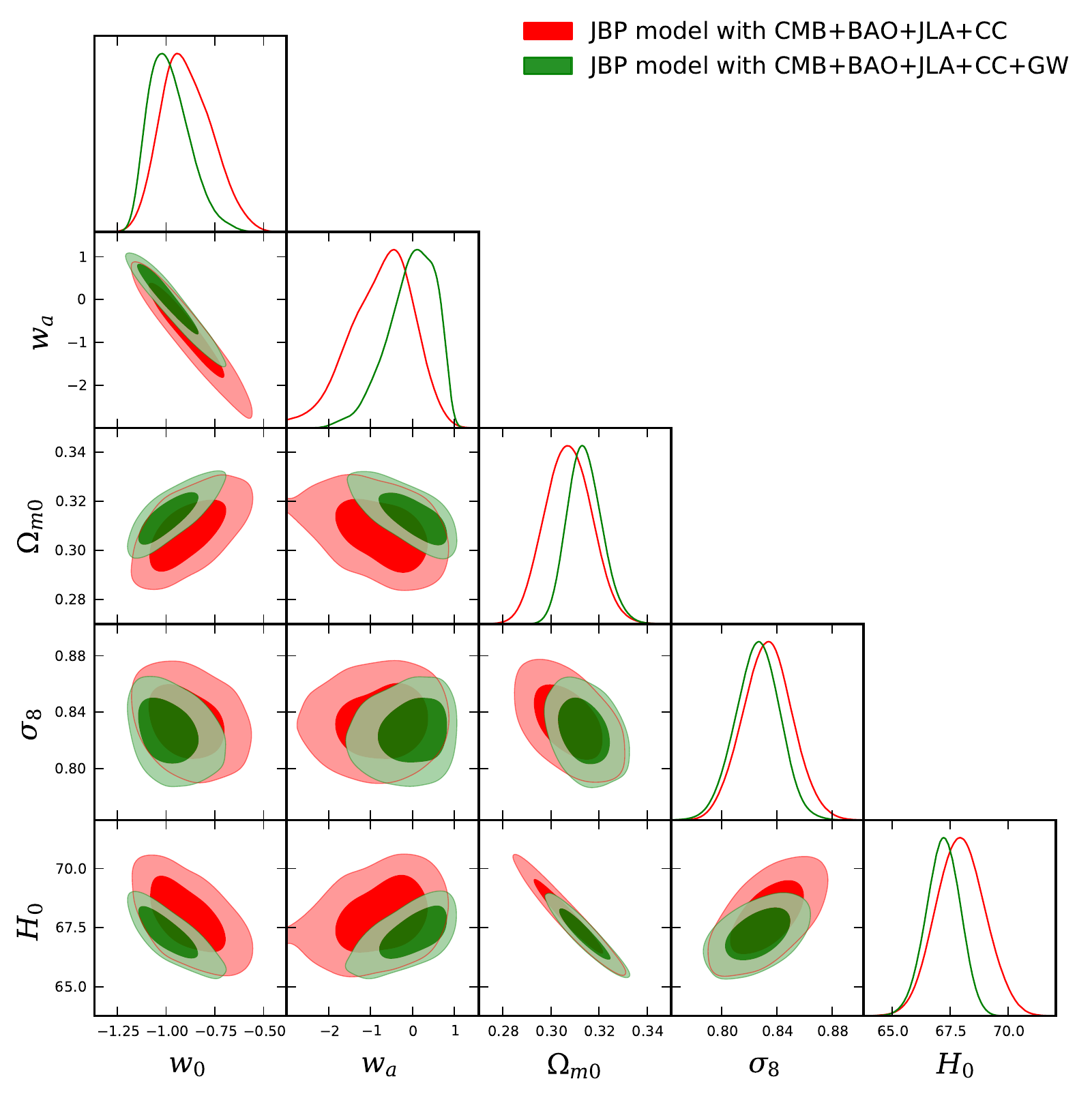}
\caption{\textit{68\% and 95\% CL contour plots for various combinations of some selected parameters of the JBP model (\ref{model-jbp}) using different  observational data in presence (absence) of the GW data.}}
\label{fig-contour-jbp}
\end{figure*}
\begin{figure*}
\includegraphics[width=0.36\textwidth]{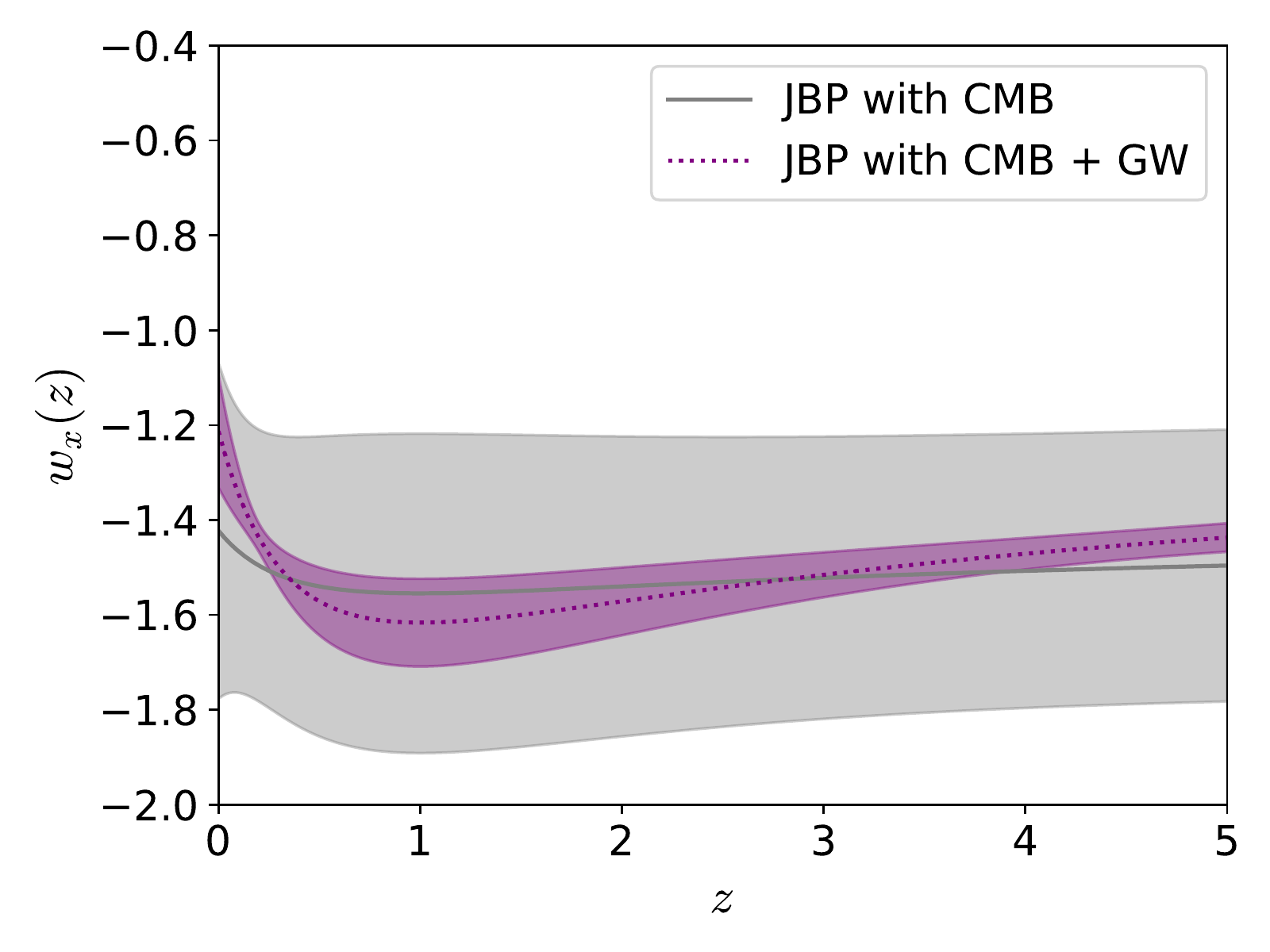}
\includegraphics[width=0.36\textwidth]{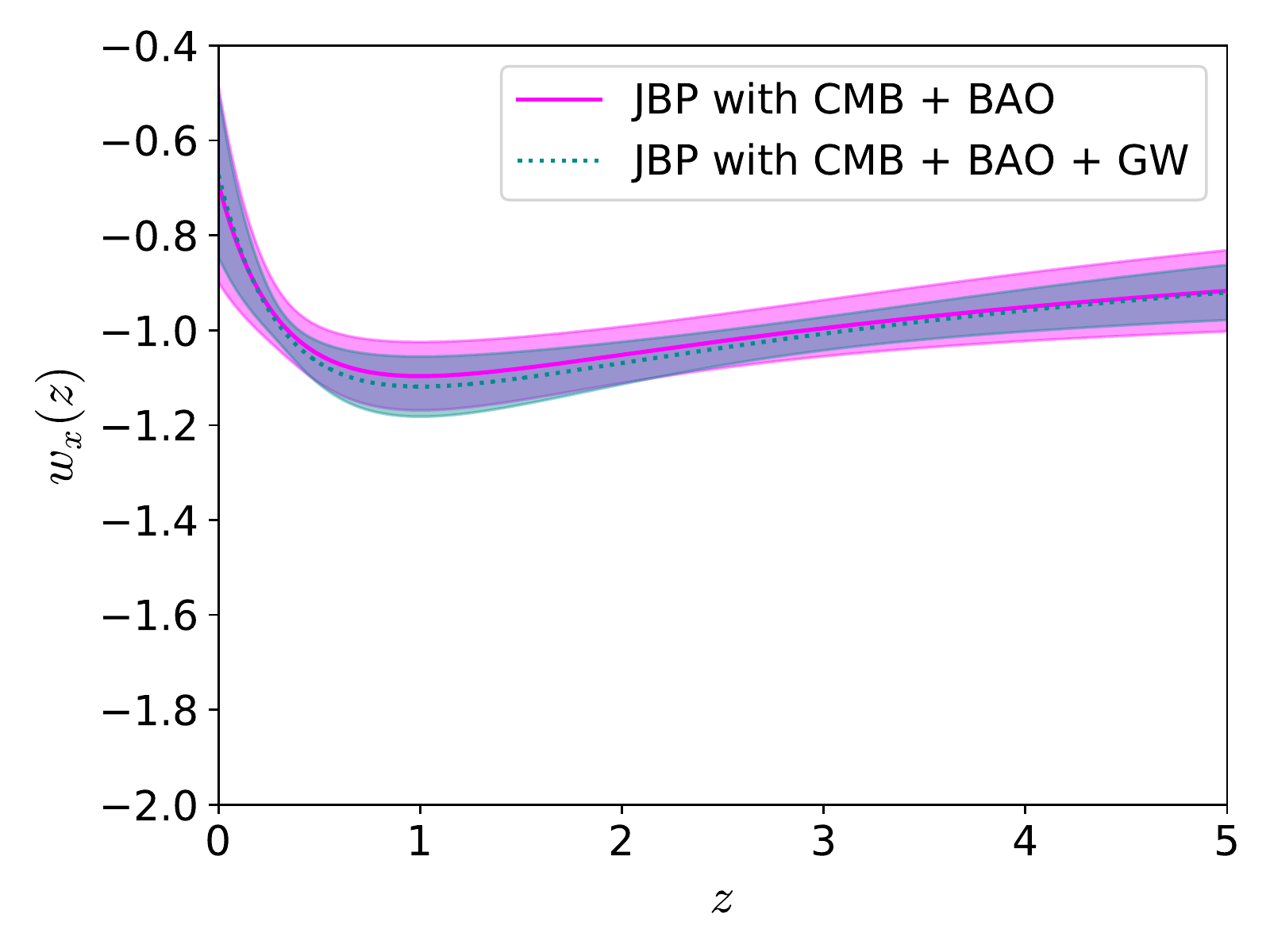}
\includegraphics[width=0.36\textwidth]{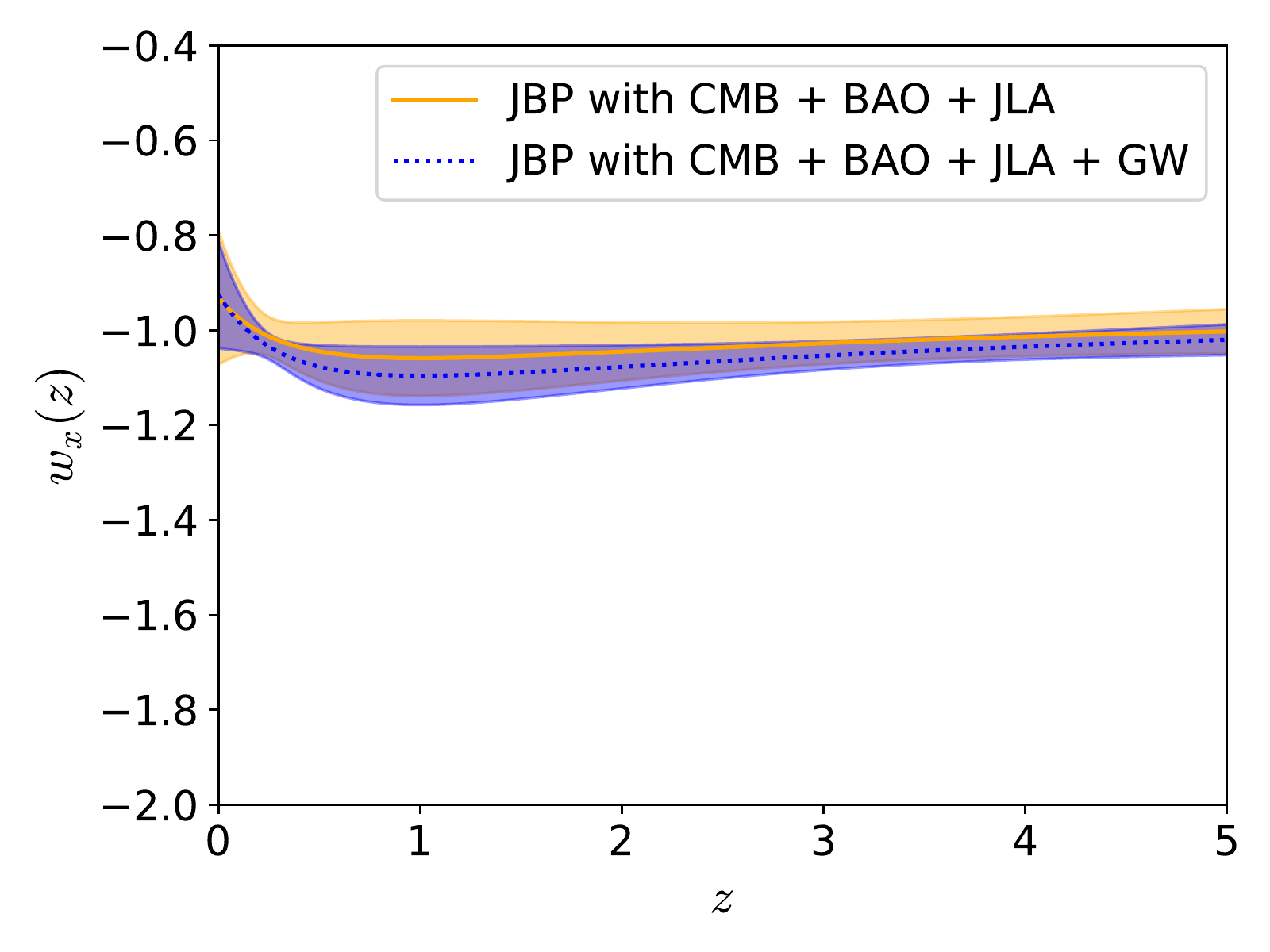}
\includegraphics[width=0.36\textwidth]{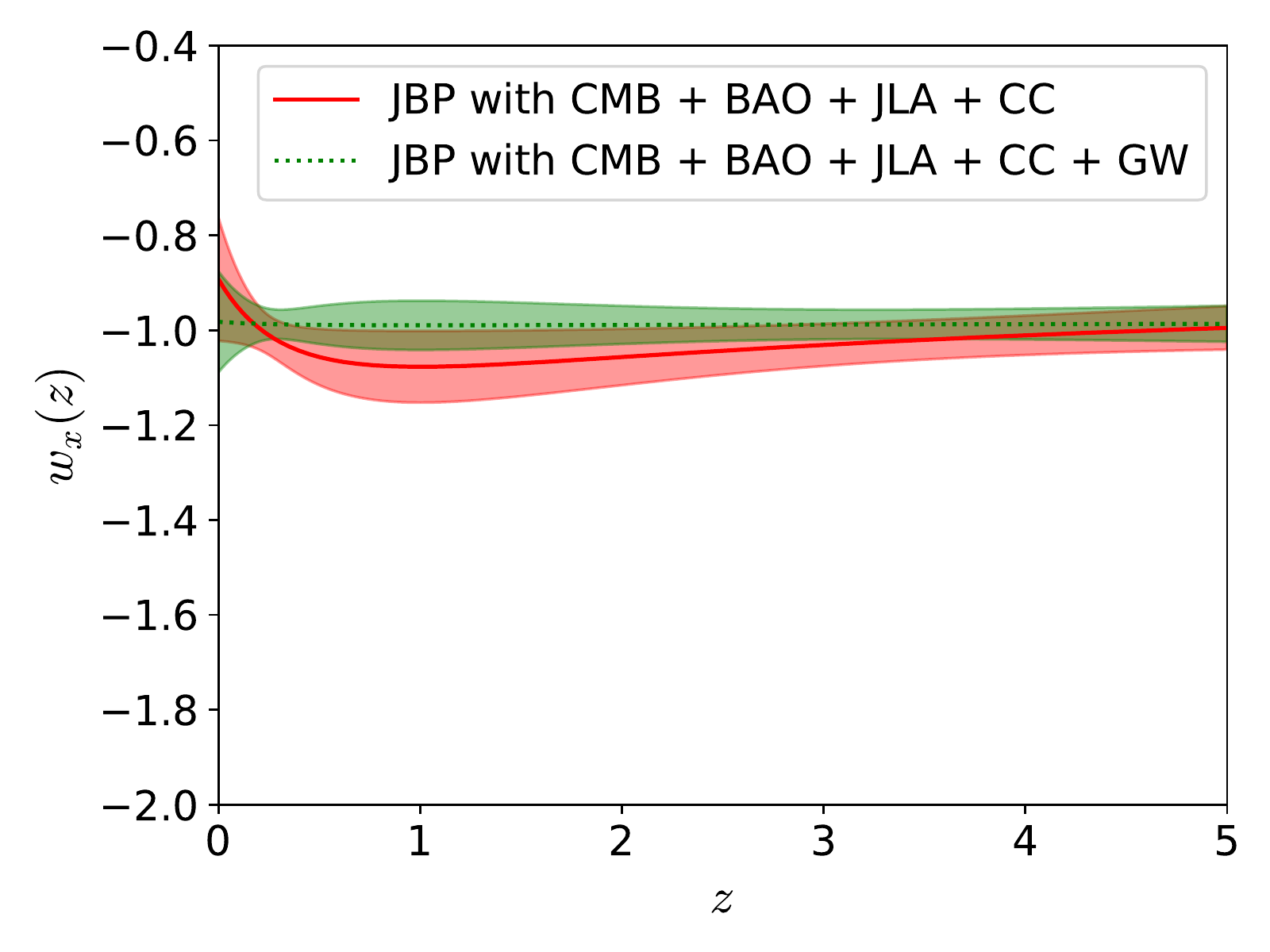}
\caption{\textit{The evolution of the dark energy equation of state for the JBP parametrziation has been shown for different datasets taking the mean values of the key parameters $w_0$ and $w_a$ from the analyses with and without the GW data. The solid curves stand for the evolution of $w_x (z)$ for the standard cosmological probes while the dotted curves for the dataset in presence of the GW data. The shaded regions show the 68\% CL constraints on these two parameters.}}
\label{fig-w-jbp}
\end{figure*}

\subsection{JBP parametrization}
\label{results-jbp}

We now discuss the observational constraints on the JBP parametrization (\ref{model-jbp}) in order to investigate the effects of GW data on the cosmological parameters of this model. We first constrain this  parametrization using the standard cosmological probes (summarized in the upper half of the Table \ref{tab:results-jbp}), and then using the best-fit values of the model parameters, we generate the GW catalogue comprising 1000 simulated GW events. In Fig. \ref{fig-dL-jbp}, we have shown the relation $d_L (z) $ vs $z$ for the 1000 simulated GW events. Now, taking into account the simulated GW events with the standard cosmological probes, we have constrained the model parameters which are summarized in the lower half of Table \ref{tab:results-cpl}. Following a similar strategy, we display Fig. \ref{fig-contour-jbp} that clearly depicts the effects of GW on the cosmological parameters.

We first discuss the constraints from the CMB data alone and the CMB+GW dataset summarized in the second column of Table \ref{tab:results-jbp}. In the top left panel of Fig. \ref{fig-contour-jbp}, we present the comparisons between the observational constraints obtained from these two datasets. Our analyses report that
for both the datasets, $H_0$ assumes very high values with
$H_0= 84.01_{- 7.82}^{+   13.21}$ (68\% CL, CMB) and $H_0= 82.73_{-0.54}^{+    0.49}$ (68\% CL, CMB+GW). Clearly, one can see that the inclusion of simulated GW data decreases the  error bars on $H_0$ by a factor of at least $15$. This is one of the interesting conclusions and similar conclusion has been found in earlier dark energy parametrizations such as CPL and Logarithmic. 
The reduction in the error bars for other cosmological parameters are equally true after the inclusion of the simulated GW data, see the top left panel of Fig. \ref{fig-contour-jbp} for a better viewing. Similar effects on the key two free parameters of this model, namely, $w_0$ and $w_a$ are observed.  An interesting remark might be the allowance of phantom  nature of $w_0$ by both the datasets, namely, CMB and CMB+GW at more than 68\% CL. Moreover, the highest peaks of the 1D posteriror distributions of $w_0$ for both the datasets, namely, CMB and CMB+GW, the phantom nature of $w_0$ is strongly suggested.
Concerning the $w_a$ parameter, we note that the addition of GW to CMB, in a similar fashion improves its parameter space (see the top left panel of Fig. \ref{fig-contour-jbp}). 

The inclusion of BAO to the former datasets works in a similar fashion as observed in the previous two parametrizations. The results of the analyses are summarized in the third column of Table \ref{tab:results-jbp} while we compare the constraints in the top right panel of Fig. \ref{fig-contour-jbp}. From this figure (i.e., the top right panel of Fig. \ref{fig-contour-jbp}), we can see that the parameters shown there are correlated with each other, where one may perhaps recognize that the correlation between $w_a$ and $\sigma_8$ is relatively low compared to other combinations in this figure. There is one more point that we should remark here.  Although $H_0$ assumes slightly lower values that the estimation from Planck's team but compared to the previous two dynamical DE parametrizations, $H_0$ is  relatively higher, $H_0 = 66.29_{- 2.26}^{+    1.58}$ and $H_0 = 66.54_{-1.41}^{+    0.89}$ (68\% CL, CMB+BAO). Similar effects are seen in the estimations of the present day matter density parameter $\Omega_{m0}$.

Let us now discuss the observational constraints from CMB+BAO+JLA and its companion CMB+BAO+JLA+GW. The fourth column of Table \ref{tab:results-jbp} summarizes the constraints on the parameters and the bottom left panel of Fig. \ref{fig-contour-jbp} corresponds to the comparison of the datasets. From the figure (i.e., bottom left panel of Fig. \ref{fig-contour-jbp}) we see that although the correlations between the parameters are present, but $\sigma_8$ does not seem to be correlated with $w_0$, $w_a$, at least, a very mild correlation might be present which is not pronounced from the plots. The Hubble constant is shifted towards the higher values after the inclusion of GW and the error bars are reduced slightly. This feature has been observed in other parametrizations. Looking   at the dark energy equation of state at present, $w_0$, we see that indeed the inclusion of GW improves the constraints on $w_0$ but not significantly. From the 1D posterior distributions of $w_0$ (see the bottom left panel of Fig. \ref{fig-contour-jbp}), it is indeed seen that the highest peaks of $w_0$ for both the datasets are very very close to $-1$, that means the cosmological constant is favoured. Furthermore, when GW data are added, a very minimal shift of the higest peak of $w_0$ towards quintessence regime is observed. On the other hand, from the  numerical estimations of the $w_a$ parameter, $w_a = -0.508_{- 0.622}^{+    1.017}$ (68\% CL, CMB+BAO+JLA) and $w_a  =  -0.683_{-0.549}^{+ 0.828} $ (68\% CL, CMB+BAO+JLA+GW), one can clearly see  that the inclusion of JLA to the previous dataset CMB+BAO reduces the magnitude of $w_a$. However, due to the very large error bars on $w_a$,
the case $w_a =0$ is definitely allowed within 68\% CL. We also notice that for CMB+BAO+JLA+GW, the highest peak of the 1D posterior distribution of $w_a$ (see the bottom left panel of Fig. \ref{fig-contour-jbp}) is shifted towards more nagative values compared to the highest peak of $w_a$ for CMB+BAO+JLA. This is true because here $w_0$ and $w_a$ are negatively correlated, see again the $(w_0, w_a)$ plane shown in the bottom left panel of Fig. \ref{fig-contour-jbp}.

Finally, we come up with the last two analyses, namely,  CMB+BAO+JLA+CC and CMB+BAO+JLA+CC+GW. The results are summarized in the last column of Table \ref{tab:results-jbp} and in the bottom right corner of Fig. \ref{fig-contour-jbp}, we compare these datasets.
The only surprising result is observed in the constraints on one of the key parameters of this model, namely, $w_a$  where we find that the 68\% constraints on this parameter are, $w_a = -0.737_{-0.689}^{+    0.839}$ (CMB+BAO+JLA+CC) and $w_a =  -0.029_{-0.391}^{+    0.755}$. One can clearly visualize the effect of GW onto this parameter where precisely the estimated value of $w_a$ is remarkably lowered together with significant reduction of its error bars. 
This is an interesting result and we remark that the previous two dynamical DE parametrizations, namely, CPL and Logarithmic did not exhibit such behaviour. We further note that although the mean values of $w_0$ attained from both the datasets are non-phantom, however, from the 1D posterior distributions of $w_0$ (bottom right corner of Fig. \ref{fig-contour-jbp}), we see that the inclusion of GW to CMB+BAO+JLA+CC shifts the highest peak of $w_0$ towards the phantom regime.

We close this analysis with the  Fig. \ref{fig-w-jbp}, where we present the qualitative evolution of the dark energy equation of state for this parametrization using the mean values of $w_0$ and $w_a$ obtained from the observational datasets employed in the work.  The solid lines in each plot stand for the $w_x (z)$ curve using the usual cosmological probe and the dotted lines represent the evolution of $w_x (z)$ in presence of the GW data. In each plot the shaded regions (with similar colors to the corresponding curves) present the 68\% regions for the parameters $w_0, w_a$ corresponding to each dataset (with or without the GW data). From the graphs, one can clearly see how GW data affect the cosmological parameters. The maximum effects of GW are seen from the CMB alone case.  We note that the dotted curve for CMB+BAO+JLA+CC+GW (see the bottom right graph of Fig. \ref{fig-w-jbp}) is almost a straight line $w_x (z) = -1$.

\begingroup
\squeezetable
\begin{center}
\begin{table*}
\begin{tabular}{ccccccccc}
\hline\hline
Parameters & CMB & CMB+BAO & CMB+BAO+JLA & CMB+BAO+JLA+CC &\\ \hline

$\Omega_c h^2$ & $    0.1191_{-    0.0014-    0.0027}^{+    0.0014+    0.0028}$ & $    0.1190_{-    0.0014-    0.0025}^{+    0.0013+    0.0026}$ & $    0.1192_{-    0.0013-    0.0027}^{+    0.0013+    0.0027}$ & $    0.1189_{-    0.0013-    0.0025}^{+    0.0012+    0.0024}$ \\

$\Omega_b h^2$ & $    0.02228_{-    0.00015-    0.00030}^{+    0.00016+    0.00031}$ & $    0.02227_{-    0.00015-    0.00028}^{+    0.00015+    0.00029}$ & $    0.02226_{-    0.00015-    0.00029}^{+    0.00015+    0.00030}$ &  $    0.02229_{-    0.00015-    0.00028}^{+    0.00015+    0.00030}$  \\

$100\theta_{MC}$ & $ 1.04079_{-    0.00033-    0.00065}^{+    0.00033+    0.00063}$ & $    1.04080_{-    0.00031-    0.00061}^{+    0.00032+    0.00061}$ & $    1.04077_{-    0.00032-    0.00064}^{+    0.00032+    0.00065}$ & $    1.04081_{-    0.00031-    0.00061}^{+    0.00031+    0.00061}$  \\

$\tau$ & $    0.076_{-    0.017-    0.033}^{+    0.017+    0.034}$  & $    0.079_{-    0.017-    0.034}^{+    0.017+    0.033}$ & $    0.079_{-    0.018-    0.034}^{+    0.018+    0.033}$ & $ 0.082_{-    0.019-    0.033}^{+    0.017+    0.034}$  \\

$n_s$ & $    0.9665_{-    0.0045-    0.0089}^{+    0.0046+    0.0087}$ & $    0.9668_{-    0.0045-    0.0084}^{+    0.0042+    0.0090}$ & $    0.9662_{-    0.0044-    0.0087}^{+    0.0044+    0.0090}$ & $0.9673_{-    0.0041-    0.0085}^{+    0.0042+    0.0087}$  \\

${\rm{ln}}(10^{10} A_s)$ & $    3.085_{-    0.033-    0.065}^{+    0.033+    0.066}$ & $    3.091_{-    0.033-    0.066}^{+    0.033+    0.064}$ & $    3.091_{-    0.034-    0.067}^{+    0.034+    0.065}$  & $    3.097_{-    0.034-    0.064}^{+    0.033+    0.067}$  \\

$w_0$ & $   -1.386_{-    0.556-    0.614}^{+    0.203+    0.761}$ & $   -0.692_{-    0.374-    0.486}^{+    0.215+    0.589}$ & $   -0.898_{-    0.090-    0.174}^{+    0.093+    0.182}$ & $ -0.933_{-    0.066-    0.139}^{+    0.064+    0.142}$  \\

$w_a$ & $  < -0.038 < 0.613 $ & $   -0.509_{-    0.282-    0.920}^{+    0.577+    0.722}$ & $   -0.263_{-    0.165-    0.388}^{+    0.211+    0.361}$  & $-0.173_{-    0.109-    0.281}^{+    0.137+    0.235}$  \\

$\Omega_{m0}$ & $    0.215_{-    0.075-    0.092}^{+    0.024+    0.136}$ & $    0.334_{-    0.038-    0.055}^{+    0.024+    0.061}$ & $    0.308_{-    0.010-    0.018}^{+    0.009+    0.020}$ & $    0.307_{-    0.010-    0.019}^{+    0.009+    0.019}$  \\

$\sigma_8$ & $    0.964_{-    0.058-    0.178}^{+    0.114+    0.146}$ & $    0.811_{-    0.027-    0.055}^{+    0.027+    0.053}$ & $    0.835_{-    0.018-    0.036}^{+    0.018+    0.037}$  & $    0.835_{-    0.017-    0.034}^{+    0.017+    0.034}$  \\

$H_0$ & $   83.55_{-    7.20-   20.82}^{+   14.43+   17.75}$ &  $   65.41_{-    3.03-    5.33}^{+    3.00+    5.39}$  & $   67.91_{-    1.06-    2.10}^{+    1.05+    2.07}$ & $67.98_{-    0.99-    2.05}^{+   1.00 +  2.03}$ \\

\end{tabular}
\begin{tabular}{ccccccccc}
\hline
Parameters & CMB+GW & CMB+BAO+GW & CMB+BAO+JLA+GW & CMB+BAO+JLA+CC+GW & \\ \hline

$\Omega_c h^2$ & $    0.1185_{-    0.0012-    0.0024}^{+    0.0013+    0.0024}$ & $    0.1193_{-    0.0013-    0.0027}^{+    0.0013+    0.0026}$ & $    0.1191_{-    0.0012-    0.0025}^{+    0.0013+    0.0026}$ & $    0.1190_{-    0.0013-    0.0024}^{+    0.0013+    0.0025}$\\

$\Omega_b h^2$ & $    0.02234_{-    0.00014-    0.00027}^{+    0.00014+    0.00027}$ & $    0.02226_{-    0.00015-    0.00029}^{+    0.00015+    0.00029}$ & $    0.02226_{-    0.00015-    0.00029}^{+    0.00014+    0.00030}$ & $    0.02224_{-    0.00014-    0.00028}^{+    0.00015+    0.00028}$ \\

$100\theta_{MC}$ & $    1.04090_{-    0.00031-    0.00061}^{+    0.00031+    0.00060}$ & $    1.04077_{-    0.00033-    0.00063}^{+    0.00032+    0.00064}$ & $    1.04078_{-    0.00031-    0.00062}^{+    0.00031+    0.00060}$ & $    1.04076_{-    0.00031-    0.00060}^{+    0.00031+    0.00062}$ \\

$\tau$ & $    0.079_{-    0.017-    0.032}^{+    0.016+    0.032}$ & $    0.078_{-    0.017-    0.035}^{+    0.018+    0.033}$ & $    0.080_{-    0.016-    0.032}^{+    0.016+    0.032}$ & $    0.081_{-    0.017-    0.034}^{+    0.017+    0.033}$ \\

$n_s$ & $    0.9679_{-    0.0042-    0.0079}^{+    0.0042+    0.0081}$ & $    0.9660_{-    0.0044-    0.0088}^{+    0.0044+    0.0086}$ & $    0.9665_{-    0.0043-    0.0084}^{+    0.0043+    0.0084}$ & $    0.9666_{-    0.0044-    0.0082}^{+    0.0043+    0.0084}$ \\

${\rm{ln}}(10^{10} A_s)$ & $    3.089_{-    0.032-    0.063}^{+    0.032+    0.062}$ & $    3.089_{-    0.033-    0.068}^{+    0.034+    0.065}$ & $    3.091_{-    0.032-    0.064}^{+    0.032+    0.063}$  & $    3.094_{-    0.034-    0.066}^{+    0.033+    0.065}$\\

$w_0$ & $   -1.253_{-    0.153-    0.255}^{+    0.131+    0.281}$ & $   -0.711_{-    0.155-    0.256}^{+    0.128+    0.284}$ & $   -0.925_{-    0.070-    0.136}^{+    0.070+    0.142}$ & $   -1.000_{-    0.075-    0.144}^{+    0.074+    0.150}$ \\

$w_a$ & $   -0.516_{-    0.295-    0.669}^{+    0.372+    0.637}$ & $   -0.523_{-    0.213-    0.517}^{+    0.296+    0.479}$ & $   -0.186_{-    0.130-    0.287}^{+    0.151+    0.258}$ & $    0.004_{-    0.123-    0.274}^{+    0.142+    0.247}$ \\

$\Omega_{m0}$ & $    0.210_{-    0.008-    0.014}^{+    0.007+    0.015}$ & $    0.328_{-    0.013-    0.025}^{+    0.012+    0.026}$ & $    0.309_{-    0.006-    0.013}^{+    0.006+    0.013}$ & $    0.312_{-    0.007-    0.014}^{+    0.007+    0.014}$ \\

$\sigma_8$ & $    0.955_{-    0.019-    0.036}^{+    0.019+    0.037}$ & $    0.818_{-    0.017-    0.033}^{+    0.017+    0.034}$ & $    0.833_{-    0.015-    0.029}^{+    0.015+    0.030}$ & $    0.828_{-    0.015-    0.029}^{+    0.015+    0.031}$ \\

$H_0$ & $   82.12_{-    1.32-    2.69}^{+    1.34+    2.54}$ & $   65.88_{-    1.17-    2.31}^{+    1.20+    2.35}$ & $   67.83_{-    0.65-    1.27}^{+    0.64+    1.30}$ & $   67.46_{-    0.70-    1.36}^{+    0.69+    1.42}$ \\

\hline\hline
\end{tabular}
\caption{Observational constraints at 68\% and 95\% confidence levels for the Barboza-Alcaniz parametrization (\ref{model-ba}) using various combinations of the observational data with and without the GW data.
The upper panel represents the constraints on the model without the GW data while in the lower panel we present the corresponding constraints using the GW data. For the CMB only case the upper limits of the $w_a$ parameter at 68\% and 95\% CL are reported.
Here, $\Omega_{m0}$ is the present value of $\Omega_m = \Omega_b +\Omega_c$ and $H_0$ is in the units of km s$^{-1}$Mpc$^{-1}$. }
\label{tab:results-ba}
\end{table*}
\end{center}
\endgroup
\begin{figure*}
\includegraphics[width=0.35\textwidth]{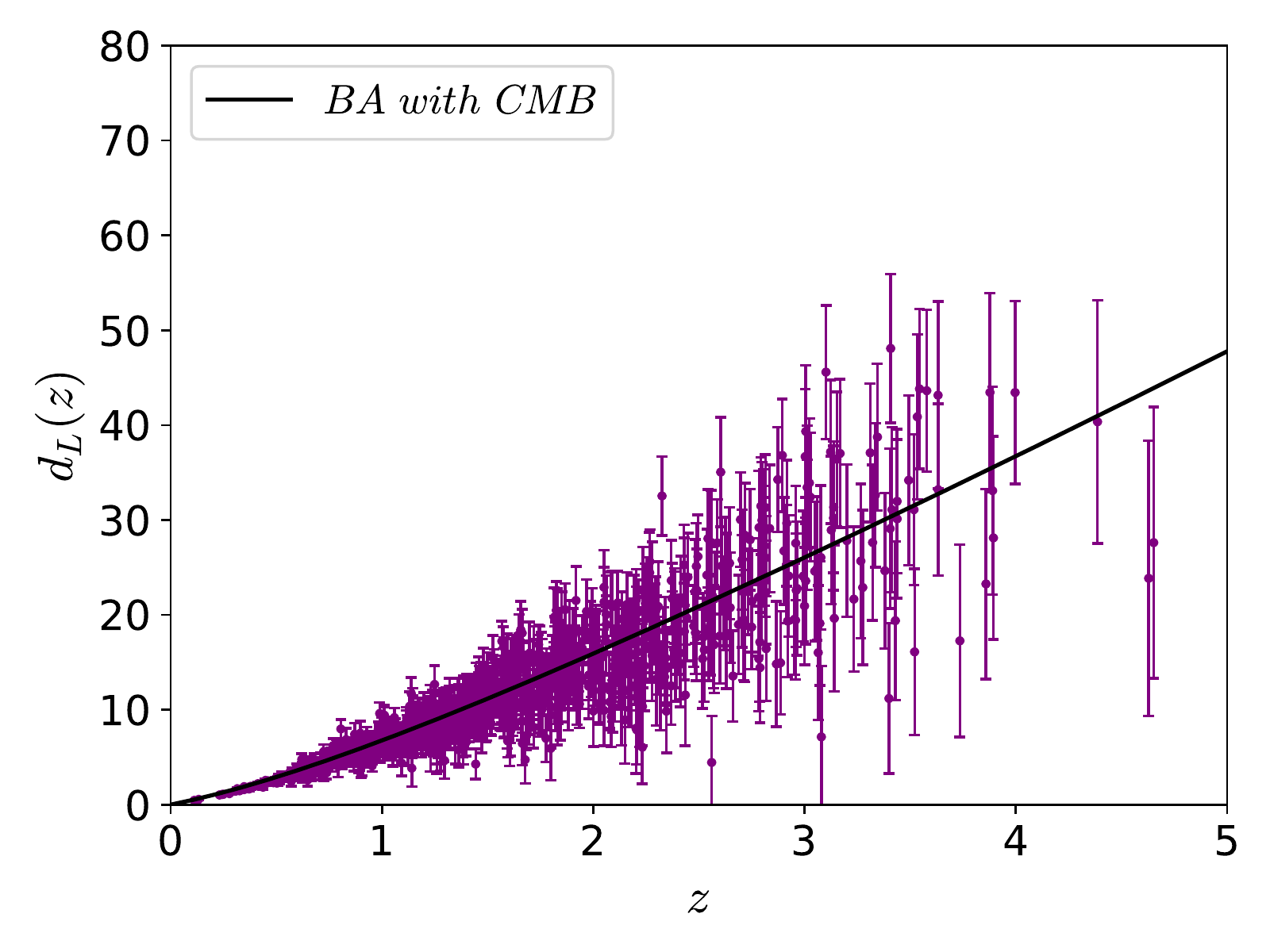}
\includegraphics[width=0.35\textwidth]{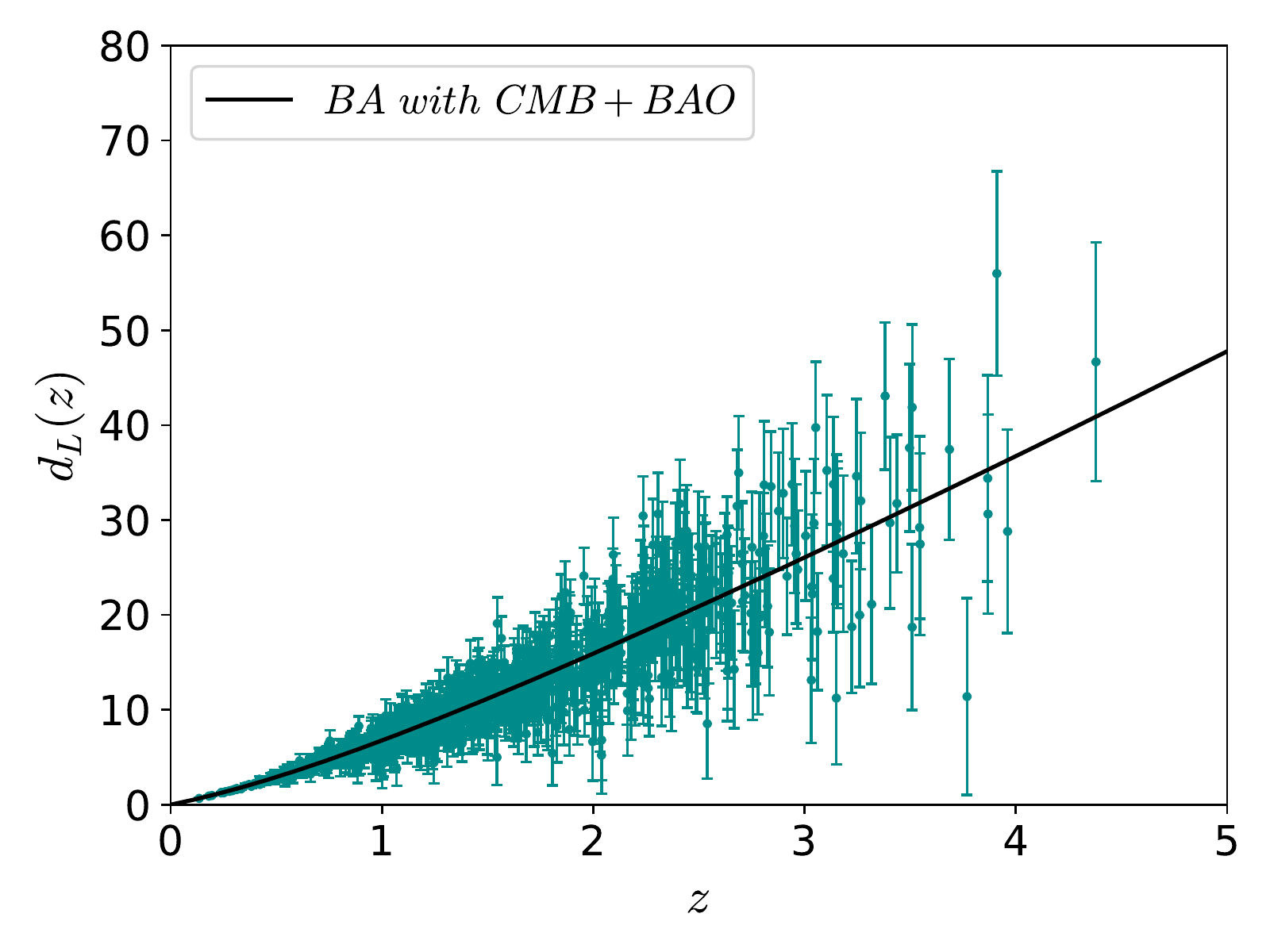}
\includegraphics[width=0.35\textwidth]{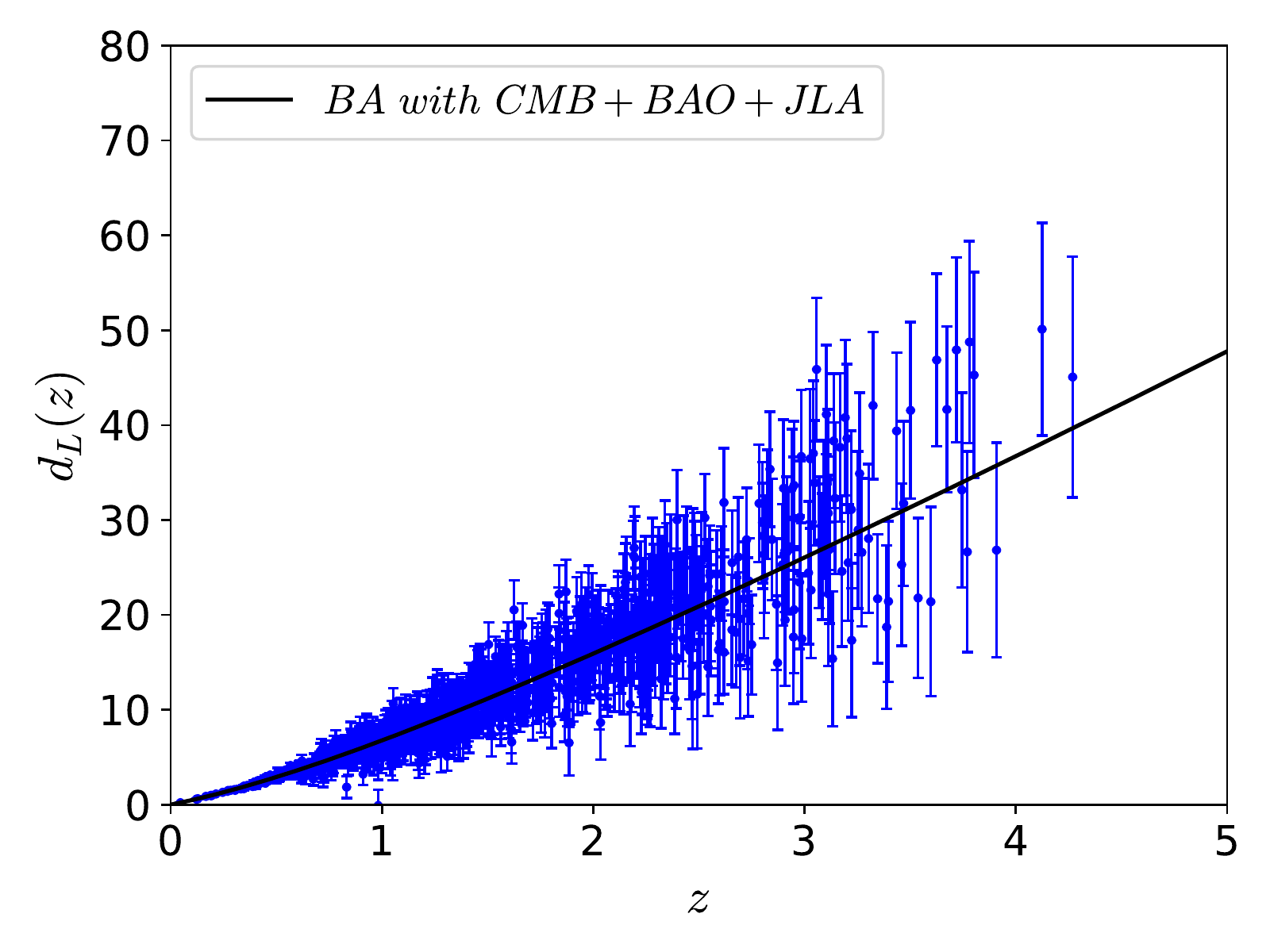}
\includegraphics[width=0.35\textwidth]{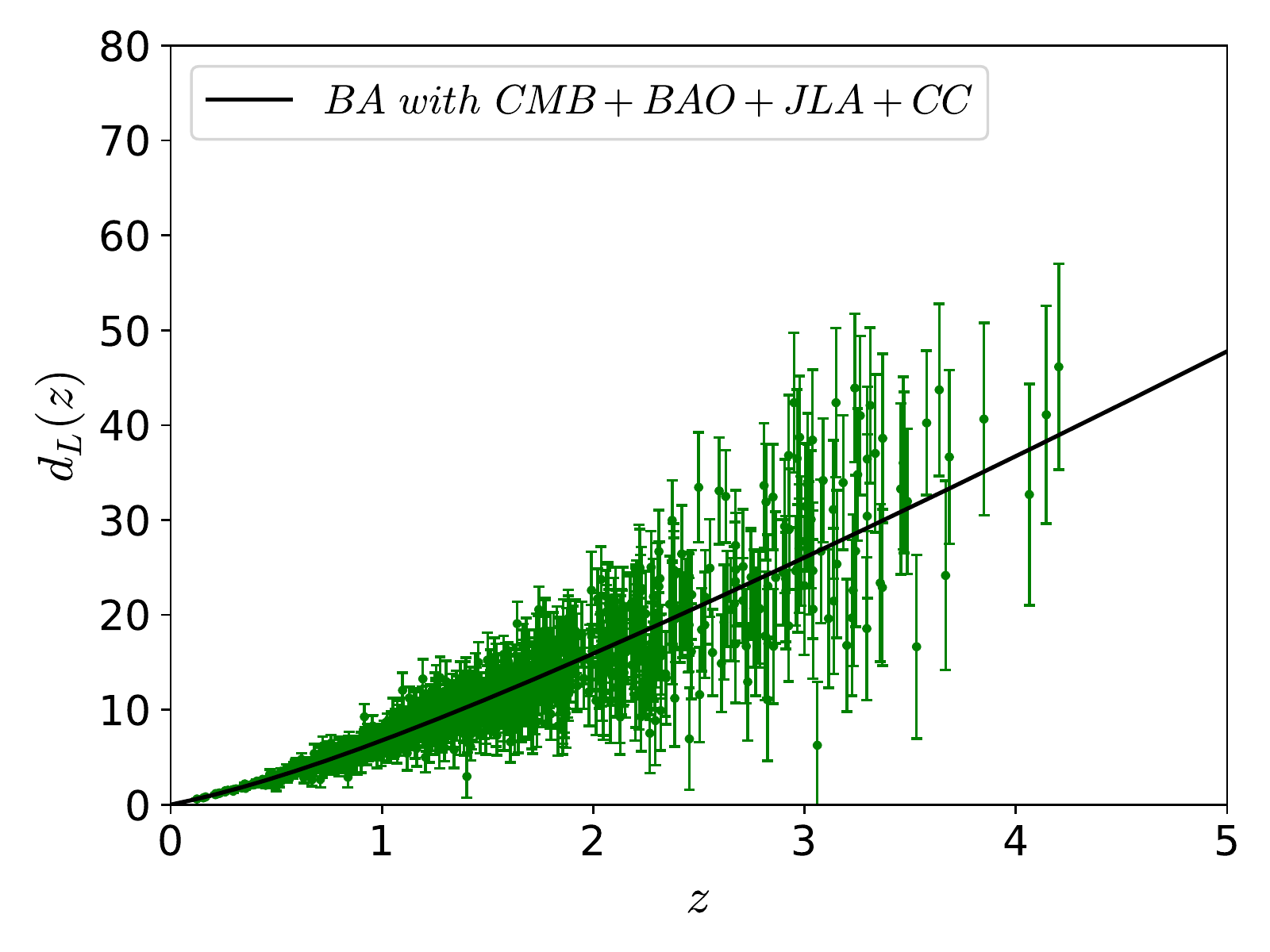}
\caption{\textit{For the fiducial BA model, we first constrain the cosmological parameters using the datasets CMB, CMB+BAO, CMB+BAO+JLA and CMB+BAO+JLA+CC and then we use the best-fit values of the parameters for ``each dataset'' to generate the corresponding GW catalogue. Following this, in each panel we show $d_L (z)$ vs $z$ catalogue with the corresponding error bars for 1000 simulated GW events. The upper left and upper right panels respectively present the catalogue ($z$, $d_L (z)$) with the corresponding error bars for 1000 simulated GW events derived using the CMB alone and CMB+BAO dataset. The lower left and lower right panels respectively present the catalogue ($z$, $d_L (z)$) with the corresponding error bars for 1000 simulated GW events derived using the CMB+BAO+JLA and CMB+BAO+JLA+CC datasets.}}
\label{fig-dL-ba}
\end{figure*}
\begin{figure*}
\includegraphics[width=0.45\textwidth]{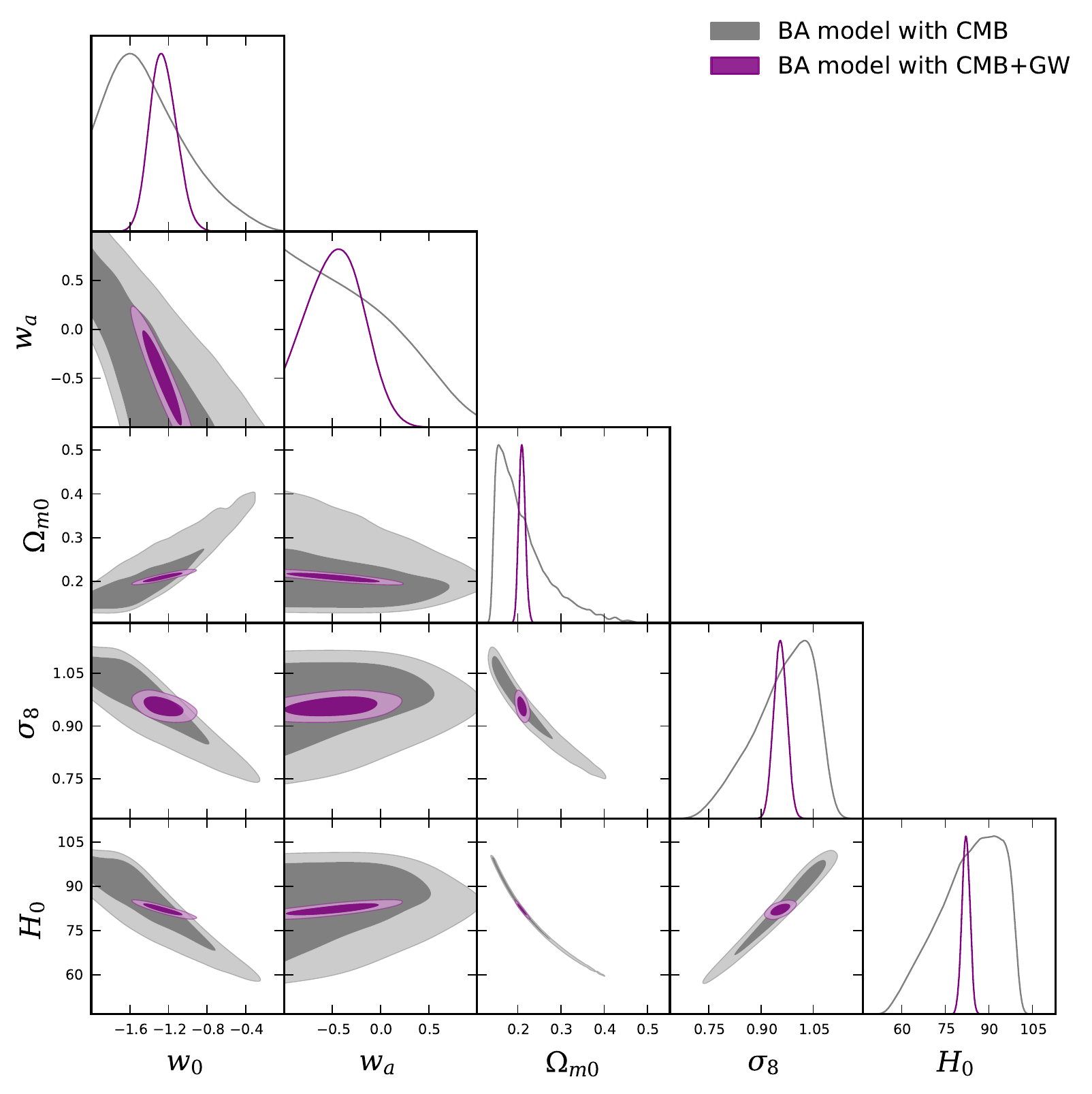}
\includegraphics[width=0.45\textwidth]{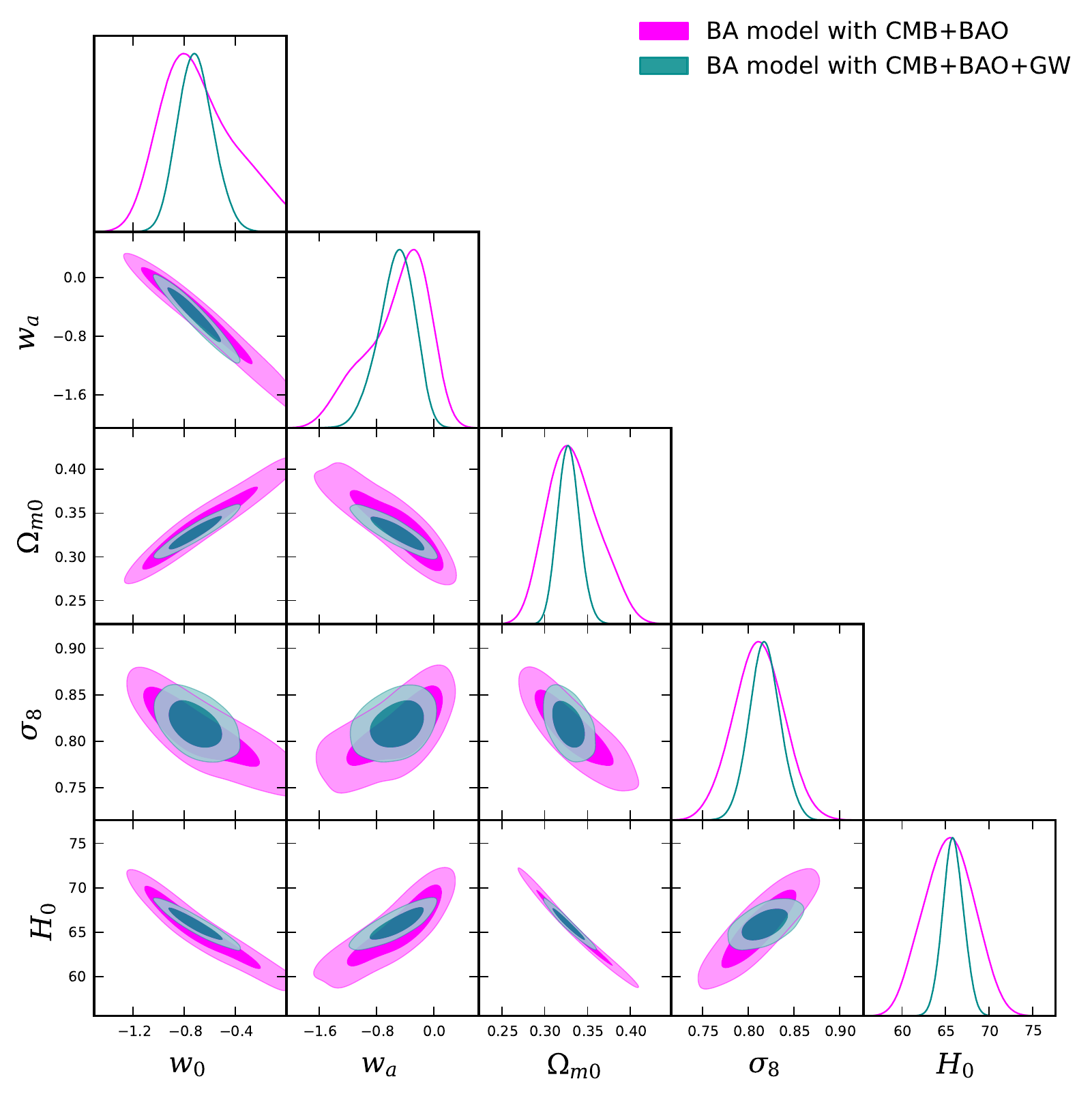}
\includegraphics[width=0.45\textwidth]{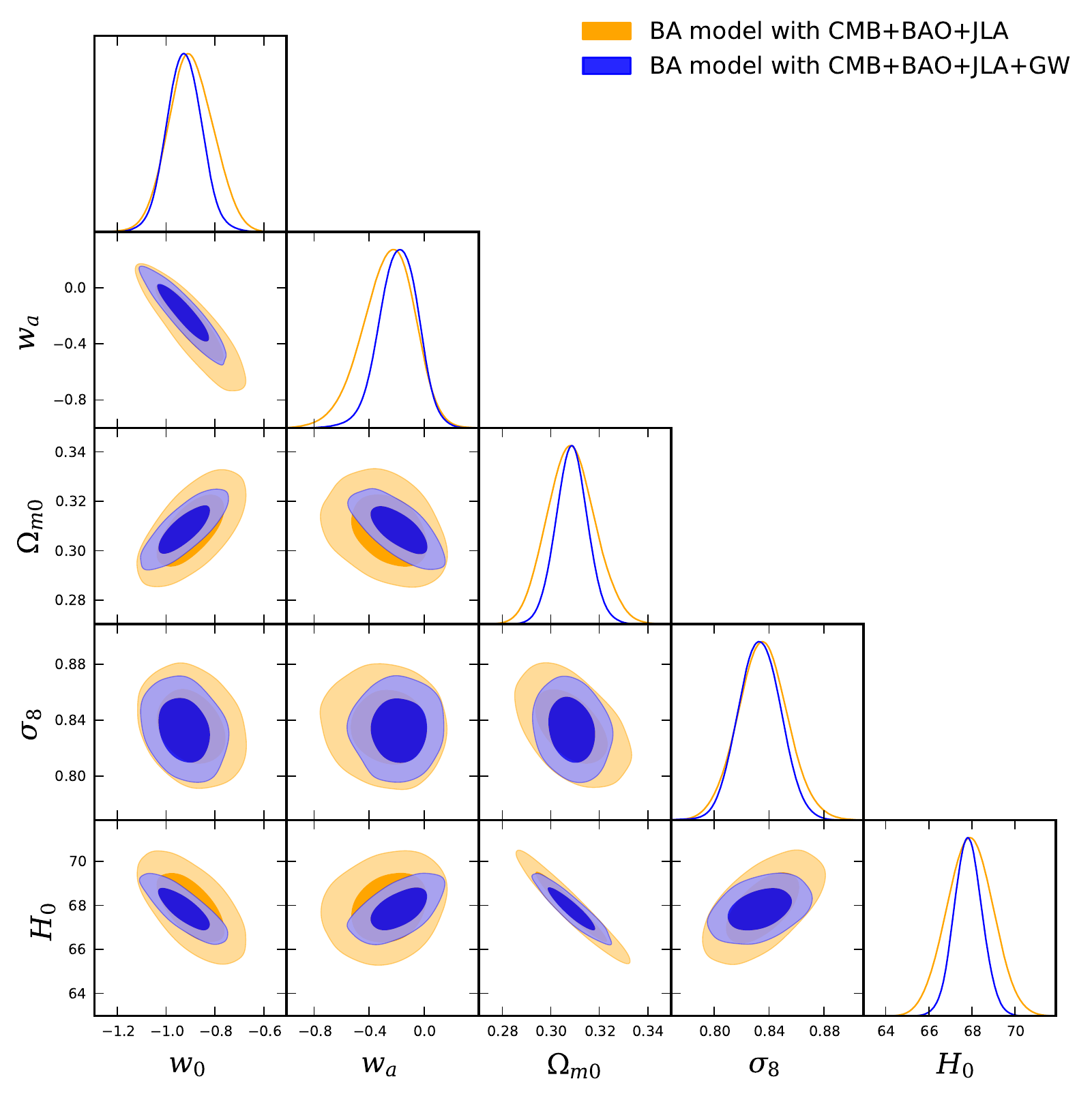}
\includegraphics[width=0.45\textwidth]{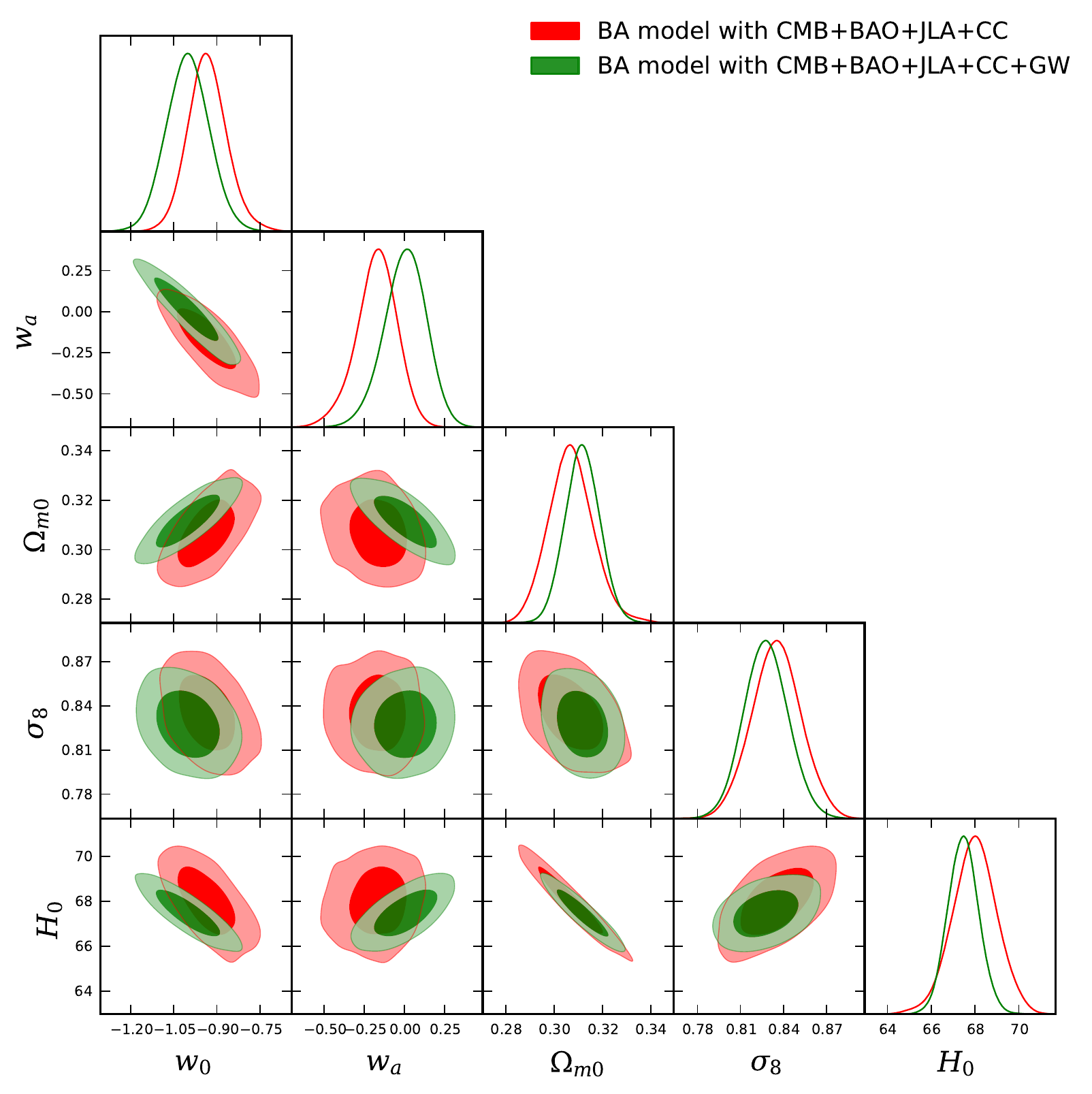}
\caption{\textit{68\% and 95\% CL contour plots for various combinations of some selected parameters of the BA model (\ref{model-ba}) using different  observational data in presence (absence) of the GW data.}}
\label{fig-contour-ba}
\end{figure*}
\begin{figure*}
\includegraphics[width=0.36\textwidth]{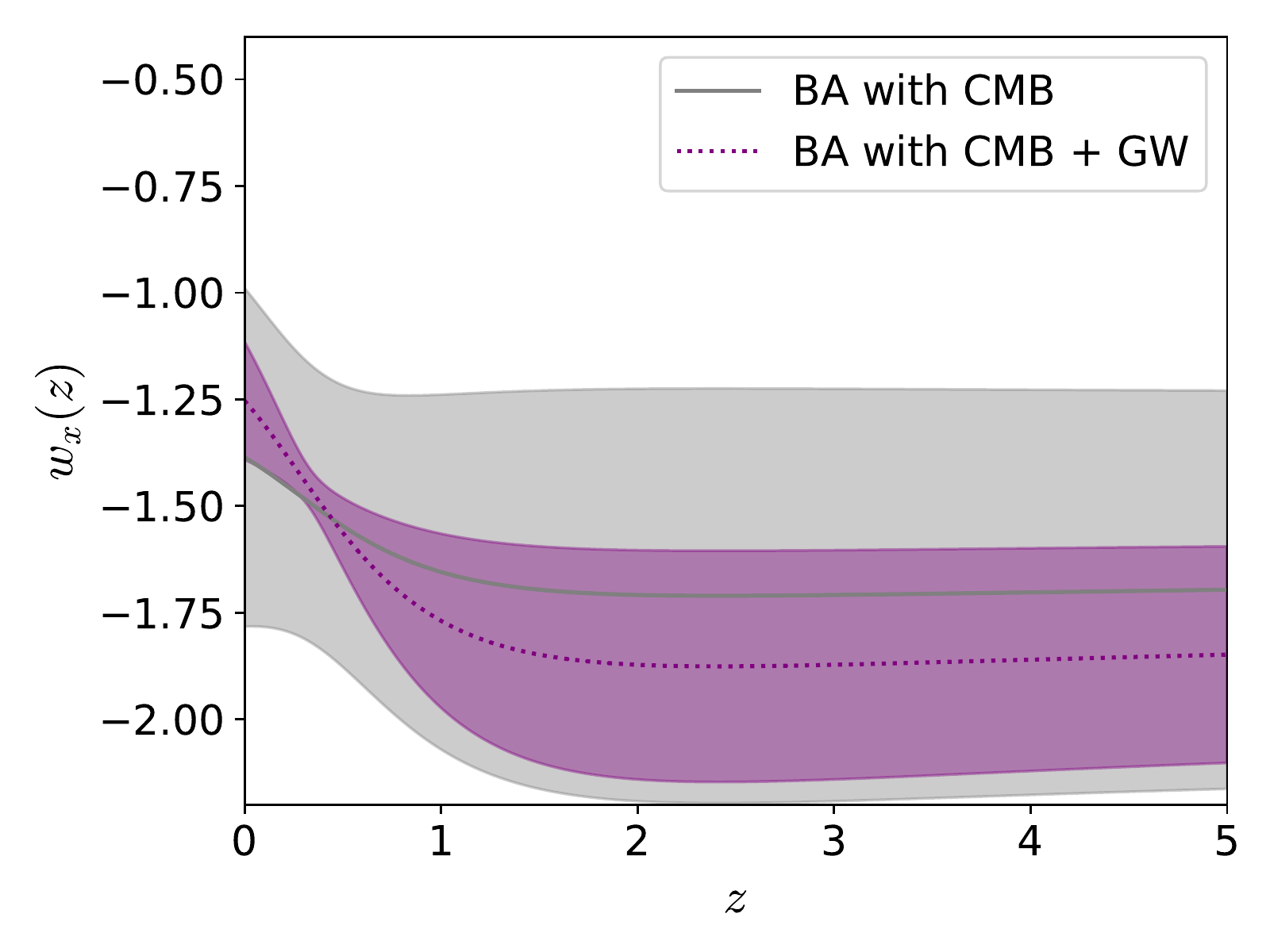}
\includegraphics[width=0.36\textwidth]{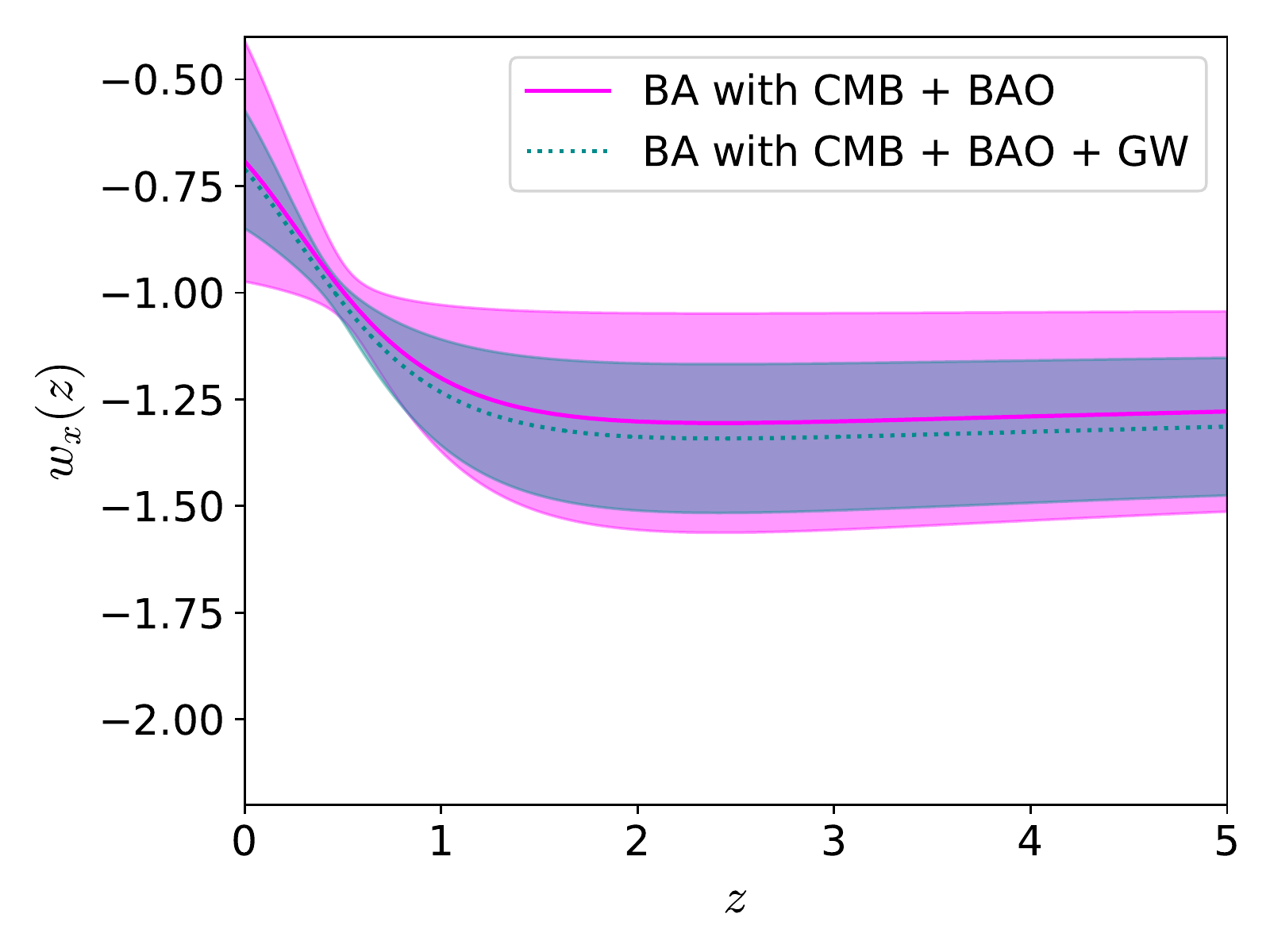}
\includegraphics[width=0.36\textwidth]{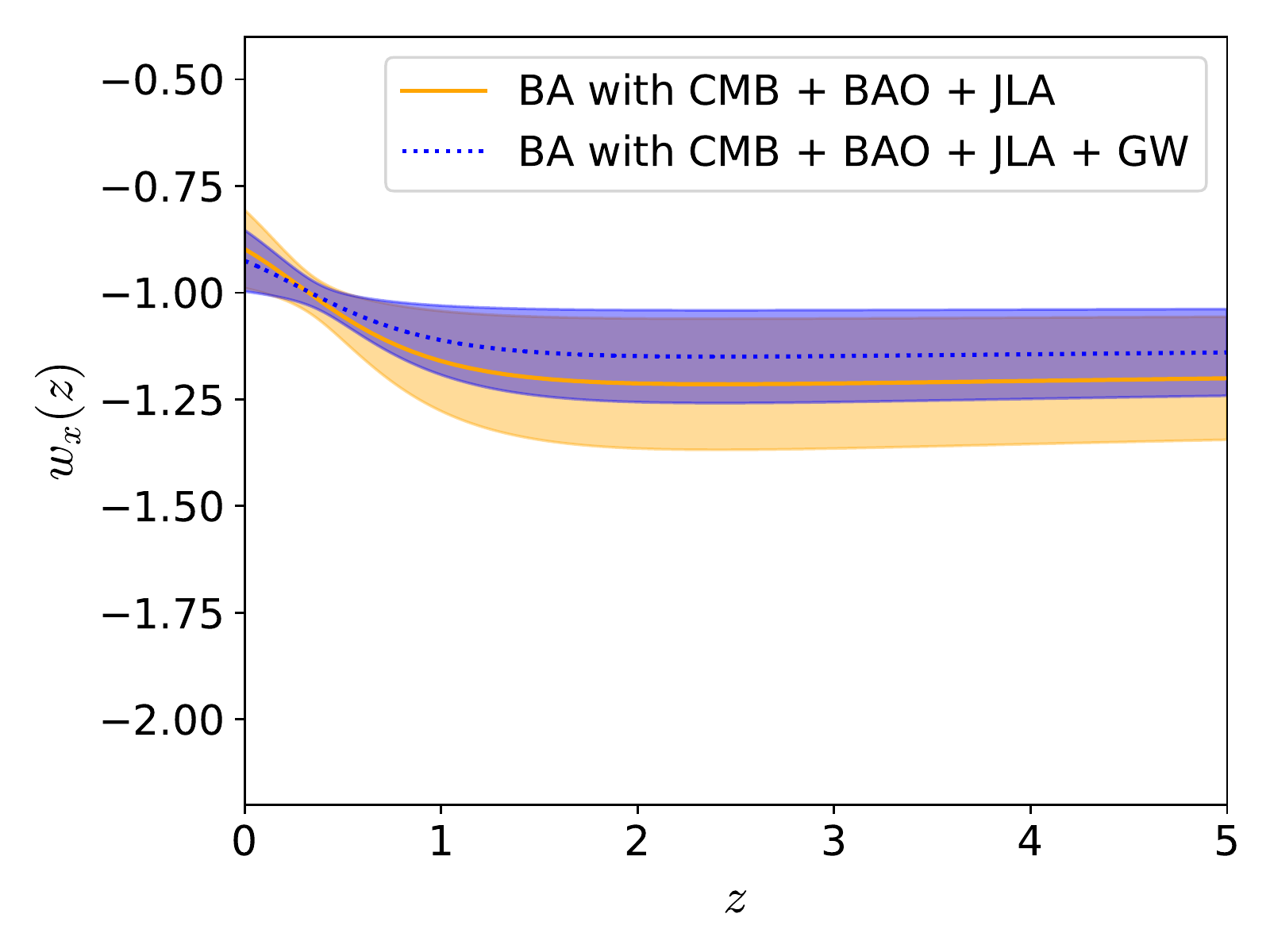}
\includegraphics[width=0.36\textwidth]{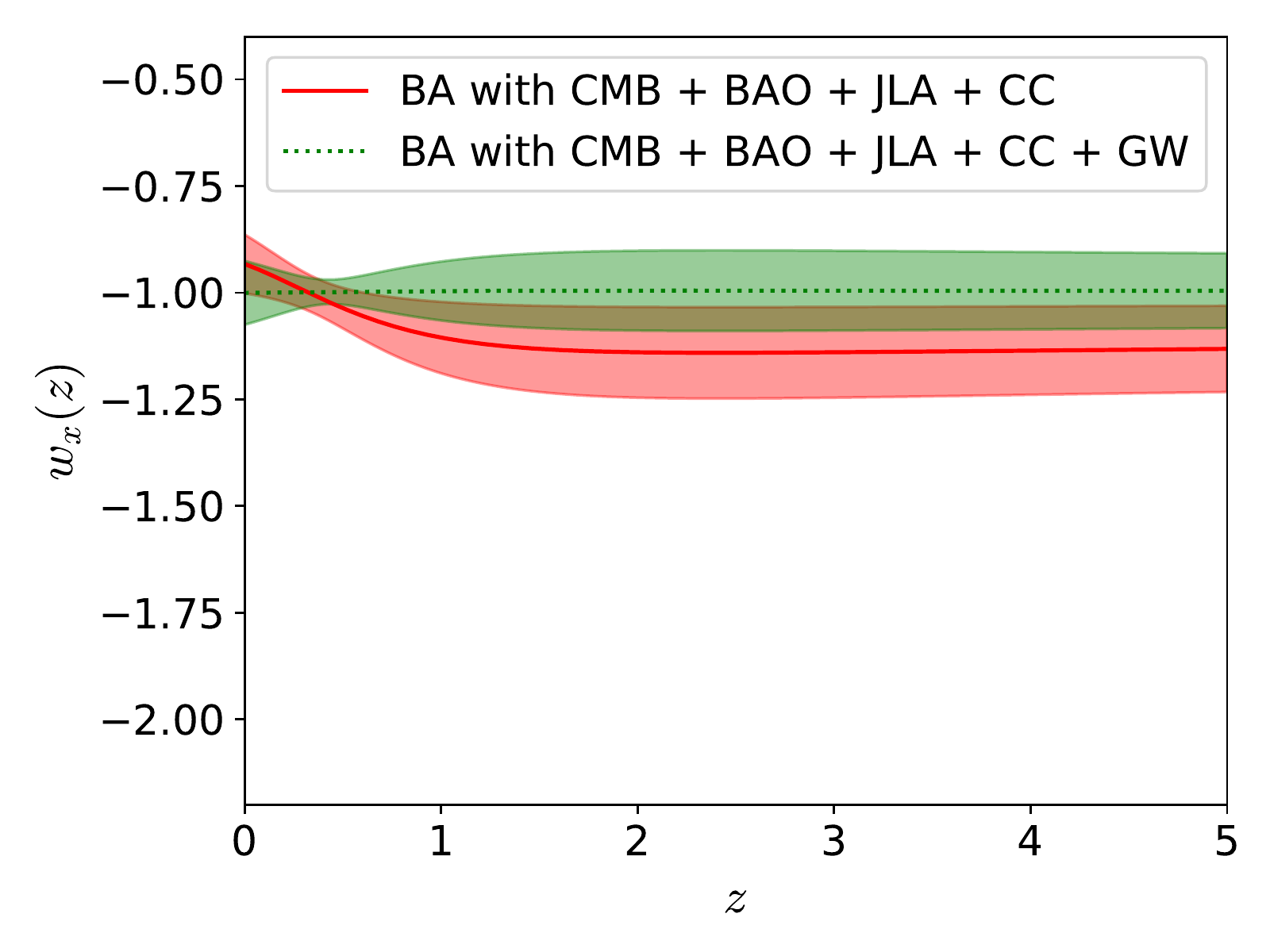}
\caption{\textit{The evolution of the dark energy equation of state for the BA parametrziation has been shown for different datasets taking the mean values of the key parameters $w_0$ and $w_a$ from the analyses with and without the GW data. The solid curves stand for the evolution of $w_x (z)$ for the standard cosmological probes while the dotted curves for the dataset in presence of the GW data. The shaded regions show the 68\% CL constraints on these two parameters.}}
\label{fig-w-ba}
\end{figure*}

\subsection{Barboza-Alcaniz parametrization}

Finally, we confront the Barboza-Alcaniz parametrization (\ref{model-ba}) following a similar pattern performed for other three dynamical DE parametrizations.

We use the standard cosmological probes, such as CMB, BAO, JLA and CC to constrain the parameter space of this model (summarized in the upper half of the Table \ref{tab:results-ba}), and then using the best-fit values of the model parameters, we have generated the GW catalogue comprising 1000 simulated GW events. In Fig. \ref{fig-dL-ba}, we have shown the relation $d_L (z) $ vs $z$ for the 1000 simulated GW events. Now, using the simulated GW events with the standard cosmological probes, we have constrained this parametrization. The summary of the observational constraints on this  model after the inclusion of the simulated GW data are shown in the lower half of Table \ref{tab:results-ba}.

In Fig. \ref{fig-contour-ba} we present
the graphical behaviour between the free parameters of the model aiming to display the effects of GW on the cosmological parameters.  From Table~\ref{tab:results-ba}, we again see that the inclusion of simulated GW data remarkably decrease the error bars on the parameters. Apart from that, this parametrization gives some interesting features that will be described soon.

Following the similar pattern, we begin the analyses using the CMB data and CMB+GW data. The results can be found in the second column of Table~\ref{tab:results-ba} and for these datasets, we have made a comparison in the top left panel of Fig.~\ref{fig-contour-ba}.
From the analyses, one can visualize that the estimations of the Hubble constant from both the datasets are quite high, similar to what we have found in previous three dynamical dark energy parametrizations. For this parametrization,  we find that, $H_0 =  83.55_{- 7.20}^{+ 14.43}$ (68\% CL, CMB) and
$H_0= 82.12_{- 1.32}^{+1.34} $ (68\% CL, CMB+GW). As one can see that inclusion of GW to CMB reduces the error bars almost by a factor of $6$ (for 68\% lower error bar on $H_0$) and almost by a factor of $10$ (for 68\% upper error bar on $H_0$). The reduction of error bars is also true for other model parameters as well. We note also that in a similar fashion, the constraints on the two key parameters of this model, namely, $w_0$ and $w_a$ are equally improved after the inclusion of GW data. However, for both the datasets, that means, CMB and CMB+GW, $w_0$ remains in the phantom regime (i.e., $w_0 < -1$) at more than 68\% CL. In fact, the highest peaks of the 1D posterior distributions for both the datasets are in the phantom regime. For a better visualization on the improvements of the parameters,   we refer to the top left panel of Fig.~\ref{fig-contour-ba}. From this figure (top left panel of Fig.~\ref{fig-contour-ba}), one can clearly see how the inclusion of GW to CMB improves the parameter space.  We also comment that the estimation of $\Omega_{m0}$ is small for both the datasets and this is also found for other three models as well.

For the next analyses with BAO, that means focusing on the combined analyses CMB+BAO and CMB+BAO+GW, we do not find anything that is worth reporting. The results can be found from the third column of Table~\ref{tab:results-ba} and the comparison between the constraints on the model parameters using various datasets are shown in the top right plot of Fig.~\ref{fig-contour-ba}.

After the inclusion of JLA, results summarized in the fourth column of Table \ref{tab:results-ba}, we find that quintessence DE (i.e., $w_0 > -1$) is preferred by the dataset CMB+BAO+JLA,  and this remains so at 68\% CL. In addition, the inclusion of GW to this dataset does not alter this conclusion meaning that within 68\% CL, $w_0 > -1$. We note that the highest peaks of the 1D posterior distributions of $w_0$ are bent towards the quintessence regime. Concerning the $w_a$ parameter, we find that the constraints are small (in magnitude) compared to the previsous datasets and also compared to previous three models as well. We see that, $w_a =  -0.263_{-0.165}^{+    0.211} $ (68\% CL, CMB+BAO+JLA) and $w_a =  -0.186_{- 0.130}^{+ 0.151}$ (68\% CL, CMB+BAO+JLA+GW). Thus, we see that both the datasets prefer $w_a \neq 0$, at least in 68\% CL. We refer to the bottom left panel of Fig. \ref{fig-contour-ba} for a comparison of the model parameters constraints obtained from different datasets.

Finally, we consider the last two combinations, namely, CMB+BAO+JLA+CC and CMB+BAO+JLA+CC+GW. In the last column of Table \ref{tab:results-ba}, we have summarized the results and in the bottom right panel of Fig. \ref{fig-contour-ba} we have compared the cosmological constraints for this parametrization obtained from both the datasets.
These analyses give some interesting results.  Concerning the present value of the dark energy equation of state, $w_0$, we see that for the dataset CMB+BAO+JLA+CC, $w_0 > -1$ and it remains so at more than 68\% CL, while after the inclusion of GW, this result is completely changed with the possibility of $w_0 < -1$ at more than 68\% CL.
More interetsingly, for the dataset CMB+BAO+JLA+CC+GW,  $w_0 = -1.000_{-    0.075}^{+    0.074}$ at 68\% CL. In adition to that, after the inclusion of GW, that means, for the dataset CMB+BAO+JLA+CC+GW, $w_a = 0.004_{-    0.123}^{+    0.142}$ (68\% CL). Thus, from the overall constraints on both $w_0$ and $w_a$ for this model, one can clearly say that forecasting with future gravitational waves data strongly hints towards the $\Lambda$CDM type cosmology, however, looking at the constraints on $w_a$ for CMB+BAO+JLA+CC+GW, the presence of large error bars imply that the observational data allow $w_a \neq 0$ as well.

Finally, using the mean values of $(w_0, w_a)$ from all the datasets,
in Fig. \ref{fig-w-ba} we depict the evolution of the dark energy equation of state $w_x (z)$ for this model. The solid lines in each plot stand for the $w_x (z)$ curve for the usual cosmological probe and the dotted lines depict the evolution of $w_x (z)$
in presence of the GW data. In each plot the shaded regions (with similar colors to the corresponding curves) present the 68\% regions for the parameters $w_0, w_a$ corresponding to each dataset (with or without the GW data). {\it A quite interesting scenario we observe from the right plot of the lower panel of Fig. \ref{fig-w-ba} (see the dotted curve in this plot) is that the mean-curve for $w_x (z)$ is exactly equal to the cosmological constant, $w_x (z) = -1$. }

\begingroup
\squeezetable
\begin{center}
\begin{table*}
\begin{tabular}{ccccccccc}
\hline\hline
Parameters & [C]~[CG] & [CB]~[CBG] & [CBJ]~[CBJG] & [CBJC]~[CBJCG] &\\ \hline

$w_0$ (CPL) & $   [-1.218_{-    0.597}^{+ 0.302}$]~[$ -1.168_{- 0.212}^{+    0.180}$] & [$   -0.524_{-    0.236}^{+    0.374}$]~[$   -0.465_{-    0.200}^{+    0.189}$]  & [$   -0.909_{-    0.123}^{+    0.095}$]~[$   -0.904_{-    0.080}^{+    0.070}$] &  [$   -0.909_{-    0.116}^{+    0.099}$]~[$ -0.902_{- 0.062}^{+    0.064}$] & \\

$w_a$ (CPL) &   [$< -0.446$]~[$ -1.081_{- 0.640}^{+    0.842}$] & [$ -1.403_{- 1.021}^{+    0.731}$]~[$   -1.523_{-    0.562}^{+    0.642}$] & [$ -0.409_{-    0.277}^{+    0.517}$]~[$   -0.256_{-0.227}^{+    0.263}$] &  [$-0.399_{-    0.297}^{+    0.423}$]~[$ -0.373_{-    0.226}^{+    0.263}$] & \\

$H_0$ (CPL) & [$ 83.06_{-    7.98}^{+   15.10}$]~[$   80.75_{-    1.92}^{+    1.71}$] & [$   64.36_{-    3.23}^{+ 2.05}$]~[$   63.77_{-    1.52}^{+    1.37}$ ] &  [$   67.94_{-    1.08}^{+    1.09}$]~[$   66.98_{-    0.55}^{+    0.55}$] &  [$   67.92_{-    1.09}^{+    1.09}$]~[$   67.72_{-    0.35}^{+    0.36}$]  \\

\hline 
\hline

$w_0$ (Log) & [$-1.058^ {+0.354}_{-0. 550}$]~[$-1.056_{-    0.196}^{+    0.179}$] & [$   -0.429_{-    0.223}^{+    0.265}$]~[$   -0.607_{-    0.186}^{+    0.172}$] & [$   -0.895_{-    0.098}^{+    0.084}$]~[$   -0.919_{-    0.085}^{+    0.071}$] & [$   -0.894_{-    0.097}^{+    0.072}$]~[$   -0.902_{-    0.078}^{+    0.057}$]  \\

$w_a$ (Log) & [$\mbox{Unconstrained}$]~[$-1.500_{-    0.571}^{+    0.718}$] & [$   -1.301_{-    0.570}^{+    0.549}$]~[$   -0.955_{-    0.388}^{+    0.543}$] & [$   -0.365_{-    0.083}^{+    0.365}$]~[$   -0.399_{-    0.172}^{+    0.251}$] & [$   -0.352_{-    0.137}^{+    0.293}$]~[$   -0.252_{-    0.089}^{+    0.220}$] \\

$H_0$ (Log) & [$   82.78_{- 8.34}^{+ 15.48}$]~[$   82.57_{-    1.65}^{+    1.66}$] & [$   63.30_{- 2.52}^{+ 1.87}$]~[$   65.19_{-    1.48}^{+    1.31}$] & [$67.93_{-    1.19}^{+    1.11}$]~[$   68.88_{-    0.76}^{+    0.73}$] & [$   67.84_{-    1.14}^{+    1.05}$]~[$   67.11_{-    0.69}^{+    0.68}$] \\

\hline 
\hline

$w_0$ (JBP) & [$   -1.423_{-    0.491}^{+    0.220}$]~[$   -1.213_{-    0.097}^{+    0.152}$] & [$   -0.692_{-    0.144}^{+    0.279}$]~[$   -0.672_{-    0.106}^{+    0.234}$] & [$   -0.932_{-    0.177}^{+    0.115}$]~[$   -0.925_{-    0.131}^{+    0.108}$] & [$   -0.893_{-    0.148}^{+    0.120}$]~[$   -0.982_{-    0.132}^{+    0.080}$ ]  \\

$w_a$ (JBP) & [$  < 0.19$]~[$   -1.614_{-    1.100}^{+    0.593}$ ] & [$   -1.618_{-    1.382}^{+    0.417}$]~[$   -1.786_{-    1.214}^{+    0.321}$] & [$-0.508_{-0.622}^{+    1.017}$]~[$   -0.683_{-    0.549}^{+    0.828}$] & [$   -0.737_{-    0.689}^{+    0.839}$]~[$   -0.029_{-    0.391}^{+    0.755}$]  \\

$H_0$ (JBP) & [$84.01_{-    7.82}^{+   13.21}$]~[$   82.73_{-    0.54}^{+    0.49}$] & [$   66.29_{-    2.26}^{+    1.58}$]~[$   66.54_{-    1.41}^{+    0.89}$] & [$ 68.07_{-    1.09}^{+    1.08}$]~[$   68.60_{-    0.60}^{+    0.61}$] & [$   67.95_{-    1.04}^{+    1.05}$]~[$   67.19_{-    0.71}^{+    0.79}$] \\

\hline
\hline

$w_0$ (BA) & [$-1.386_{-    0.556}^{+    0.203}$]~[$   -1.253_{-    0.153}^{+    0.131}$] & [$ -0.692_{-    0.374}^{+    0.215}$]~[$   -0.711_{-    0.155}^{+    0.128}$] & [$   -0.898_{-    0.090}^{+    0.093}$]~[$   -0.925_{-    0.070}^{+    0.070}$] & [$ -0.933_{-    0.066}^{+    0.064}$]~[$   -1.000_{-    0.075}^{+    0.074}$]  \\

$w_a$ (BA) & [$  < -0.038 $]~[$   -0.516_{-    0.295}^{+    0.372}$] & [$   -0.509_{-    0.282}^{+    0.577}$]~[$   -0.523_{-    0.213}^{+    0.296}$] & [$   -0.263_{-    0.165}^{+    0.211}$]~[$   -0.186_{-    0.130}^{+    0.151}$]  & [$-0.173_{-    0.109}^{+    0.137}$]~[$    0.004_{-    0.123}^{+    0.142}$]  \\

$H_0$ (BA) & [$   83.55_{-    7.20}^{+   14.43}$]~[$   82.12_{-    1.32}^{+    1.34}$] &  [$   65.41_{-    3.03}^{+    3.00}$]~[$   65.88_{-    1.17}^{+    1.20}$]  & [$   67.91_{-    1.06}^{+    1.05}$]~[$   67.83_{-    0.65}^{+    0.64}$] & [$67.98_{-    0.99}^{+   1.00}$]~[$   67.46_{-    0.70}^{+    0.69}$] \\

\hline 
\hline 
\end{tabular}
\caption{Reporting the 68\% CL constraints on some key parameters of all the dark energy parametrizations namely, $w_0$, $w_a$ and $H_0$  before and after the inclusion of GW to the standard cosmological probes. Let us note that here C $ =$ CMB, CG $=$ CMB+GW, CB $=$ CMB+BAO, CBG $=$ CMB+BAO+GW, CBJ $=$ CMB+BAO+JLA, CBJG $=$ CMB+BAO+JLA+GW, CBJC $=$ CMB+BAO+JLA+CC and CBJCW $=$ CMB+BAO+JLA+CC+GW. }
\label{tab:results-common-table}
\end{table*}
\end{center}
\endgroup

\section{Concluding remarks}
\label{sec-discuss}

The detection of gravitational waves has thrilled the scientific community by offering a new window of tests that may shine some light on the nature of gravity, dark matter and dark energy. In the present work we  investigate how GW data could bring further cosmological constraints to a class of dynamical dark energy models. In particular,  we use 1000 simulated GW data from the Einstein Telescope (we refer the readers to section \ref{sec-gw} for detailed discussions on how the GW catalogue can be generated for any fiducial cosmological model).

In order to proceed we first consider four dynamical dark energy models characterized by their equation of state, namely, Chevallier-Polarski-Linder parametrization [eqn. (\ref{model-cpl})], Logarithmic paramerization [eqn. (\ref{model-log})], Jassal-Bagla-Padmanabhan parametrization [eqn. (\ref{model-jbp})], Barboza-Alcaniz parametrization [eqn. (\ref{model-ba})] and constrain them using the standard cosmological probes such as CMB, BAO, JLA, and CC. Then considering the corresponding parametrizations as the fiducial models and using the best-fit values of the model parameters from each dataset (for a quick look at the best-fit values of the model parameters, see the tables \ref{tab:app-cpl}, \ref{tab:app-log}, \ref{tab:app-jbp} and \ref{tab:app-ba} given in Appendix A) we generate the corresponding GW catalogue for the next step. In this way, we generate 1000 simulated GW events from the present dynamical models. Now, along with the 1000 simulated GW data from the Einstein Telescope, we include the aforementioned standard cosmological probes, namely, CMB, BAO, JLA, and CC, in order to  understand the constraining power of GW data in the context of dynamical dark energy models. 

The results of these four dynamical DE parametrizations are shown in Table \ref{tab:results-cpl} (CPL), Table \ref{tab:results-log} (Logarithmic), Table \ref{tab:results-jbp} (JBP), Table \ref{tab:results-ba} (BA). Further, the comparisons between the  results of the observational constraints obtained from different cosmological datasets (with and without the GW data) have been  graphically shown in Fig. \ref{fig-contour-cpl} (CPL), Fig. \ref{fig-contour-log} (Logarithmic), Fig. \ref{fig-contour-jbp} (JBP) and Fig. \ref{fig-contour-ba} (BA).  For a quick review on some key parameters of these DE parametrizations, namely, $w_0$, $w_a$ and $H_0$, one can see Table  \ref{tab:results-common-table}.

Concerning the CPL parametrization [eqn. (\ref{model-cpl})], we find that the effects of GW are clearly pronounced, and such effects are clearly recognized if one looks at the constraints from CMB and CMB+GW datasets. In particular, one can see that, $H_0 = 83.06_{- 7.98}^{+   15.10}$ (68\% CL, CMB) and $H_ 0 = 80.75_{-1.92}^{+ 1.71}$ (68\% CL, CMB+GW). This clearly shows that the addition of GW to CMB reduces the error bars on $H_0$ in a significant way. Regarding other parameters, we also find that the dataset CMB+GW improves them compared to their constraints from CMB alone. The improvements are significantly visible from some parameters, see the 1D posterior distributions as well as the 2D contour plots shown in the upper left panel of Fig. \ref{fig-contour-cpl}. However, for other datasets, such as CMB+BAO+GW, CMB+BAO+JLA+GW and CMB+BAO+JLA+CC+GW, the effects of GW is of course seen but not much compared to what we observed for the CMB+GW dataset. Now, concerning the present value of the dark energy equation of state, $w_0$, we see that the inclusion of GW significantly improves its parameter space by reducing its error bars. For the CMB alone case we see that the highest peaks of the 1D marginalized posterior distributions of $w_0$ (upper left panel of Fig. \ref{fig-contour-cpl}) are bent towards the phantom regime while for the remaining three cases, the highest peaks of the 1D marginalized posterior distributions of $w_0$ are bent towards the quintessence regime. While statistically, within 68\% CL, all four combinations, allow $w_0 > -1$ phase. 
Regarding the remaining free parameter of this model, namely, $w_a$, its improvements after the inclusion of GW are similarly visible, see for instance the 1D posterior distributions for $w_a$ for all the datasets (see Fig. \ref{fig-contour-cpl}). We would like to remark that the highest peaks of the posterior distributions for $w_a$, are never zero, which go in favour of the dynamical dark energy equation of state. 

Now, for the Logarithmic parametrization [eqn. (\ref{model-log})] we have almost similar behaviour to the CPL model [(\ref{model-cpl})]. The improvements of the parameters after the inclusion of GW are clearly visible, see the 1D posterior distributions of various parameters and the 2D contour plots shown in Fig. \ref{fig-contour-log}. One important remark for this parametrization is that for CMB data, $w_a$ is unconstrained while the addition of GW to CMB becomes able to constrain it.

For JBP parametrization [eqn. (\ref{model-jbp})] the reduction of error bars in presence of GW follows a similar pattern as observed in CPL and Logarithmic parametrizations. We refer to Fig. \ref{fig-contour-jbp} (JBP parametrization) for a better understanding on how GW improves the cosmological constraints.
Similar effects on the two key parameters of this model, namely, $w_0$ and $w_a$ are observed and we again refer to 1D posterior distributions of the parameters in Fig. \ref{fig-contour-jbp}. For the $w_0$ parameter, its highest peaks in the 1D posterior distributions are bent towards the quintessence regime. For the $w_a$ parameter we would like to note that the final combination with GW, that means the dataset CMB+BAO+JLA+CC+GW returns its extremely lower mean value with $w_a =   -0.029_{-    0.391}^{+    0.755}$ (68\% CL) compared to its constraint obtained from the usual dataset CMB+BAO+JLA+CC: $w_a  = -0.737_{-    0.689}^{+    0.839}$ (68\% CL).

For the last parametrization, i.e., BA parametrization [eqn. (\ref{model-ba})], the improvements of the entire parameter space due to the inclusion of GW are evident from Table \ref{tab:app-ba} and from Fig. \ref{fig-contour-ba}. We also refer to Table \ref{tab:results-common-table} in order to see how the key parameters $w_0$ and $w_a$ are affected due to the inclusion of GW. We see that the behaviour of $w_a$ is different which  distinguishes it from other DE parametrizations.  We notice that due to the addition of GW to the usual cosmological probes, the mean values of $w_a$ obtained from the standard cosmological probes,
become half with improvements in the parameter space (68\% constraints: $w_a = -0.263_{-    0.165}^{+    0.211}$ for CMB+BAO+JLA, while $w_a  = -0.186_{-    0.130}^{+    0.151}$ for CMB+BAO+JLA+GW) or even smaller than half (68\% constraints: $w_a = -0.173_{-    0.109}^{+    0.137}$ for CMB+BAO+JLA+CC, whereas $w_a = 0.004_{-    0.123}^{+    0.142} $ for CMB+BAO+JLA+CC+GW). 
Finally, we would like to remark that for the last observational combination with GW, that means for CMB+BAO+JLA+CC+GW, the 68\% CL constraints on $w_0$ and $w_a$ are, $w_0 = -1.000_{-0.075}^{+    0.074}$, $w_a = 0.004_{-0.123}^{+    0.142}$. This reflects its closeness to the $\Lambda$ type cosmology, however, due to the large error bars on $w_a$, its dynamical character is certainly allowed within 68\% CL.

Thus, our results clearly indicate that the future GW data may {\it significantly affect} the cosmological parameters providing stringent constraints on them by reducing their error bars in  a remarkable way.  Although one may argue that for the above DE parametrizations, the future constraints from GW seem to prefer a  quintessential DE (for BA parametrization, the last combination CMB+BAO+JLA+CC+GW supports towards $\Lambda$-cosmology), however, this result holds as long as the underlying cosmological model corresponds (or, is close enough) to the model adopted for generating the GW mock data. 

Last but not least, considering the upcoming cosmological surveys, such as Simons Observatory Collaboration, CMB Stage-4, DESI, LSST, weak lensing, galaxy clusters, it is quite reasonable to examine the constraining power of the Einstein Telescope compared to others. A systematic and dedicated analysis for dynamical dark energy is the subject of a forthcoming work.

\bigskip

\begin{acknowledgments}
The authors gratefully acknowledge the referee for several essential comments that improved the manuscript significantly.
WY was supported by the National Natural Science Foundation of China under Grants No.
11705079 and No.  11647153. SP would like to acknowledge the financial support from the Faculty Research and Professional Development Fund (FRPDF) Scheme of Presidency University, Kolkata, India. 
DFM thanks the Research Council of Norway and the NOTUR computing facilities. This paper is based upon work from COST action CA15117 (CANTATA), supported by COST (European Cooperation in Science and Technology).
\end{acknowledgments}

\bigskip 

\appendix
\section{Best-fit values of free and derived parameters of the DE parametrizations}
\label{appendix}

In this section we show the best-fit values of the free and derived parameters of all the dark energy parametrizations that have been investigated in this work. The tables \ref{tab:app-cpl}, \ref{tab:app-log}, \ref{tab:app-jbp}, \ref{tab:app-ba} correspond to the CPL, Logarithmic, JBP and BA parametrization, respectively.  We again note that the best-fit values of those parameters summarized in the aforementioned tables were used to generate the GW catalogue.

\begingroup                                                                                                                     
\squeezetable                                                                                                                   
\begin{center}                                                                                                                  
\begin{table}                                                                                                                   
\begin{tabular}{ccccccc}                                                                                                            
\hline\hline                                                                                                                    
Parameters & CMB & CB & CBJ & CBJC \\ \hline

$\Omega_c h^2$ & $    0.1203$ & $0.1194$ & $    0.1193$ & $ 0.1189 $  \\

$\Omega_b h^2$ &  $    0.02232$ & $    0.02218$ & $    0.02227$ & $ 0.02214$ \\

$100\theta_{MC}$ &  $    1.041$  & $    1.041$  & $ 1.04055$ & $ 1.04100$ \\

$\tau$ &  $    0.060$ & $    0.051$ & $    0.0775$ & $    0.089 $ \\

$n_s$ &  $    0.9628$ & $    0.964$ & $    0.9657$ & $    0.967 $ \\

${\rm{ln}}(10^{10} A_s)$ & $    3.060$  & $    3.038$  & $    3.082$ & $ 3.108 $  \\

$w_0$ &  $   -1.905$ & $   -0.284$ & $   -0.885$ & $   -0.873 $ \\

$w_a$ &  $   -0.083$ & $   -1.927$  & $   -0.518$ & $   -0.510 $  \\

$\Omega_{m0}$ &  $    0.149$ & $    0.376$  & $    0.308$ & $    0.308$  \\

$\sigma_8$ &  $    1.072$ & $    0.760$  & $    0.834$ & $    0.841$  \\

$H_0$ &  $   97.89$ & $   61.50$ & $   67.99$ & $   67.80$ \\

\hline\hline                                                                                                                    
\end{tabular}                                                                                                                   
\caption{The table summarizes the best-fit values of the free and derived parameters of the CPL parametrization (\ref{model-cpl}). Here,  CB $=$ CMB+BAO, CBJ $=$ CMB+BAO+JLA and CBJC $=$ CMB+BAO+JLA+CC. }
\label{tab:app-cpl}                                                                                                   
\end{table}                                                                                                                     
\end{center}                                                                                                                    
\endgroup

\begingroup                                                                                                                     
\squeezetable                                                                                                                   
\begin{center}                                                                                                                  
\begin{table}                                                                                                                   
\begin{tabular}{ccccccc}                                                                                                            
\hline\hline                                                                                                                    
Parameters & CMB & CB & CBJ & CBJC \\ \hline

$\Omega_c h^2$ & $    0.1190$  & $    0.1198$  &  $    0.1193$ & $0.1194$ \\

$\Omega_b h^2$  & $    0.02215$ & $    0.02211$  & $    0.02237$  & $    0.02234$ \\

$100\theta_{MC}$ & $    1.04082$ & $    1.04056$ & $    1.04058$ & $1.04047$ \\

$\tau$ &  $    0.083 $  & $    0.080 $  & $    0.089 $ & $    0.084$  \\

$n_s$ &  $    0.9650$ & $    0.9654 $  & $    0.9688 $  & $    0.9659$ \\

${\rm{ln}}(10^{10} A_s)$  & $    3.098$ & $  3.094$ & $ 3.118 $ & $    3.099$ \\

$w_0$ &  $   -1.413$ & $   -0.651$  & $   -0.872$ & $   -0.908$ \\

$w_a$ &  $   -2.363$ &  $   -0.790$  & $   -0.392$ & $   -0.299$ \\

$\Omega_{m0}$ &  $    0.145$ &  $ 0.341$ & $ 0.311$ & $ 0.311$ \\

$\sigma_8$ & $    1.096$ & $    0.815 $ & $    0.845 $  & $ 0.838 $ \\

$H_0$ &   $   98.89$ & $   64.70$  & $   67.65 $  & $   67.67$  \\

\hline\hline                                                                                                                    
\end{tabular}                                                                                                                   
\caption{The table summarizes the best-fit values of the free and derived parameters of the Logarithmic parametrization (\ref{model-log}). Here,  CB $=$ CMB+BAO, CBJ $=$ CMB+BAO+JLA and CBJC $=$ CMB+BAO+JLA+CC. }
\label{tab:app-log}                                                                                                   
\end{table}                                                                                                                     
\end{center}                                                                                                                    
\endgroup

\begingroup                                                                                                                     
\squeezetable                                                                                                                   
\begin{center}                                                                                                                  
\begin{table}                                                                                                                   
\begin{tabular}{ccccccc}                                                                                                            
\hline\hline                                                                                                                    
Parameters & CMB & CB & CBJ & CBJC \\ \hline

$\Omega_c h^2$  & $    0.1197$  & $    0.1203$  & $    0.1175$ & $    0.1188$  \\

$\Omega_b h^2$  & $    0.02234$ & $    0.02218$  & $0.02242$ & $    0.02235$ \\

$100\theta_{MC}$  & $    1.04103$ & $1.04084$  & $1.04095$ & $    1.04082$  \\

$\tau$  & $    0.078$ & $    0.091$ & $    0.082$ & $    0.085$ \\

$n_s$ & $    0.965$ & $    0.964$  & $    0.973$  & $    0.966$  \\

${\rm{ln}}(10^{10} A_s)$  & $    3.086$ & $    3.120$ & $    3.090$ & $    3.107$ \\

$w_0$  & $   -1.687$ & $   -0.538$ & $   -0.824$  & $   -0.915$ \\

$w_a$  & $   -0.752$ & $   -2.335$  & $   -0.870$ & $   -0.499$ \\

$\Omega_{m0}$ &  $    0.162$ & $ 0.344$ & $    0.311$  & $0.310$\\

$\sigma_8$ & $    1.052$ & $    0.823$ & $    0.815$  & $    0.832$ \\

$H_0$  & $   93.96$ & $   64.55$   & $   67.21$ & $   67.58$ \\

\hline\hline                                                                                                                    
\end{tabular}                                                                                                                   
\caption{The table summarizes the best-fit values of the free and derived parameters of the JBP parametrization (\ref{model-jbp}). Here,  CB $=$ CMB+BAO, CBJ $=$ CMB+BAO+JLA and CBJC $=$ CMB+BAO+JLA+CC. }
\label{tab:app-jbp}                                                                                                   
\end{table}                                                                                                                     
\end{center}                                                                                                                    
\endgroup

\begingroup                                                                                                                     
\squeezetable                                                                                                                   
\begin{center}                                                                                                                  
\begin{table}                                                                                                                   
\begin{tabular}{ccccccc}                                                                                                            
\hline\hline                                                                                                                    
Parameters & CMB & CB & CBJ & CBJC \\ \hline

$\Omega_c h^2$ & $    0.1191$ & $    0.1183$ & $    0.1198$ & $    0.1191$ \\

$\Omega_b h^2$  & $    0.02239$ & $0.02236$ & $    0.02222$ & $    0.02237$ \\

$100\theta_{MC}$ & $    1.04095$ & $ 1.04087$ & $    1.04100$ & $    1.04094$ \\

$\tau$  & $    0.096$ & $    0.076$ & $    0.076$ & $    0.089$ \\

$n_s$ &  $    0.970$ & $    0.969$ & $    0.967$ & $    0.972$ \\

${\rm{ln}}(10^{10} A_s)$ & $    3.118$ & $    3.082$ & $    3.084$ & $    3.118$ \\

$w_0$ & $   -1.660$ & $   -0.743$ & $   -0.744$ & $   -0.910$ \\

$w_a$ &  $   -0.486$ & $   -0.395$ & $   -0.511$ & $   -0.133$ \\

$\Omega_{m0}$  & $    0.156$ & $    0.327$  & $0.322$ & $    0.318$  \\

$\sigma_8$  & $    1.078$ & $    0.806$ & $ 0.826$ & $    0.834$ \\

$H_0$ &  $   95.44$ & $   65.76$ & $   66.56$ & $   66.86$ \\

\hline\hline                                                                                                                    
\end{tabular}                                                                                                                   
\caption{The table summarizes the best-fit values of the free and derived parameters of the BA parametrization (\ref{model-ba}). Here,  CB $=$ CMB+BAO, CBJ $=$ CMB+BAO+JLA and CBJC $=$ CMB+BAO+JLA+CC. }
\label{tab:app-ba}                                                                                                   
\end{table}                                                                                                                     
\end{center}                                                                                                                    
\endgroup

%---------------------------------------------------------------------------

\end{document}